\documentclass[showpacs,preprintnumbers,prd]{revtex4}

\input{epsf}

\newcommand{\be}{\begin{equation}}
\newcommand{\ee}{\end{equation}}
\newcommand{\bea}{\begin{eqnarray}}
\newcommand{\eea}{\end{eqnarray}}
\newcommand{\bef}{\begin{figure}}
\newcommand{\eef}{\end{figure}}

\newcommand{\sm}[1]{\mbox{{\scriptsize #1}}}

\renewcommand{\phi}{\varphi}
\renewcommand{\rho}{\varrho}

\newcommand{\s}{\,{\rm s}}

\newcommand{\simge}{\gtrsim}
\newcommand{\simle}{\lesssim}
\def\trho{\tilde{\rho}}

\def\tp{\tilde{p}}

\newcommand{\laplace}{\nabla^2}

\newcommand{\half}{\frac{1}{2}}
\newcommand{\eV}{\mbox{eV}}
\newcommand{\keV}{\mbox{keV}}
\newcommand{\MeV}{\mbox{MeV}}
\newcommand{\GeV}{\mbox{GeV}}
\newcommand{\cm}{\mbox{cm}}

\renewcommand{\sec}{\mbox{s}}
\newcommand{\pc}{\mbox{pc}}
\newcommand{\kpc}{\mbox{kpc}}
\newcommand{\Mpc}{\mbox{Mpc}}
\newcommand{\G}{\mbox{Gauss}}

\newcommand{\di}{{\rm d}}

\newcommand{\Rey}{{R_e}}

\newcommand{\vA}{{\bf A}}
\newcommand{\vB}{{\bf B}}

\newcommand{\ve}{{\bf e}}
\newcommand{\vf}{{\bf f}}
\newcommand{\vk}{{\bf k}}

\newcommand{\vs}{{\bf s}}

\newcommand{\vv}{{\bf v}}
\newcommand{\vvA}{{\vv_{\sm{A}}}}
\newcommand{\vx}{{\bf x}}

\newcommand{\vAl}{{v_{\sm{A}}}}

\newcommand{\tH}{\tilde{H}}

\newcommand{\tit}{\tilde{t}}
\newcommand{\tT}{\tilde{T}}

\newcommand{\tvB}{\tilde{\vB}}

\newcommand{\tvs}{\tilde{\vs}}
\newcommand{\tvv}{\tilde{\vv}}

\newcommand{\Ah}{\hat{\vA}}
\newcommand{\Ak}{{\Ah_{\vk}}}
\newcommand{\Bh}{\hat{\vB}}
\newcommand{\Bk}{{\Bh_{\vk}}}
\newcommand{\curl}{\nabla\times}
\newcommand{\diver}{\nabla\cdot}
\newcommand{\dt}{\partial t}

\newcommand{\dfdt}[1]{\frac{\partial #1}{\dt}}

\newcommand{\dtt}{\partial\tit}
\newcommand{\dfdtt}[1]{\frac{\partial #1}{\dtt}}

\newcommand{\He}{{\cal H}}
\newcommand{\Od}{{\cal O}}

\newcommand{\Mpl}{M_{\sm{Pl}}}

\def\eps@scaling{0.70}

\def\showone#1{
  \centering
  \leavevmode
  \epsfxsize=\eps@scaling\linewidth
  \epsfbox{#1.eps}
}

\def\epstwo@scaling{0.48}

\def\showtwo#1#2{
  \centering
  \leavevmode
  \epsfxsize=\epstwo@scaling\linewidth
  \epsfbox{#1.eps} \hfil
  \epsfxsize=\epstwo@scaling\linewidth
  \epsfbox{#2.eps}
}

\begin{document}

\title{The Evolution of Cosmic Magnetic Fields:\\ From the Very Early
Universe, to Recombination, to the Present} 
\author{Robi Banerjee$^1$ and Karsten Jedamzik$^{2}$} 
\affiliation{$^1$Sharcnet Postdoctoral Fellow, Department of Physics
  and Astronomy, McMaster University, Hamilton, Ontario L8S 4M1,
  Canada} 
\affiliation{$^2$Laboratoire de Physique Math\'emathique et
Th\'eorique, Universit\'e de Montpellier II, 34095 Montpellier Cedex
5, France}

\begin{abstract}
A detailed numerical and analytical examination of the evolution of
stochastic magnetic fields between a putative magnetogenesis era
at high cosmic temperatures $T\sim 100\,\MeV$ -- $100\,\GeV$
and the present epoch is presented. 
The analysis includes all
relevant dissipation processes, such as neutrino- and photon- induced
fluid viscosities as well as ambipolar- and hydrogen- diffusion.
A simple and intuitive analytical model matching the results of 
the three-dimensional
MHD simulations allows for the prediction of pre-recombination and
present day magnetic field
correlation lengths and energy densities as a function of initial magnetic
field energy density, helicity, and spectral index.
Our conclusions are multi fold. (a) Initial primordial fields with only
a small amount of helicity are evolving into maximally helical fields
at the present. Furthermore, the simulations show a self-similarity in the
evolution of maximally helical fields implying a seemingly acausual
amplification of magnetic fields on large scales is observed.
(b) There exists a correlation between the strength of 
the magnetic field $B$
at the peak of it's spectrum and the location of the peak, given at the
present epoch by: $B\approx 5\times 10^{-12}\,\G\,(L/\kpc)$, 
where $L$ is the magnetic field
correlation length determined by the initial properties of the magnetic field.
(c) Concerning studies of generation of
cosmic microwave background (CMBR) anisotropies due
to primordial magnetic fields of $B\sim 10^{-9}\,\G$ on $\simge 10\,\Mpc$
scales, such fields are not only impossible to generate in early causal
magnetogenesis scenarios but also seemingly ruled out by
distortions of the CMBR spectrum due to magnetic field
dissipation on smaller scales and the overproduction of cluster
magnetic fields.
(d) The most promising detection possibility of CMBR distortions due
to primordial magnetic fields may be on much smaller scales at
higher multipoles $l\sim 10^6$ where the signal is predicted to be
the strongest
(e) It seems possible that magnetic fields in clusters of galaxies are
entirely of primordial origin, without invoking dynamo amplification.
Such fields would be of (pre-collapse) strength $10^{-12}-10^{-11}\,\G$ with
correlation lengths in the kpc range, and would also exist in voids of
galaxies. 
\end{abstract}


\maketitle

\setlength{\baselineskip}{12pt}

\section{Introduction}

Magnetic fields exist throughout the observable Universe. They exist
in stars, in the interstellar medium, in galaxies, and clusters of
galaxies (for a review see~\cite{beck95}), where in the latter two
environments they are often observed with $\mu\G$ strength. 
Magnetic fields likely also reside in the intergalactic medium, though
at present, their strength may only be limited by Faraday rotation
measures of distant quasars~\cite{kim91}. 
The origin of galactic- and cluster- magnetic fields is still unknown. A plausible, 
though by far not convincingly established, possibility is the
generation of magnetic seed fields and their subsequent amplification
via a galactic dynamo mechanism. Seed fields may be due to a variety of processes
(and with a variety of strengths), such as the Biermann battery within
intergalactic shocks \cite{kuls97a}, stellar magnetic fields expelled
in planetary nebulae, or during supernovae explosions,
either into the intragalactic, or in the presence of galactic outflows into
the intergalactic medium~\cite{rees87}, as well as due to quasar outflows of
magnetized plasma~\cite{furl01}. Seed fields may also be of
primordial origin with a multitude of proposed scenarios.
These include generation during first-order phase transitions (e.g.
QCD or electroweak), around cosmic defects, or during an inflationary
epoch (with, nevertheless, extremely small amplitudes), as well as
before the epoch of neutrino decoupling or recombination.  
For a review of proposed scenarios we refer the reader
to~\cite{gras01,giova03}.

The philosophy in prior studies of primordial magnetogenesis is often
(but not always) as follows.  After establishing a battery mechanism
(e.g. separation of charges and production of currents) and a \lq\lq
prescription\rq\rq\, or estimate for the final, non-linearly evolved
magnetic field strength (e.g. equipartition with turbulent flows),
subsequent evolution is approximated by simply assuming frozen-in
magnetic field lines into the plasma. Though this may be appropriate
on the very largest scales, it should be clear, that this may not be
the case on the {\em fundamental} coherence scale of the field. Here,
coupling of the magnetic fields to the gas induces non-linear cascades
of energy in Fourier space.  The characteristics of initially created
magnetic field are thus vastly modified during cosmic evolution
between the epoch of magnetogenesis and the present.  The final step
in such studies is then often to determine field strengths on some
prescribed scale (e.g. 10 Mpc) typically falling in the range
$10^{-30}\,\G \simle B \simle 10^{-20}\,\G$, inferring that this may
be sufficient to seed an efficient dynamo for the production of
galactic- and cluster- magnetic fields. This is observed in negligence
of the fact that much stronger fields on smaller scales, result not
only from a variety of astrophysical seeds, but from these very same
primordial scenarios.

Considering the likelihood of a magnetized early Universe (i.e. due to
the large number of charged particles and the multitude possible of
{\em out-of-equilibrium} processes) it should be instructive to develop
a somewhat complete picture of magnetic field evolution in the early
Universe, subsequent to the epoch of magnetogenesis. This should be
accomplished {\it irrespective} of such fields providing the seeds for
galactic fields, or not.  For example, it may be that at some later
time relatively weak field strength in galactic voids are measurable
via the propagation of the highest energy- cosmic
rays~\cite{lemo97,bert02}, or via accurate measurements of
$\gamma$-ray bursts ~\cite{plaga95}. The interpretation of such
putative measurements, which could hint to fields of primordial
origin, is then possible only if one understands the evolution of
these fields between magnetogenesis and the present.

One step in this direction has been performed by Dimopoulos \&
Davis~\cite{dimo96} as well as Son~\cite{son99} who exactly considered
such non-linear processing of magnetic fields due to
magnetohydrodynamic cascades in the early Universe. Another step has
been provided by Jedamzik {\it et al.}~\cite{jeda98} (hereafter;
JKO98) and shortly after Subramanian \& Barrow~\cite{subr98} who
considered fluid-viscosity (due to neutrinos and photons) induced
damping of magnetic fields. The study by Son, though describing
appropriately the gross non-linear features of MHD evolution, does not
properly deal with the effect of fluid viscosity~\footnote{In
particular, magnetic fields are simply assumed frozen in after the
epoch of $e^{\pm}$ annihilation.}, with the net effect of estimates of
present day coherence lengths being orders of magnitude smaller than
those we find. Moreover, this study, as most others, does not provide
explicit expressions for the final magnetic field energy. The study by
Ref.~\cite{dimo96}, on the other hand, though examining the effect of
photon drag before recombination, employs a somewhat particular model
of magnetic field coherence length growth which is not supported by
results of numerical simulations. The study by JKO98 is strictly
speaking applicable only in the linear regime (i.e. under the
assumption of a homogeneous background magnetic field), whereas
Subramanian \& Barrow also considered a limited class of non-linear
configurations. Results of both studies on the magnetic field
coherence length at recombination are identical to those found in the
non-linear analysis attempted here. Nevertheless, at intermediate
stages of evolution (i.e. well before recombination) the predicted
magnetic coherence length in these studies deviates from that found
here. Moreover, none of these works discusses the effects of ambipolar
diffusion after the epoch of recombination, neither verifies claims by
complete three-dimensional numerical simulations.

Two-dimensional- (e.g.~\cite{brand96b}) and three-dimensional-
(e.g.~\cite{chris01}) numerical simulations of magnetohydrodynamics in
the early Universe were performed in the context of maximal helical
fields. Similarly, effective three-dimensional cascade
models~\cite{son99} have also been employed. Helicity of primordial
magnetic fields could play an important role, as noted by a number of
authors~\cite{corn97,oles97,son99,field00,vach01,sigl02}, as it may
significantly speed up the growth of magnetic field coherence length,
thereby leading to potentially large magnetic fields strengths on
comparatively large scales ($\sim 1-10\,\kpc$ depending on the amount
of initial helicity). It has also been argued that a net primordial
magnetic helicity may be potentially linked to the cosmic
baryon-to-entropy ratio (e.g.~\cite{corn97,vach01}).  Adopted models
of field evolution are either appropriate to turbulent
evolution~\cite{field00}, or to viscous evolution (i.e. assuming a
drag force due to photons~\cite{sigl02}).  Before passing, we also
note studies of the effects of magnetic fields on the cosmic microwave
background radiation (CMBR), as for example, the generation of
temperature anisotropies below~\cite{subr98b,subr02} or
above~\cite{mack02} the Silk damping scale, as well as the distortions
of the CMBR Planck spectrum by magnetic field dissipation
\cite{jeda00}. These studies have also to include certain evolutionary
features of magnetic fields, but due to the largeness of the Silk
scale ($\sim 10\,\Mpc$ comoving) back reaction of the peculiar flows
generated by magnetic fields on the magnetic fields themselves. This
is in contrast to the importance of back reaction on the typically
much smaller magnetic field coherence scale (such as for the analysis
in~\cite{jeda00}) .

In this paper we attempt to provide a unified picture for the gross
features of magnetic field evolution in the early Universe. As a
function of initial conditions for the magnetic fields generated
during a putative magnetogenesis era, we predict the magnetic field
coherence length and magnetic energy density for all subsequent epochs
for fields of arbitrary strength and helicity. Our treatment
incorporates all the relevant dissipative processes, in particular,
due to photon- and neutrino- diffusion as well as free-streaming, and
due to ambipolar- and hydrogen- diffusion.

The outline of the paper is as follows. While many of the
preliminaries to the discussion, such as the equations, treatment of
Hubble expansion, and magnitudes of dissipative terms, are deferred to
the appendices, Sec.~\ref{sec:turbulence} immediately commences with a
discussion of turbulent MHD cascades and the presentation of results
of three-dimensional numerical simulations. In Sec.~\ref{sec:viscous}
magnetic field evolution in the viscous regime before recombination
(which is a regime particular to the early Universe) is discussed and
numerically simulated, whereas Sec.~\ref{sec:ambipolar} discusses the
effects of ambipolar diffusion after recombination.  The general
picture and detailed analytical results for cosmic magnetic field
evolution is developed in Sec.~\ref{sec:evolution}, whereas
Sec.~\ref{sec:summary} provides a discussion of the highlights of our
findings. In the Appendices~\ref{apx:general_MHD}
and~\ref{apx:FRW_MHD} we compile the MHD equations appropriate for the
study of magnetohydrodynamics in the expanding Universe, whereas
Appendix~\ref{apx:dissipation} complies the various dissipation terms
in the early Universe.  Details on the generation of helical fields
are given in Appendix~\ref{apx:helicity} and details on the numerical
simulations in Appendix~\ref{apx:numerics}.

\section{Turbulent Magnetohydrodynamics}
\label{sec:turbulence}

In this section we discuss general features of the evolution of
magnetized fluids in the turbulent regime (Reynolds number $\Rey \gg
1$ as applicable well before neutrino coupling and recombination),
such as the decay of energy density as well as the growth of magnetic
field coherence length. The exceedingly large Prandtl numbers
(cf. Appendix~\ref{apx:dissipation} in the early Universe allow one to
neglect dissipative effects due to finite conductivity. Further, the
generation of primordial magnetic fields in magnetogenesis scenarios
is generally believed to occur during well-defined periods
(e.g. QCD-transition). Subsequent evolution of these magnetic fields
is therefore described as a {\em free decay} without any further input
of kinetic or magnetic energy, i.e. as freely decaying MHD.  Due to
the largeness of the speed of sound in a relativistic plasma $v_s =
1/\sqrt{3}$, the assumption of incompressibility of the fluid is
appropriate during most epochs, as well as for a large range of
initial magnetic field configurations and energy densities. Exception
to the incompressibility may occur for initial conditions which result
in magnetic fields of strength $B\simge 6\times 10^{-11}\,\G$
(comoving to the present epoch, cf. Sec.~\ref{sec:evolution}) and only
after the decoupling of photons from the flow.

To verify theoretical expectations we have performed numerical
simulations of incompressible, freely decaying, ideal, but viscous
MHD.  These simulations are performed with the help of a modified
version of the code ZEUS-3D~\cite{stone92a,stone92b,clar94} in a
non-expanding (Minkowski) background. Modifications lie in the
inclusion of fluid viscosities, e.g a drag coefficient $\alpha$ as
given in Eqs.~(\ref{eq:dissipation_term}),~(\ref{eq:drag_nu}),
and~(\ref{eq:drag_gamma}).  From the discussion in
Appendix~\ref{apx:FRW_MHD} it should be clear, that for most purposes
results of numerical simulations with existing (or slightly extended)
codes with Minkowski metric, may be re-interpreted into results of MHD
in an expanding Universe with FRW metric, when rescaled variables as
given in the Appendix are considered. For details of the numerical
simulations the reader is referred to Appendix~\ref{apx:numerics}.

Incompressible MHD is described by the following equations
(for the equations of compressible MHD the reader is referred to
Appendix~\ref{apx:general_MHD} and~\ref{apx:FRW_MHD}):
\be
\dfdt{\vv} + \left(\vv\cdot\nabla\right)\vv  
  - \left(\vvA\cdot\nabla\right)\vvA =
 \vf
 \,\, ,
\label{eq:Euler_inc}
\ee
\be
\dfdt{\vvA} + \left(\vv\cdot\nabla\right)\vvA  
  - \left(\vvA\cdot\nabla\right)\vv = \nu\laplace\,\vvA\,\, ,
\label{eq:induction_inc}
\ee
where we have defined a {\em local} Alfv\'en velocity $\vvA (x) = \vB
(x) /\sqrt{4\pi (\rho + p)}$, and where $\vv$, $\vB$, $\rho$ and $p$
are the velocity, magnetic field, mass-energy density, and pressure,
respectively.  Here fluid dissipative terms in the Euler equation are
given by
\be
\vf = \left\{
  \begin{array}{lcl}
\displaystyle
\eta \laplace\,\vv & &{l_{\sm{mfp}} \ll l}
\\*[4mm]
\displaystyle 
-\alpha \vv & &  {l_{\sm{mfp}} \gg l}
  \end{array} \right.
\quad ,
\label{eq:dissipation_term}
\ee
where there exists a distinction between dissipation due to diffusing
particles, with mean free path smaller than the characteristic
scale $l_{\sm{mfp}} \ll L$, or dissipation due to a free-streaming 
(i.e. $l_{\sm{mfp}} \gg L$) {\em background} component exerting drag on
the fluid by occasional scatterings with fluid particles. Both regimes
are of importance in the early Universe as already noted in
JKO98. Note that in the computation of $\vvA$ only those particles
with $l_{\sm{mfp}} \ll L$  contribute to $\rho$ and $p$. 
An important characteristics of
the fluid flow is given by it's local kinetic Reynolds number 
\be
\Rey (l) = \frac{v^2/l}{|\vf|} = \left\{
  \begin{array}{lcl}
\displaystyle
     \frac{v\,l}{\eta}    &  \qquad & {l_{\sm{mfp}} \ll l} 
\\*[4mm]
\displaystyle
     \frac{v}{\alpha\,l} &  \qquad &  {l_{\sm{mfp}} \gg l}
  \end{array} \right.
 \quad,
\label{eq:Reynolds_number} 
\ee
with $l$ some length scale. The Reynolds number is a measure of the relative
importance of fluid advective terms and dissipative terms in the Euler
equation, given by the ratio of 
a typical dissipative time scale $\tau_d = ({l^2/\eta, 1/\alpha})$ to 
the {\it eddy turnover time scale} $\tau_{\sm{eddy}} = l/v$.
For most magnetic field configurations it is possible to define an
{\em integral scale}, $L$, i.e. the scale which contains most of the
magnetic- and fluid kinetic- energy. We will frequently refer to this
scale as the {\it coherence scale} or {\it coherence length} of the
magnetic field.  In the case of {\it turbulent} flow, with $\Rey (L)
\gg 1$ on this scale, the decay rate of the total energy is
independent of dissipative terms and only depends on the flow
properties on the integral scale.
This is in contrast to the 
decay of magnetic- and fluid- energy in the {\it viscous}
regime, $\Rey (L) \ll 1$, where the total decay rate depends on the
magnitude of viscosities. 
In the following, the dynamic evolution of magnetic fields
in the former (turbulent) regime will be studied.

\subsection{Nonhelical fields}

Consider Eqs.~(\ref{eq:Euler_inc}) and (\ref{eq:induction_inc}) with a
stochastic, statistically isotropic, magnetic field and, for the
purpose of illustration, with initially zero fluid velocities. For the
moment we will also assume that the magnetic field does not possess
any net helicity. In the limit of large Reynolds numbers on the
coherence scale, the dissipative term may be neglected on this
scale. Magnetic stresses (the third term on the LHS of
Eq.~(\ref{eq:Euler_inc})) will establish fluid motions of the order
$v\approx \vAl$ within an Alfv\'en crossing time $\tau_{\sm{A}} \approx
l/\vAl$, at which point back reaction of the fluid flow on the magnetic
fields will prevent further conversion of magnetic field energy into
kinetic energy. The resultant fully turbulent state is characterized
by close-to-perfect equipartition (in the absence of net helicity)

\be
  \langle\vv^2\rangle\,\, \approx\,\, \langle\vvA^2\rangle\,\, ,
\label{eq:equipart_inc}
\ee
between magnetic and kinetic energy. This may be seen in
Fig.~\ref{fig:ekin_emag}, which shows the decay of magnetic- and
kinetic- energy in freely decaying turbulent MHD.

Non-linear MHD processes quickly establish turbulence on scales below
the integral scale (cf. Fig.~\ref{fig:e_k_turb}). Working with Fourier
transforms (assuming statistical isotropy and homogeneity) and
defining the total magnetic- and kinetic- energy density
\be  
E \approx 
\int \di\ln\, k\, k^3 \Bigl(\langle |v_k|^2\rangle + 
\langle |v_{A,k}|^2\rangle\Bigr)
\equiv \int \di\ln\, k\, E_l\, ,
\ee
one finds that a typical
root-mean-square velocity perturbation on scale $l = 2\pi /k$ is
$v_l \approx \sqrt{k^3\langle |v_k|^2\rangle }\approx \sqrt{E_l}$. 
Note that in the above and for the remainder of this section we 
set $(\rho + p)/2 = 1$ (cf. Eq.~(\ref{eq:total_energy})), as frequently done
in studies of incompressible MHD, such that energy density has the dimension
of velocity square. 
By inspection of the Fourier transformed Eqs.~(\ref{eq:energy_decay})
and (\ref{eq:helicity_decay}) it may be seen that dissipation of
energy is dominated by flows on the smallest scales (largest $k$),
given that energy spectra $E_l$ fall not too steeply with growing
$k$. Dissipation of energy into heat thus occurs at some much smaller
scale $l_{\sm{diss}} \ll L$ (where $\Rey (l)\approx 1$). The transport
of the fluid energy from the integral scale $L$ to the dissipation
scale $l_{\sm{diss}}$ occurs via a cascading of energy from large
scales to small scales, referred to as {\it direct cascade}.

Ever since the work of Kolmogorov, as well as Iroshnikov and Kraichnan
it is known that this cascading of energy occurs as a quasi-local
process in $k$-space, with flow eddies on a particular scale $l$
breaking up into eddies of somewhat smaller scale $\sim l/2$. This
continuous flow of energy through k-space
\be
\frac{ \di E_l}{\di t} \approx \frac{E_l}{\tau_l} 
  \approx {\rm const}(k)\,\, ,
\label{eq:loc_energy_flow}
\ee
results in a quasi-stationary energy spectrum on scales $l\simle L$,
with energy flow rates approximately independent of wave vector. We
remind the reader that throughout $L$ denotes the integral scale.
Typical dissipation times $\tau_l$ are given by an eddy-turnover time
scale $\tau_l \sim \tau_{\sm{eddy},l}\approx l/v_l\sim l/(\sqrt{E_l})$
in the unmagnetized case, and $\tau_l \sim \tau_{IK,l}\approx
(l/v_l)(\vAl_{,L}/v_l)\propto l/(E_l)$ in the MHD case.  Here the latter
time scale takes account of the {\it Alfv\'en effect}, in particular,
a prolonging of energy cascading time in the case of scattering of
oppositely traveling Alfv\'en waves~\cite{irosh64,krai65}.  A
description of the turbulence as a superposition of Alfv\'en waves
seems appropriate, as the magnetic field on the integral scale
($\vAl_{,L} \gg \vAl_{,l}$) effectively provides a strong and locally
homogeneous background field on the scale $l\ll L$.  Note that,
$\tau_{\sm{eddy},L} = \tau_{\sm{IK},L}$ on the integral scale itself.

Equation~(\ref{eq:loc_energy_flow}) may thus be solved yielding
\be
\tilde{E}_k \equiv E_k/k \propto \left\{
  \begin{array}{lcl}
\displaystyle
     k^{-5/3}    & : \qquad & {\rm unmagnetized} 
\\*[4mm]
\displaystyle
     k^{-3/2}    & : \qquad &  {\rm magnetized}
  \end{array} \right.
 \quad,
\label{eq:Kolmogorov} 
\ee
where $E_k \equiv E_l$, to yield the familiar Kolmogorov- and
Iroshnikov-Kraichnan (IK) spectra on scales below the integral
scale. It is matter of current debate if spectra as proposed by IK
indeed result during MHD turbulence. Goldreich \&
Sridhar~\cite{gold95} have established that MHD turbulence is
intrinsically anisotropic with eddies elongated in the direction of
the background (integral scale) magnetic field (i.e. $k_{\parallel}
\approx k_{\perp}^{2/3}L^{-1/3}$ where $k_{\parallel}$ and $k_{\perp}$
are wave vectors parallel and perpendicular to the background magnetic
field $\vAl_{,L}$) and energy cascading more rapidly in $k_{\perp}$-space
orthogonal to the magnetic field. Though the predicted anisotropy has
been observed in numerical simulations, the predicted modification of
IK spectra, in particular, the existence of ``one-dimensional''
Kolmogorov-type spectra $\tilde{E_k}\propto k_{\perp}^{-5/3}$ has been
not~\cite{maron01}.  Rather, these spectra seem to follow the one
proposed by IK.  In contrast, M\"uller \& Biskamp~\cite{muell00} find
``three-dimensional'' energy spectra consistent with Kolmogorov but
inconsistent with IK. Inspection of Fig.~\ref{fig:e_k_turb}, which
shows such ``three-dimensional'' spectra in our $256^3$ numerical
simulations of freely decaying MHD, illustrates how difficult it is to
distinguish between exponents $-5/3$ and $-3/2$.  This is due to the
{\it inertial range} between the dissipation scale (here given by a
few times the Nyquist frequency due to numerical dissipation) and the
integral scale being rather small. Moreover, both scales do not seem
to be well defined, resulting in a small-scale spectrum more
consistent with an exponential than a power-law. This is also not
significantly changed when one proceeds to $512^3$ simulations, such
that a numerical confirmation of one, or the other, spectrum may be
premature.

Evolution of global properties of the magnetic field in freely
decaying MHD, such as total energy density and
coherence length, depend on the magnetic field spectra on scales above
the integral scale, $l>L$, and are related to initial conditions. 
Consider an
initial magnetic field with blue spectrum,
\be 
E_k\approx E_0\,\left(\frac{k}{k_0}\right)^{n} = 
E_0\,\left(\frac{l}{L_0}\right)^{-n}\quad {\rm for}\,\,\, l > L_0\,\, .
\label{eq:energy_spectra}
\ee
The scale-dependent relaxation time, $\tau_l \approx l/v_{A,l} \approx
l/\sqrt{E_l}$ (with $v_{A,l} = \sqrt{k^3\langle |v_{A,k}|^2\rangle }$)
increases with scale as $\tau_l \propto l^{1+n/2}$. Transfer of
magnetic energy to kinetic energy and a fully developed turbulent
state may only occur for times $t \simge \tau_l$. When such a state is
reached the energy on this scale decays through the cascading of
large-scale eddies to smaller-scale eddies down to the dissipation
scale. Since the relaxation time for the ``next'' larger scale $l$ is
longer, this larger scale now becomes to dominate the energy density,
i.e. becomes the integral- or coherence- scale.  This is sometimes
referred to as {\it selective decay} of modes in k-space. The
remaining energy density is then the initial energy density of modes
between the very largest scales and this next larger scale.  Given
these arguments and the initial spectrum of
Eq.~(\ref{eq:energy_spectra}) one then may derive for the time
evolution of energy and coherence length of the magnetic field
\be
E\approx E_0\, \left(\frac{t}{\tau_0}\right)^{-\frac{2n}{2+n}}\,\, 
L\approx L_0\, \left(\frac{t}{\tau_0}\right)^{\frac{2}{2+n}}
\quad {\rm no\,\,\, helicity}\, ,\quad \Rey \gg 1\,\, ,
\label{eq:turb_damp_nh}
\ee
for $t\simge \tau_0$, where $\tau_0$ is the relaxation time on the
scale $L_0$, i.e. $\tau_0\approx L_0/\sqrt{E_0} \approx L_0/v^A_{L,0}$,
and where indices $0$ denote quantities at the initial time.
For instance, for a spectral index of $n=3$ (which
corresponds to the large-scale magnetic field due to a large
number of randomly oriented and homogeneously distributed
magnetic dipoles~\cite{hoga83}) the energy density follows 
$E\propto t^{-6/5}$ which is Saffman's law known from fluid 
dynamics~\cite{saff67,lesi97}.

An increase of magnetic field coherence scale with time due to
selective decay may be observed in Fig.~\ref{fig:e_k_turb}, whereas
the decay of magnetic energy density for a variety of initial magnetic
field spectra is shown in Fig.~\ref{fig:e_on_spectra}. It can be seen
that initial spectra with larger $n$ indeed lead to a more rapid
decrease of energy with time as predicted by
Eq.~(\ref{eq:turb_damp_nh}). Nevertheless, comparison of the
theoretically expected decay exponents
(cf. Eq.~(\ref{eq:turb_damp_nh})) to the numerically found exponents
(cf. Fig.~\ref{fig:e_on_spectra}) indicate slight
differences. Generally, our numerical simulations result in a slower
energy decay than predicted by Eq.~(\ref{eq:turb_damp_nh}).  For
example, the theoretically predicted damping exponent, i.e. $E(t)
\propto t^{-\gamma}$, for a $n = 3$ initial energy spectrum is $\gamma
= 1.2$ whereas the best fit of our numerical simulations gives $\gamma
\approx 1.05$.  It is not easy to find a physical explanation for
this, as it would entail an additional {\it with time increasing}
slow-down of relaxation at large scales $l\simge L$ and/or slow-down
of energy dissipation of already turbulent modes at small scales
$l\simle L$. In either case, to explain such a phenomenon a quantity
with physical dimension of length or velocity, which has not yet
entered the analysis, should exist. Given that the assumed initial
magnetic field distribution is statistically self-similar on different
scales, and that helicity is negligible, this quantity may only be the
dissipation length and/or length of the simulation box.  Whereas the
latter is a complete numerical artifact, the former is so widely
separated from the integral scale during most periods of the high
Reynolds number flow in the early Universe, that we expect it not to
influence the dynamics on the integral scale. We have noted, that
spectra at late times show a peak region $\Delta L$ quite spread, and
are likely only marginally resolved by the simulations. In any case,
larger numerical simulations are required to address this effect seen
also by others (e.g.~\cite{MacLow:1998,chris01}).

\bef
\showone{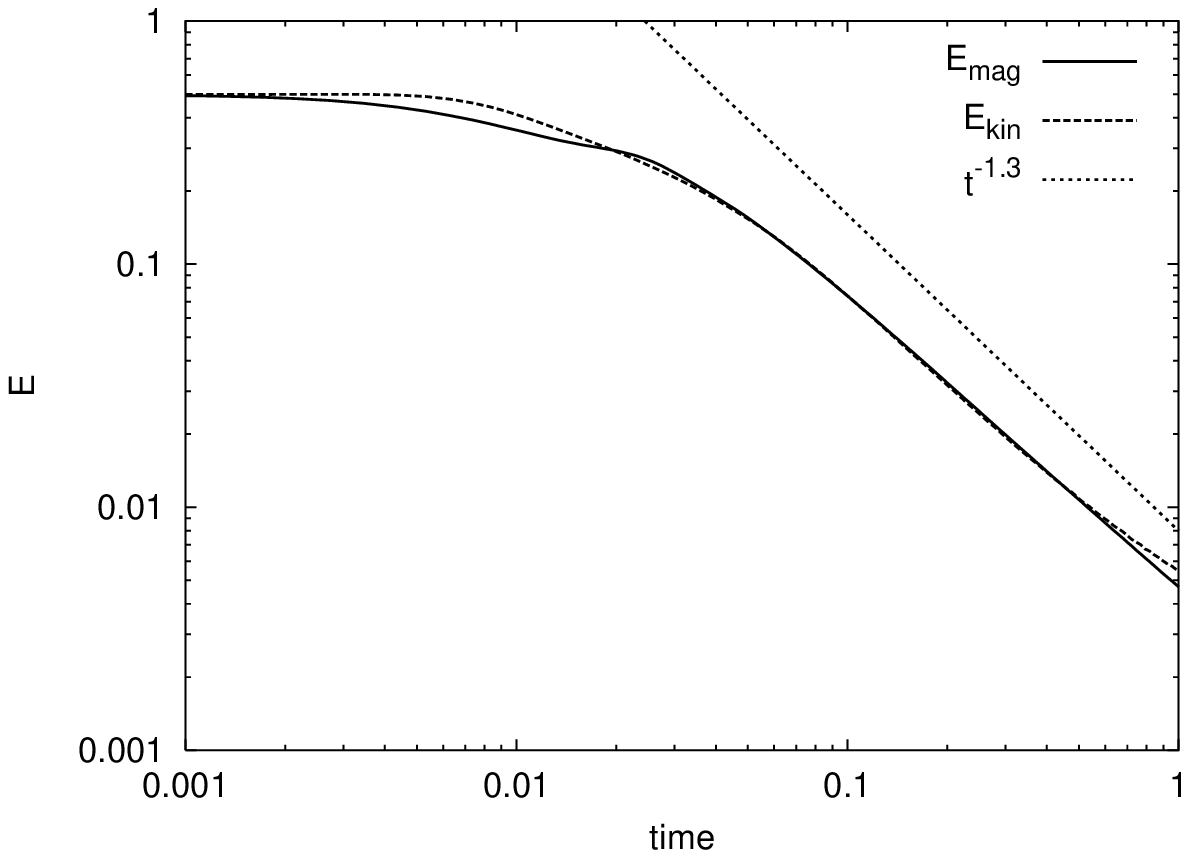}
\caption{Comparison of the time evolution of the magnetic (solid line)
and the kinetic (dashed line) energy in the turbulent regime ($\Rey
\gg 1$) for a magnetic field without initial helicity. For comparison,
also the theoretical damping law, $E \propto t^{-1.3}$, is shown
(dotted line). Here, the simulation was performed on a mesh with
$128^3$ grid points and the magnetic field is excited up to $k_c
\approx 16$ with a spectral index $n \approx 4$
(cf. Eqs.~(\ref{eq:energy_spectra}) and (\ref{eq:turb_damp_nh} and
Appendix~\ref{apx:numerics}).}
\label{fig:ekin_emag}
\eef

\bef
\showone{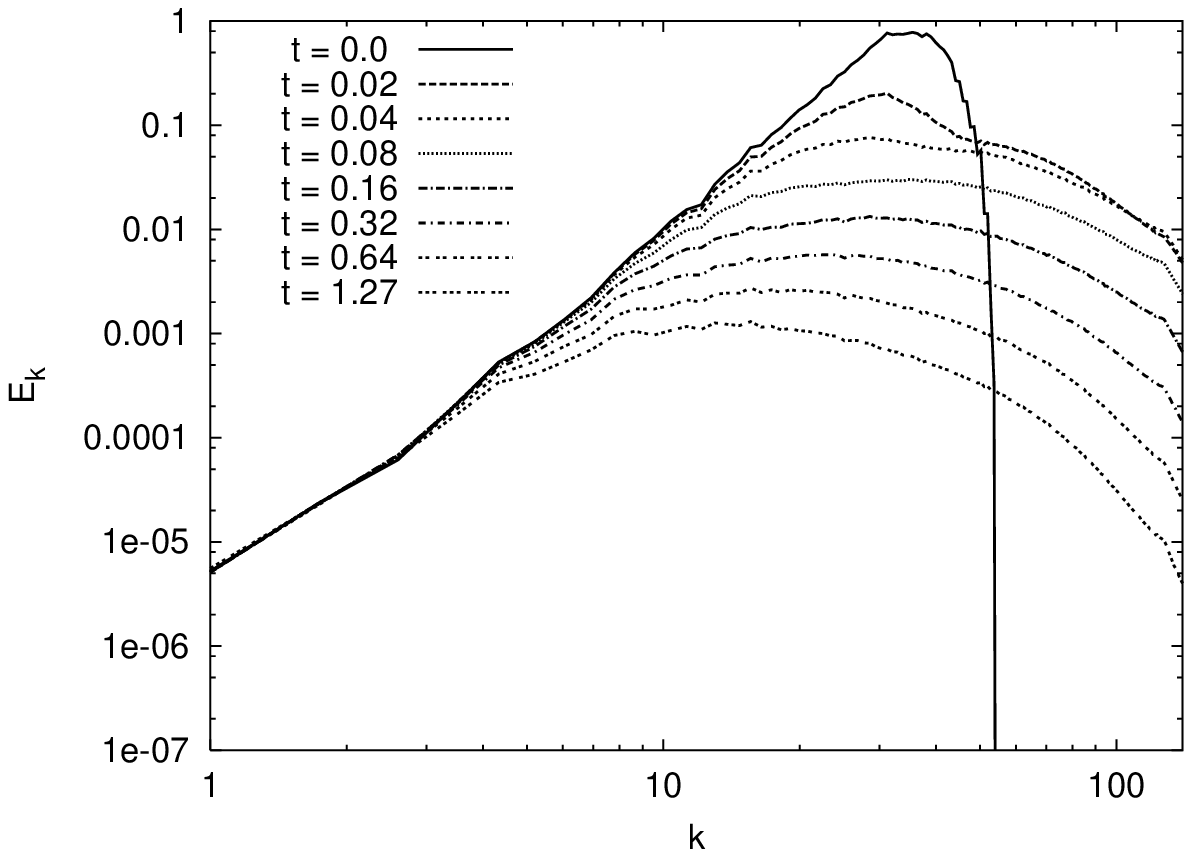}
\caption{Evolution of magnetic energy spectra in the turbulent regime
for a magnetic field with no initial helicity. Here, the spectral
index of the initial energy spectra is $n \approx 4$.  Note that $E_k$
as opposed to $\tilde{E}_k$ is shown (cf. Eq.~(\ref{eq:Kolmogorov})}
\label{fig:e_k_turb}
\eef

\bef
\showone{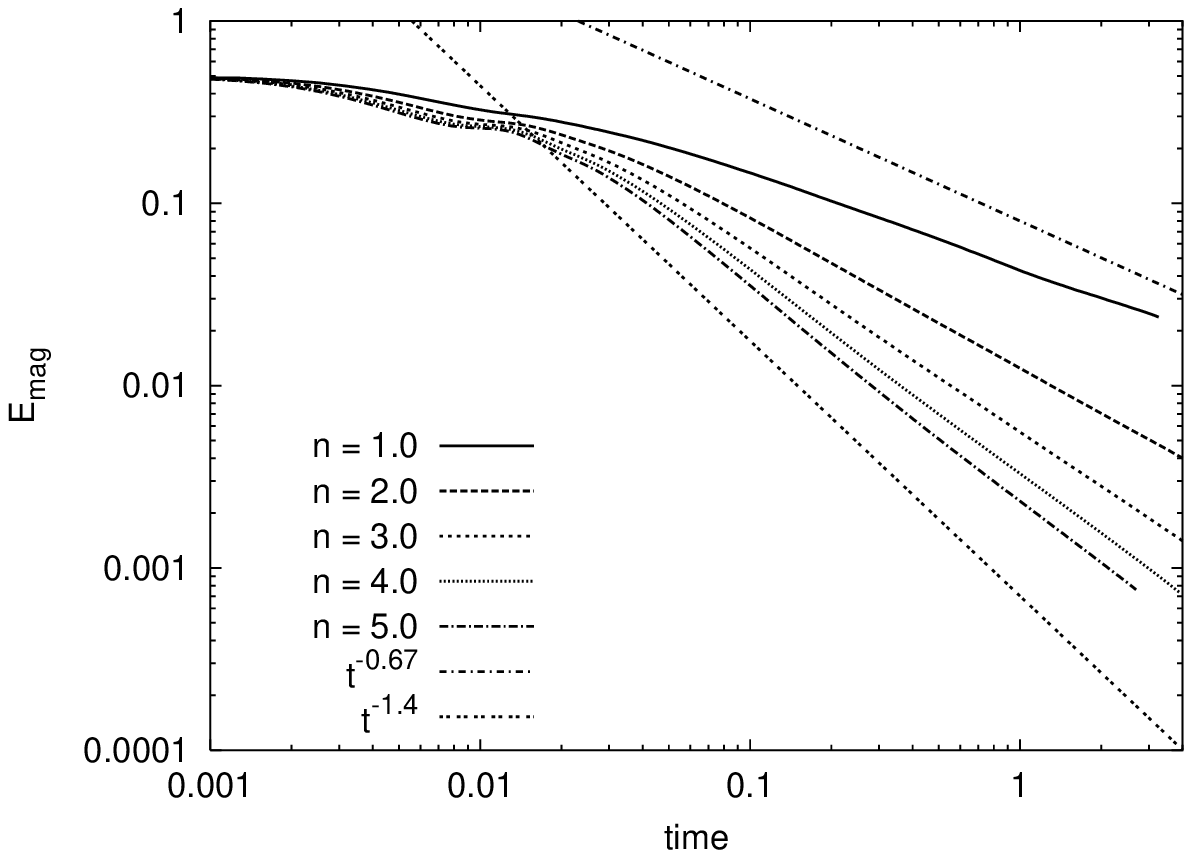}
\caption{The evolution of the magnetic energy in the turbulent regime
  for different initial energy spectra $n$, where $E_k = k^3\,|b_k|^2
  \propto k^n$ with a cut-off $k_c \approx 32$. Here, the initial
  magnetic field is non-helical. In this 
  case, the damping law depends on the spectral index
  (cf. Eq.~(\ref{eq:turb_damp_nh})). For comparison, the
  theoretical predicted damping laws for $n = 1$ ($E \propto t^{-0.67}$)
  and for $n = 5$ ($E \propto t^{-1.4}$) are also shown.}
\label{fig:e_on_spectra}
\eef

\subsection{Helical fields}
\label{sec:helical_fields}

We have so far studied the evolution of a statistically isotropic and
homogeneous magnetic field in the absence of net helicity (see
appendices~\ref{apx:dissipation} and \ref{apx:helicity} for the
definition and dissipation of magnetic helicity).  Given that magnetic
helicity should be an ideal invariant in the early Universe, and that
magnetic fields with even small initial net helicity ultimately reach
maximal helicity density
\be
\He \simle\, \He_{\sm{max}} \approx\, \langle B^2\, L\rangle\, \approx\, 
(8\pi)\,E\,L\, , 
\label{eq:Hel_by_E}
\ee
it should be of importance to also study the maximally helical
case. Note that, a maximally helical state is reached during the
course of MHD turbulent evolution due to a slower decay of the helical
component of fields as compared to the non-helical one
(cf. Eq.~(\ref{eq:turb_damp_nh}) and Eq.~(\ref{eq:turb_damp_h})
below). When maximal helicity is reached magnetic field evolution is
significantly altered with respect to the case of zero, or sub-maximal
helicity.  Fig.~\ref{fig:gamma_by_time} shows the results of $256^3$
simulations of the evolution of the ratio between kinetic- and
magnetic- energy density $\Gamma$ assuming initial conditions of a
maximally helical field and negligible velocity perturbations.  After
a relaxation time of the order of the Alfv\'en crossing time
$L/\vAl_{,L}$ over the integral scale a quasi-steady state with
constant $\Gamma\approx 0.2$ develops. Note that in contrast to the
turbulent, non-helical case, full equipartition is not reached.  The
associated spectrum of $\Gamma$ is shown in Fig.~\ref{fig:gamma_k}
showing that at the integral scale kinetic energy density is always
smaller than magnetic energy density. Though not apparent from the
figure, an integral scale (i.e. the scale of maximum energy density)
defined for kinetic energy density only $L^{\sm{kin}}$, trails the
integral scale for total energy density with time,
i.e. $L^{\sm{kin}}/L < 1$ with a ratio approximately constant in time.
Magnetic field spectra for this simulation are shown in
Fig.~\ref{fig:e_k_turb_h}. It is seen that {\em inertial range}
magnetic spectra at $l \simle L$ are well-described by power-laws over
a limited range in k-space. For this exponent ($n = 4$) we find
$E_k\propto k^{\beta}$, $\beta \approx -1.7$, significantly steeper
than either Komogorov or IK.

Fig.~\ref{fig:e_k_turb_h} also illustrates the intriguing property of 
self-similarity of spectra at different times. This phenomenon of 
self-similarity has also been observed by~\cite{chris01}. 
Magnetic field amplification on very large scales occurs even at times
much shorter than the typical relaxation time for magnetic fields
(i.e. Alfv\'en crossing time) on these scales, indicating the
topological constraint (by helicity) imposed on the field
evolution. Note that if magnetic fields on large 
scales would not be enhanced, magnetic coherence length could not grow
with time, as generally the initially existing energy density on
large scales would not suffice to keep $\He$ constant. 
Having performed simulations of maximally helical fields with different
initial spectral indices $n$  
we have noted that though the amplitude of large-scale magnetic field
grows with time, the spectral index of the magnetic field
configuration on large scales seems to be approximately preserved.

There seems to be a misconception in the literature (see,
e.g. \cite{field00,vach01}) that maximally helical fields do not
dissipate energy via excitation of fluid flows and the subsequent
dissipation of these flows due to fluid viscosities. 
It is argued, that maximally helical fields with a fairly peaked spectrum
are essentially force-free (i.e. $\vvA\times (\nabla\times\vvA )
\approx 0$) and may thus not excite fluid flows. Note that if this
indeed would be the case, Eq.~(\ref{eq:induction_inc}) would imply trivial
magnetic field evolution $\vB = {\rm const}$ for initially zero velocity
fluctuations and resistivity. 
Though the magnetic stresses in the Euler equation are indeed smaller
for a maximally helical field as compared to a non-helical field of
similar strength, an increase of magnetic coherence length
and the continuous excitation of sub-equipartition
fluid flows are observed in our simulations. 
In the limit of large Prandtl number, dissipation of these flows by
fluid viscosity will then provide the main dissipation of magnetic
field energy. 

The decay rate of total magnetic energy in freely decaying MHD
turbulence of maximally helical fields may be well approximated by the
decay rate of energy on the integral scale
\be
\frac{\di E}{\di t} \approx \frac{E}{\tau_L} 
\approx \frac{E^{3/2}}{L}\Gamma\sim  \frac{E^{5/2}}{\He}\Gamma\, ,
\ee
where $\tau_L \approx L/\vAl_{,L}$, and Eq.~(\ref{eq:Hel_by_E}) for a
maximally helical field has been employed in the second step. Since
$\He$ and $\Gamma$ (see Appendix~\ref{apx:dissipation} and
Fig.~\ref{fig:gamma_by_time}) are constant it is straight forward to
derive the power-law exponents for the decay of energy and growth of
coherence length with time
\be
E\approx E_0\,\left(\frac{t}{\tau_0}\right)^{-2/3}\,  
\quad L\approx L_0\,\left(\frac{t}{\tau_0}\right)^{2/3}\, 
\quad {\rm maximal\,\,\, helicity}\, ,\quad \Rey\gg 1\,\, ,
\label{eq:turb_damp_h}
\ee
for $t\simge \tau_0 \approx L_0/\sqrt{E_0} \approx L_0/v^A_{L,0}$,
yielding a predicted decay which is independent of the spectral index
of the large-scale magnetic field.  The correctness of
Eq.~(\ref{eq:turb_damp_h}) has been recently questioned by Biskamp \&
M\"uller~\cite{bisk99}. These authors advocate a decay of kinetic
energy with time as $\Gamma\propto E/\He$, yielding a modified
Eq.~(\ref{eq:turb_damp_h}) $\di E/\di t \sim E^3/\He^{3/2}$, and
energy decay $E\propto t^{-1/2}$. We note here that a decay of
$\Gamma$ was not found in our simulations. Moreover, a relationship
$\Gamma\propto E/\He$ is dimensionally incorrect, and must be modified
by an as yet unknown quantity of dimension length.  Due to the absence
of a physically well-motivated choice for this quantity (other than
$l_{\sm{diss}}$ or $L_{\sm{box}}$), we suspect their results to be an
artifact of limited resolution. In particular, Biskamp \&
M\"uller~\cite{bisk99} observe a decay in $\Gamma$ only at late times,
when the coherence scale has already moved dangerously close to
$L_{\sm{box}}$.  Note, that larger kinetic (numerical) viscosities
result in larger magnetic dissipation times
(cf. Sec.~\ref{sec:viscous}).  Therefore, the rather moderate Reynolds
numbers ($\Od(10^3)$) achieved in numerical simulations could be
responsible for the slower decay rates found in these simulations.

Fig.~\ref{fig:e_on_spectra_h}
shows the total magnetic energy as a function of time for a variety
of maximally helical magnetic fields of different initial spectral
index. With the exception of the rather red spectrum $n=1$, for which
the Fourier transform of helicity is not peaked in k-space, the decay
of energy seems to be indeed approximately independent of 
spectral index. Residual dependencies on $n$ may possibly be
associated with the non-conversation of helicity as shown in
Fig~\ref{fig:h_on_spectra}. This dissipation of helicity in our 
simulations is due to numerical dissipation at the Nyquist
frequency. Similar to the case of non-helical fields, the decay slopes
observed in the simulations are somewhat shallower than those predicted by
Eq.~(\ref{eq:turb_damp_h}). For example, for $n=5$ we find a damping
exponent  of $\gamma \approx 0.5$ (coincidentally agreeing with
~\cite{bisk99}). 
Arguments very similar to those presented at the end of the previous
paragraph, in particular, the absence of a quantity of dimension
length or velocity beyond those employed in
Eq.~(\ref{eq:turb_damp_h}), makes us believe this deviation to be
unphysical. 

\bef
\showone{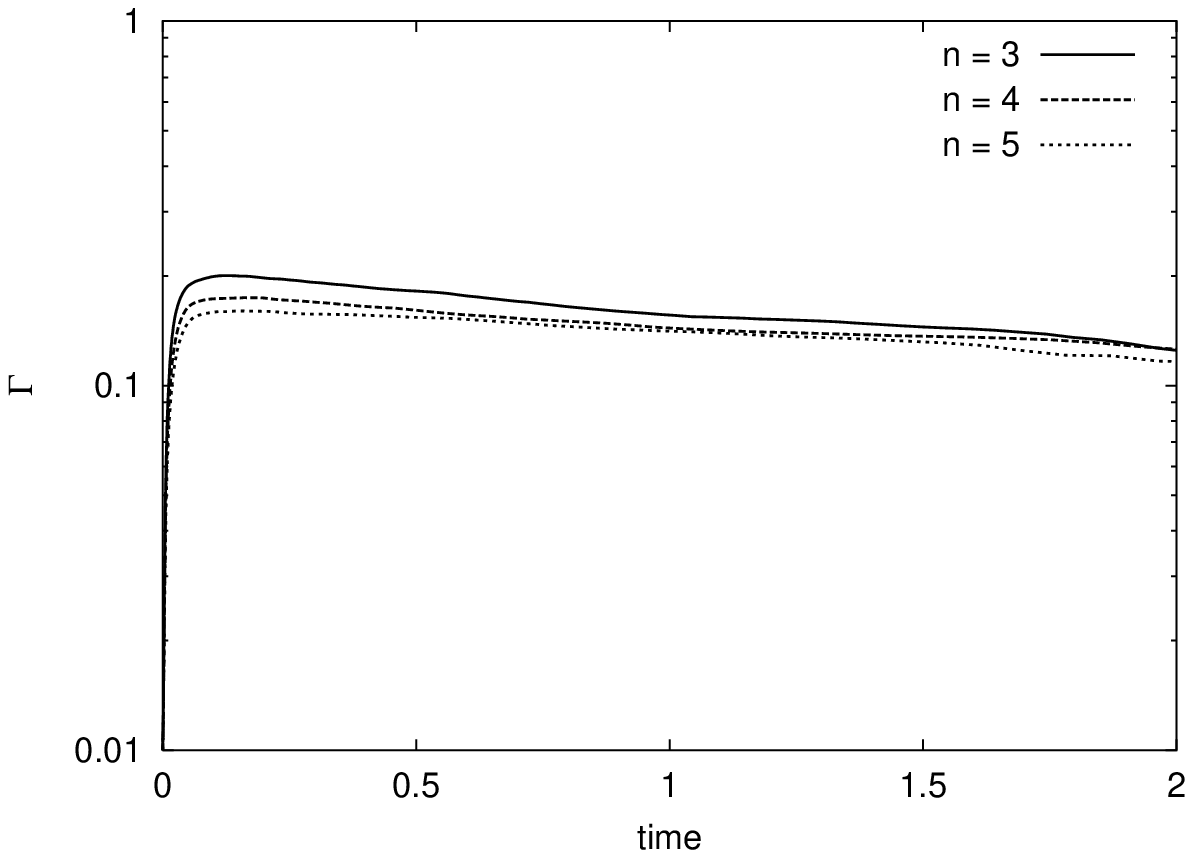}
\caption{Time evolution of 
$\Gamma = E_{\rm kin}/E_{\rm mag}$
for maximal helical magnetic fields with different spectral indices
$n$ in the turbulent regime. The initial kinetic energy is set to
$10^{-4}\,E_{\rm mag}$. The ratio $\Gamma$ is nearly constant in
time, although, equipartition of kinetic and magnetic energy is not
established for helical magnetic fields.}
\label{fig:gamma_by_time}
\eef

\bef
\showone{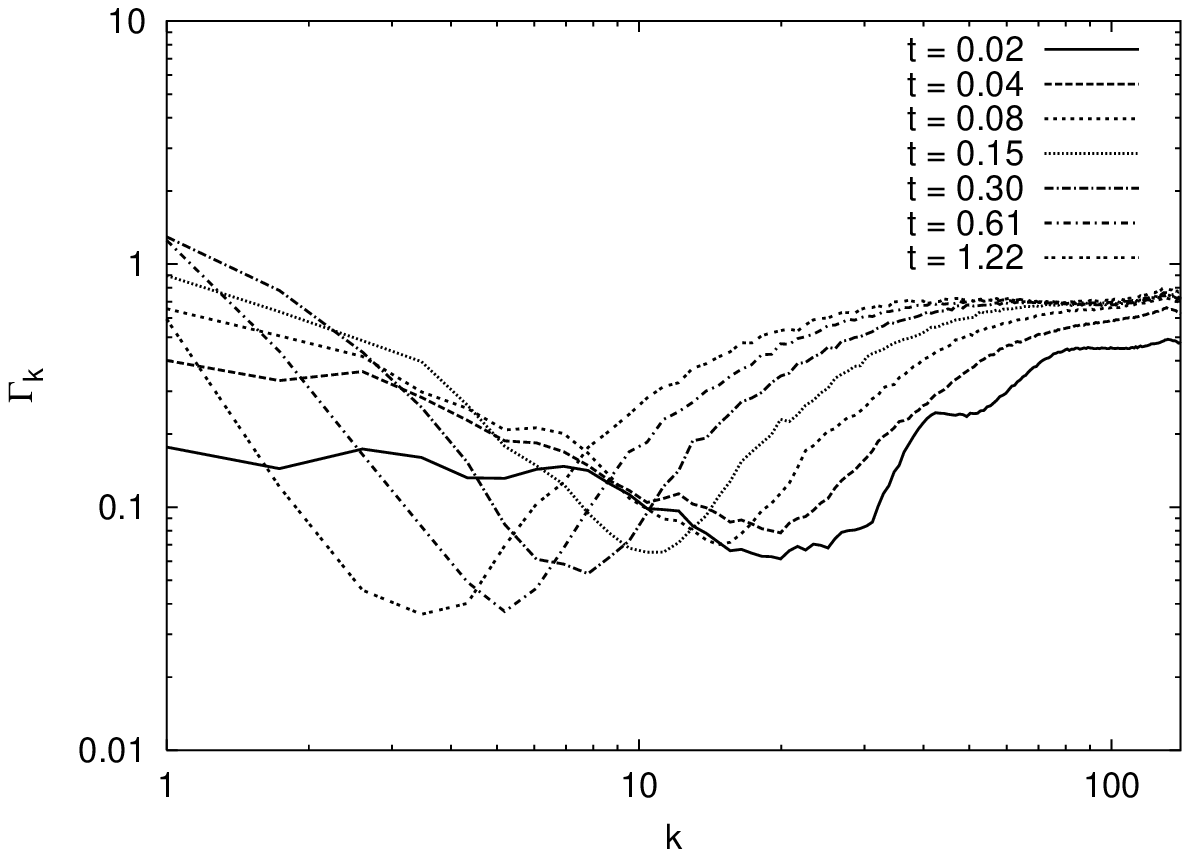}
\caption{The evolution of the ratio of the kinetic and magnetic energy
spectrum $\Gamma_k = E^{\rm kin}_k/E^{\rm mag}_k$ for a maximal
helical magnetic field in the turbulent regime. In this case
equipartition ($\Gamma_k \approx 1$) is only established on very small
scales. At the integral scale the kinetic energy is always much smaller
than the magnetic energy.}
\label{fig:gamma_k}
\eef

\bef
\showone{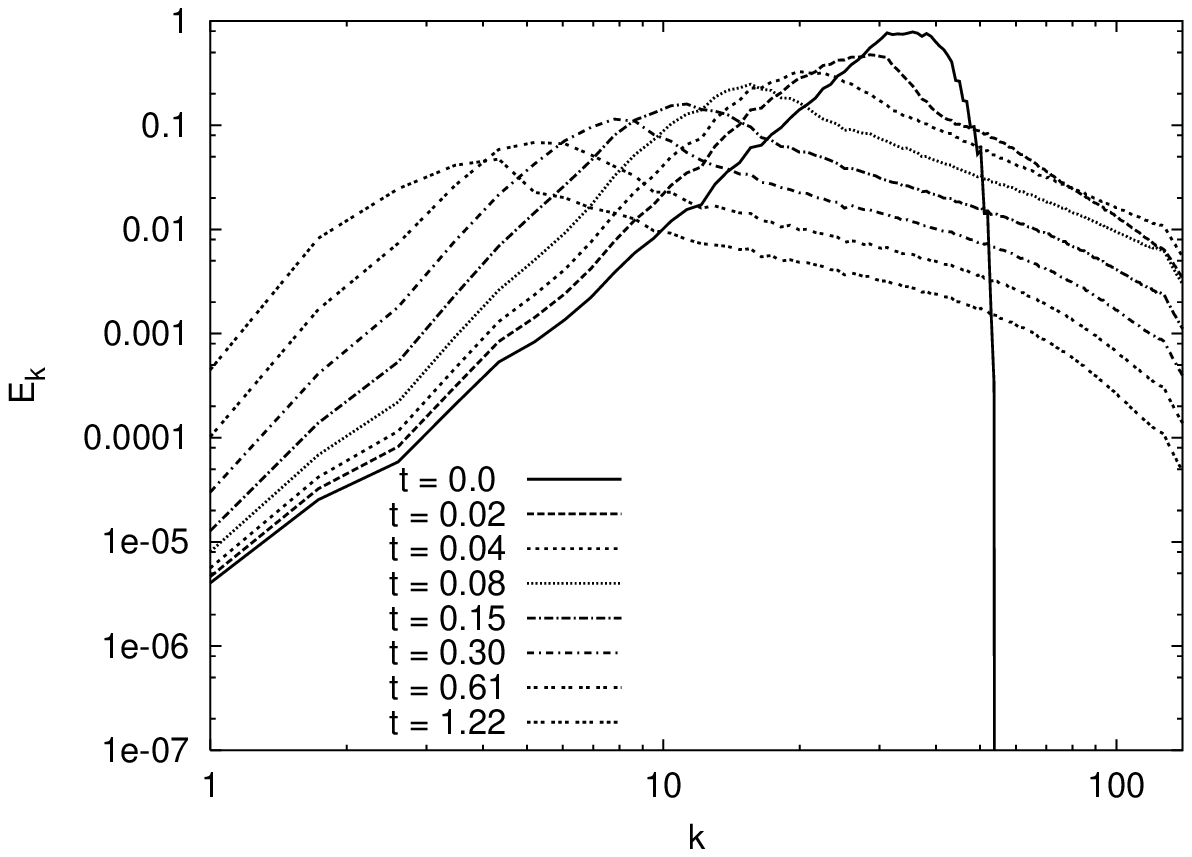}
\caption{Evolution of magnetic energy spectra in the turbulent
regime for magnetic fields with initially maximal helicity. The
spectral index of the energy spectra is $n \approx 4$.}
\label{fig:e_k_turb_h}
\eef

\bef
\showone{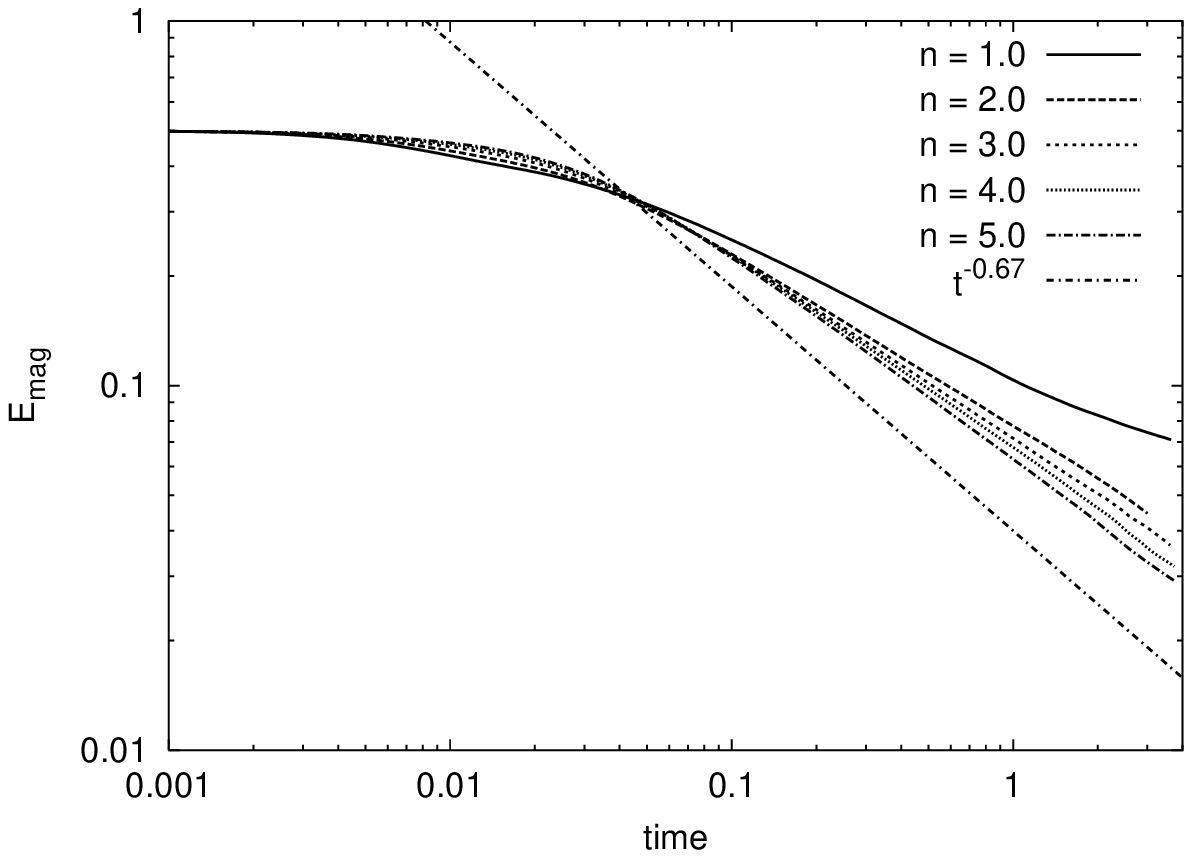}
\caption{The evolution of the magnetic energy in the turbulent regime for 
  different initial energy spectra $n$, where $E_k = k^3\,|b_k|^2
  \propto k^n$. Here, the initial magnetic field is maximal helical. For
  comparison, also the theoretical damping law, $E \propto t^{-0.67}$,
  is shown. In contrast to non helical case, the damping law for a
  helical magnetic field is nearly independent of the spectral index
  $n$ for $n < 1$.} 
\label{fig:e_on_spectra_h}
\eef

\bef
\showone{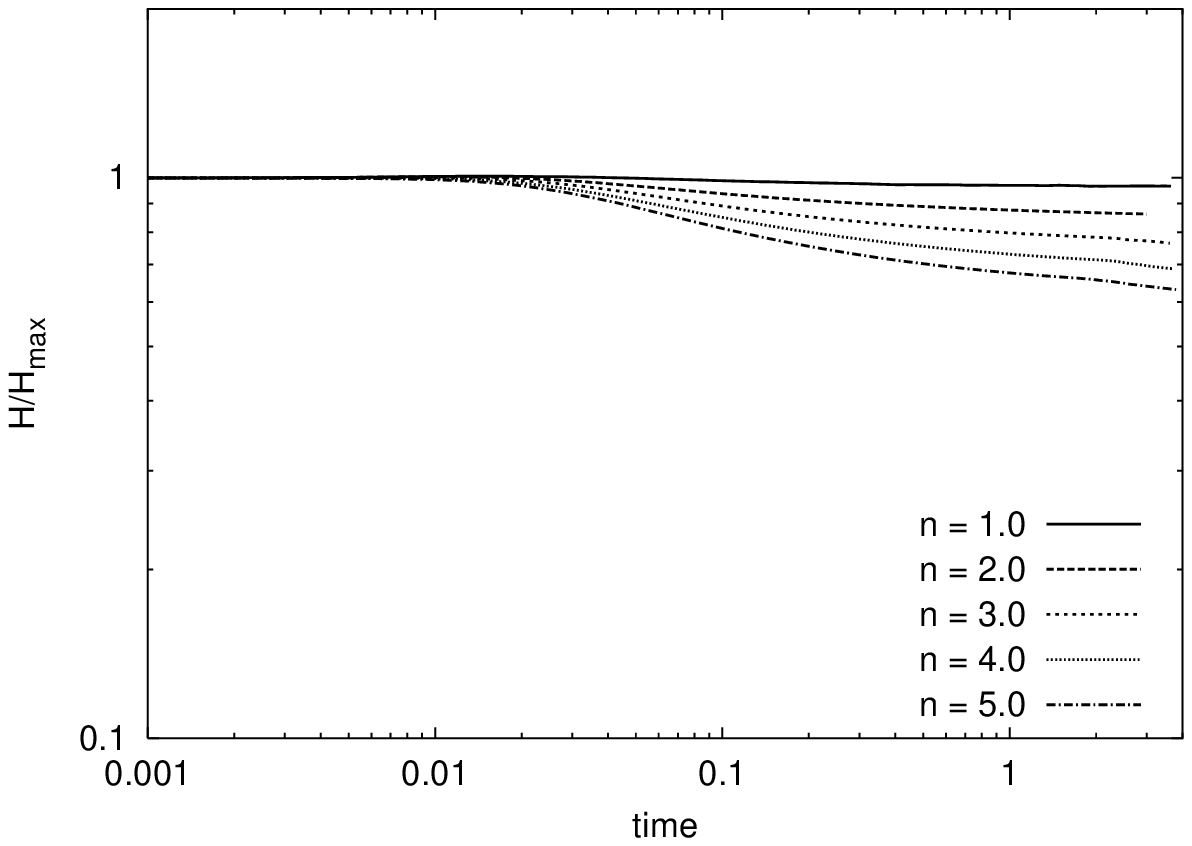}
\caption{Evolution of helicity as a function of time for different 
spectral indices $n$. In the numerically ideal case $\He/\He_{max} = 1$
should obtain.} 
\label{fig:h_on_spectra}
\eef

\section{Viscous Magnetohydrodynamics}
\label{sec:viscous}

Magnetic field dissipation in high Prandtl number fluids may also
occur in the viscous regime, where kinetic Reynolds numbers are 
much smaller than unity. Of particular importance to MHD evolution
in the early Universe is the case of photons or neutrinos 
free-streaming over the scales of interest, $l_{\sm{mfp}} \gg l$, resulting
in a drag force in Eq.~(\ref{eq:Euler_inc}) with drag coefficient $\alpha$.
To our knowledge, such a case of ``dragged'' MHD is not encountered in
other astrophysical environments, and numerical simulations 
of this case have so far not been performed.

Consider again the Euler equation Eq.~(\ref{eq:Euler_inc}). Whereas in
the turbulent case there is a balance of the terms on the
LHS, which are all of similar magnitude, in the dragged
case there is a balance between the last term on the right-hand-side
and the dissipative term $f$, with all other terms negligible. In the
terminal velocity regime one finds thus (using Eq.~(\ref{eq:dissipation_term}))
\be
\vv\approx \frac{1}{\alpha}\,\left(\vvA\cdot\nabla\right)\vvA
\ee
such that $v_l \approx \vAl_{,L} \,
(\tau_{\sm{drag}}/\tau_{\sm{A},l})\, \ll\,\vAl_{,l}$ for
$\tau_{\sm{drag}} \equiv \alpha^{-1} \ll \tau_{\sm{A},l}$.  This
yields a kinetic Reynolds number of
\be
\Rey \approx \left(\frac{\vAl_{,l}}{\alpha\, l}\right)^{2} \ll 1 \, .
\label{eq:visc_Reynolds_number}
\ee 

Though one would naively expect that at small Reynolds number the
total energy gets immediately dissipated due to viscous terms, this is
not the case (JKO98). For large Prandtl number the energy may only be
dissipated via the excitation of fluid motions. Nevertheless, due to
the strong drag, such excitation is slow and inefficient, and a system
with $\Gamma \ll 1$, i.e. well below equipartition between magnetic-
and kinetic- energy results. Since the dissipation rate is
proportional to the velocity fluctuations $v$ the net effect of strong
fluid viscosities is a delayed dissipation and quasi-frozen-in
magnetic fields. Note that in the case of viscous MHD, flows are
effectively dissipated on the integral scale, and cascading of energy
in $k$-space is not required.  One finds for the energy dissipation
rate
\be
\frac{\di E}{\di t} \approx \frac{E}{\tau_L} 
\sim \frac{E^{2}}{L^2\,\alpha}\, ,
\quad {\rm with}\quad \tau_L\approx L/v_L\,\sim\, \frac{L^2\alpha}{E}\, ,
\label{eq:visc_dissipation}
\ee
and where $\tau_L$ a formal {\em eddy} turn-over time scale identical
to the {\em overdamped} time scale for the evolution of overdamped
Alfv\'en and slow- magnetosonic modes as found in JKO98.

\subsection{Nonhelical fields}

With a blue spectrum ($n > 0$) for the magnetic fields on large scales
as given in Eq.~(\ref{eq:energy_spectra}), and with very similar
reasoning as in the turbulent case, one may compute the asymptotic
power-law for decay of energy density and growth of magnetic field
coherence length as
\be
E\approx E_0\,\left(\frac{t}{\tau^{\sm{visc}}_0}\right)^{-\frac{n}{n+2}}\,  
\quad 
L\approx L_0\,\left(\frac{t}{\tau^{\sm{visc}}_0}\right)^{\frac{1}{n+2}}\,
\quad {\rm no\,\,\, helicity}\, ,\quad \Rey \ll 1\,\, ,
\label{eq:drag_damp_nh}
\ee
for $t\simge \tau^{\sm{visc}}_0$ and where $\tau^{\sm{visc}}_0 \approx
\tau^A_{L,0}\, (\tau^A_{L,0}/\tau_{\sm{drag}}) \approx
L_0^2\alpha/E_0$. Here, in contrast to the condition in the early
Universe, a constant (in time) drag coefficient $\alpha$ has been
assumed. Note that Eq.~(\ref{eq:drag_damp_nh}) indeed predicts slower
magnetic field energy decay than it's counterpart
Eq.~(\ref{eq:turb_damp_nh}) in the turbulent case, in particular, a
longer relaxation time $\tau^{\sm{visc}}_0 \gg \tau_0$ and smaller
decay slope $\gamma_{\sm{visc}} = \gamma_{\sm{turb}}/2$ for energy
density.  In Fig.~\ref{fig:E_k_visc_fs_n5_nh} and
Fig.~\ref{fig:E_visc_fs} results of our numerical simulations of
viscous non-helical MHD in the free-streaming regime are shown. For
times longer than the relaxation time on the integral scale,
small-scale power spectra are well described by power laws $E_k\propto
k^{\beta}$ of exponent $\beta \approx -2.0$.  This power-law is
approximately consistent with a Reynolds number $\Rey_l\sim {\rm
const}(l)$ independent of scale $l$. Fig.~\ref{fig:E_visc_fs}
illustrates that Eq.~(\ref{eq:drag_damp_nh}) is a good approximation
to the numerical simulations, though numerically simulated fields tend
to decay somewhat slower than predicted, as observed in the sections
before.

\subsection{Helical fields}

In the case of maximally helical fields one may use the constancy
of helicity density in Eq.~(\ref{eq:Hel_by_E}) to find
\be
\frac{\di E}{\di t} \sim \frac{E^{4}}{\He^2\,\alpha}\, ,
\ee
yielding
\be
E\approx E_0\,\left(\frac{t}{\tau^{\sm{visc}}_0}\right)^{-1/3}\, 
L\approx L_0\,\left(\frac{t}{\tau^{\sm{visc}}_0}\right)^{1/3}\,\,
{\rm max.\,\,\, helicity}\, ,\quad \Rey\ll 1\,\, ,
\label{eq:drag_damp_h}
\ee
where $\tau^{\sm{visc}}_0$ is as before. Results of our numerical simulations
for this case may be found in Fig.~\ref{fig:E_k_visc_fs_n5_h} and
Fig.~\ref{fig:E_visc_fs}. As in
the the non-helical case these are consistent with a small-scale power
law spectrum $E_k\propto k^{\beta}$ 
with $\beta\approx -2.0$. Note that, in contrast to
before, agreement of Eq.~(\ref{eq:drag_damp_h}) and the simulation
seems excellent.

\bef
\showone{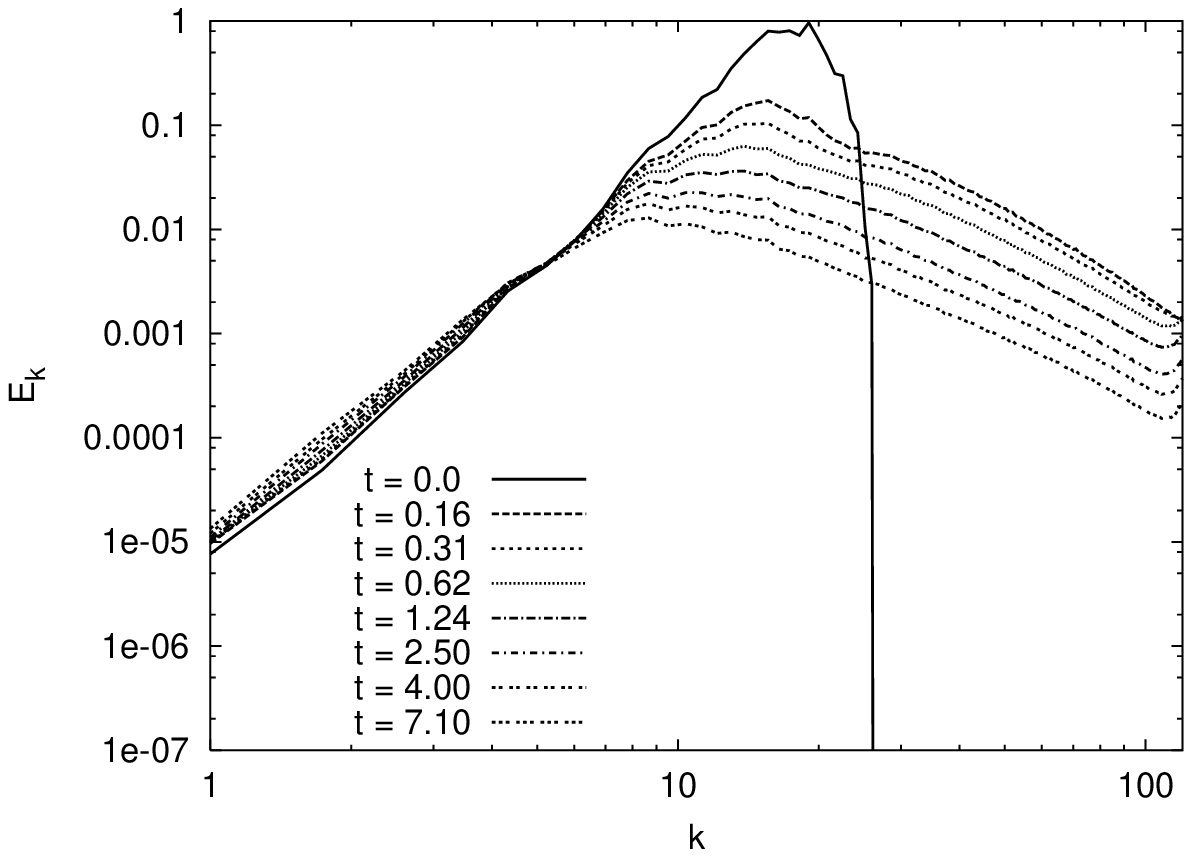}
\caption{The evolution of the magnetic energy spectra in the 
viscous free streaming regime ($\Rey \ll 1$) for a magnetic field without
initial helicity. The simulations were performed on a mesh with
$256^3$ grid points, and the cut-off is $k_c \approx 16$.}
\label{fig:E_k_visc_fs_n5_nh}
\eef

\bef
\showone{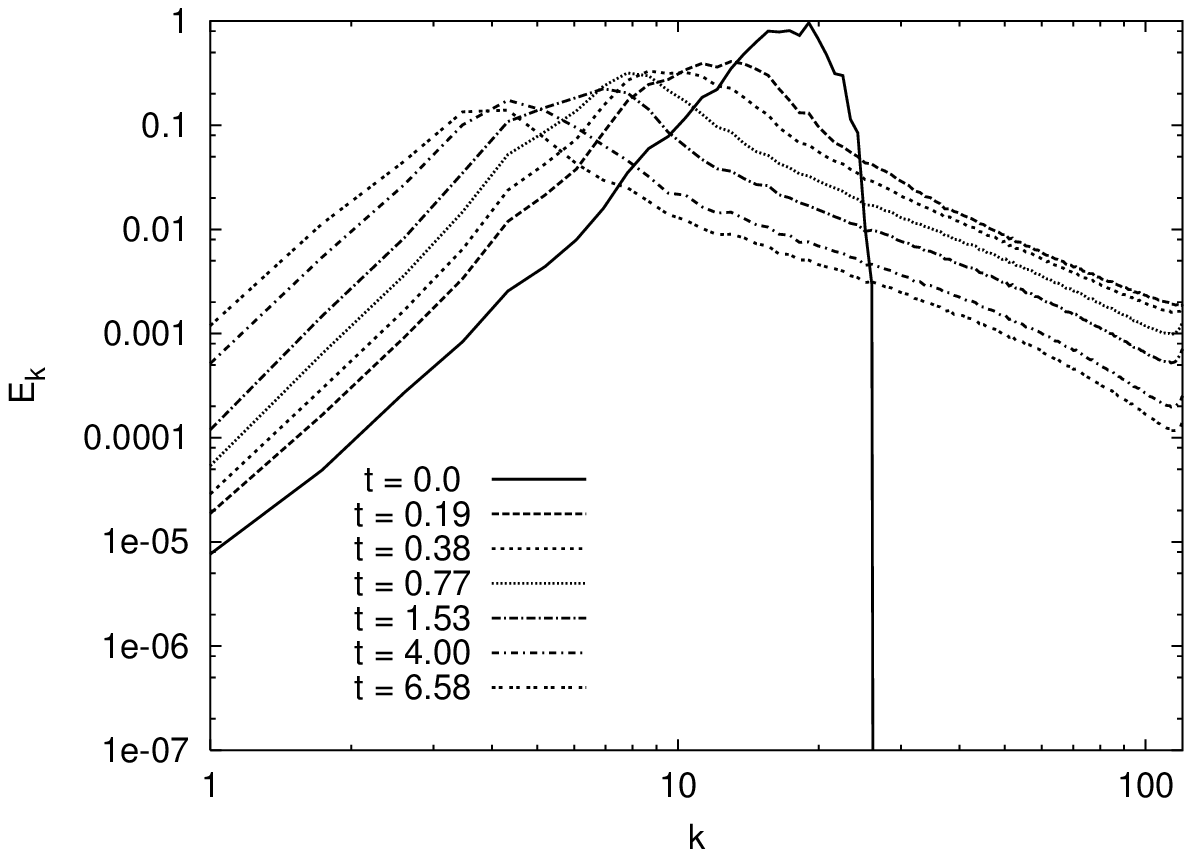}
\caption{The evolution of the magnetic energy spectra in the
  viscous free streaming regime ($\Rey \ll 1$) for a magnetic field with
  maximal helicity. The simulations were performed on a mesh with 
  $256^3$ grid points, and the cut-off is $k_c \approx 16$.}
\label{fig:E_k_visc_fs_n5_h}
\eef

\bef
\showone{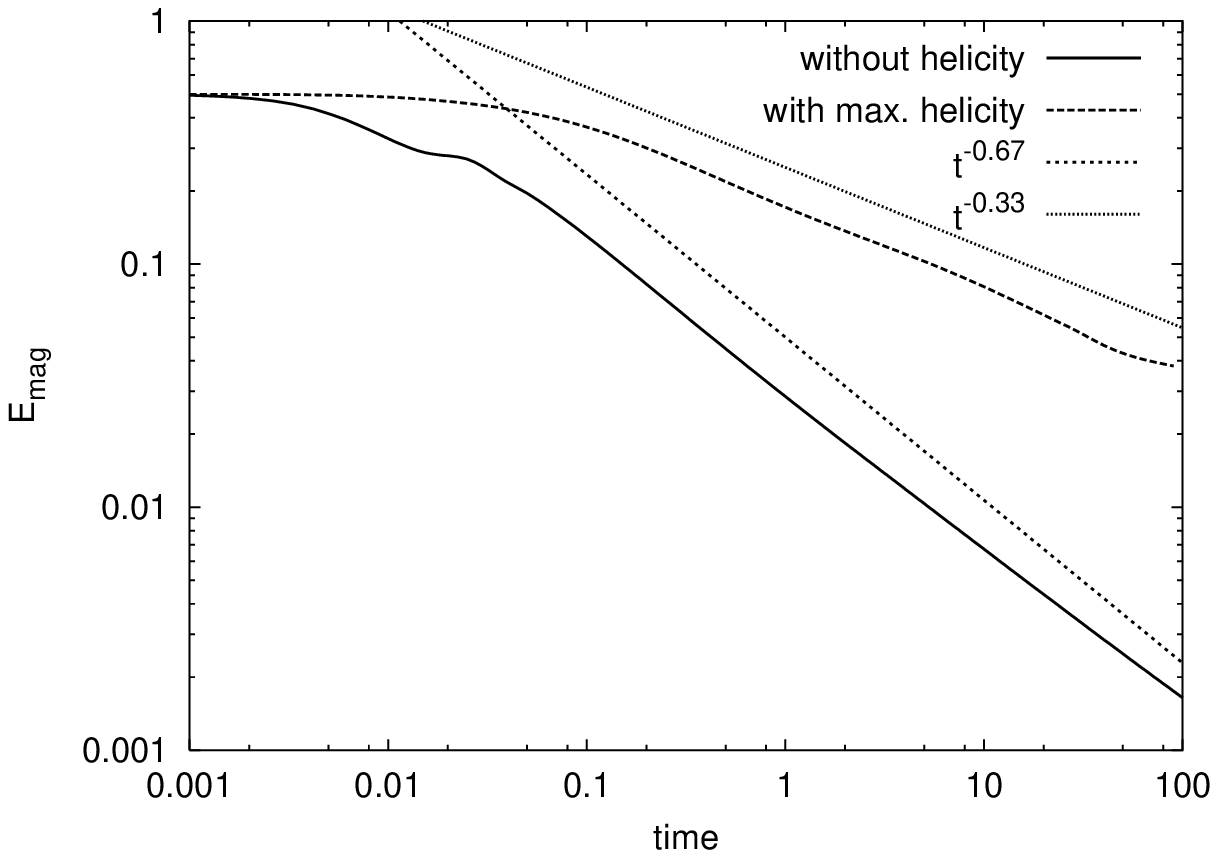}
\caption{Evolution of the magnetic energy without (solid) and
with maximal (dashed) initial helicity in the viscous free streaming
regime ($\Rey \ll 1$). The simulations were performed on a mesh with
$128^3$ grid points, the cut-off is $k_c \approx 16$ and the spectral
index $n \approx 4$. For comparison, also the theoretical expected
damping laws are shown, i.e. $E_{\rm mag} \propto t^{-0.67}$ (without
helicity) and $E_{\rm mag} \propto t^{-0.33}$ (with max. helicity).}
\label{fig:E_visc_fs}
\eef

\section{MHD With Ambipolar Diffusion}
\label{sec:ambipolar}

After recombination the Universe is only weakly ionized (i.e $X_e \ll 1$). 
Neutral particles, i.e. hydrogen atoms, do not respond to magnetic
stresses and may therefore slip by the magnetic field lines, unless
the scattering between neutral and charged particles is rapid
enough. To investigate if this is the case one has to consider the
(here assumed incompressible) equations of MHD with a significant
neutral component 
\bea
\rho_i\,\left(\dfdt{\vv_i} + \vv_i\cdot\nabla\,\vv_i\right) & = & 
\frac{\left(\curl\vB\right)\times\vB}{4\pi}
 - \rho_i\,\alpha_{\sm{in}}\,\left(\vv_i - \vv_n\right)  
\label{eq:Euler_ion}
\\
\rho_n\,\left(\dfdt{\vv_n} + \vv_n\cdot\nabla\,\vv_n\right) & = & 
- \rho_n\,\alpha_{\sm{ni}}\,\left(\vv_n - \vv_i \right)
\label{eq:Euler_neutral}
\eea
where $\rho_n$, $\rho_i$, $\vv_n$, $\vv_i$ are matter density and
velocity of neutrals and ions, respectively, and we will
assume $\rho_i \ll \rho_n$ throughout. The momentum transfer
rate due to neutral-ion collisions satisfy
\be
\alpha_{\sm{ni}} = \frac{\rho_i}{\rho_n}\,\alpha_{\sm{in}}
 \approx X_e\,\alpha_{\sm{in}}
\label{eq:drag_ni}
\ee
The equations of MHD are closed by including the induction equation
\be
\dfdt{\vB} = \curl\left(\vv_i\times\vB\right)
\label{eq:induction_ion}
\ee
for ions. The condition of tight coupling between ions and neutrals, i.e. 
$\vv_D\equiv \vv_i - \vv_n \ll \vv_i$ may be derived from
Eq.~(\ref{eq:Euler_neutral}) (noting that the first two terms are
usually of the same magnitude)
to be equivalent to 
\be
\frac{\vv_i}{L} \approx \frac{\vv_n}{L} \ll X_e\,\alpha_{\sm{in}}
\ee
One may show (cf. also ~\cite{Shu:1983,Subramanian:1997,Zweibel:2002})
self-consistently that in this limit the LHS of
Eq.~(\ref{eq:Euler_ion}) is negligible, leaving the ion-neutral
drift velocity $v_D$ in the terminal velocity regime 
\be
\vv_D = \frac{\left(\curl\vB\right)\times\vB}
               {4\pi\,\rho_{i}\, \alpha_{\sm{in}}}
\label{eq:drift_vel}
\ee
Inserting this equality into Eq.~(\ref{eq:Euler_neutral}), and for
$\vv_n\approx \vv$, where $\vv$ is the center-of-mass velocity, 
one obtains the usual Euler equation Eq.~(\ref{eq:Euler_inc}).
The induction equation~(\ref{eq:induction_ion}) is modified to include
a dissipative term. 
Replacing $\vv_i = \vv_D + \vv_n$ one finds
\be
\dfdt{\vB} = \curl\left(\vv\times\vB\right) + \curl\left(
\frac{\left(\curl\vB\right)\times\vB}
{4\pi\,\rho_{i}\, \alpha_{\sm{in}}}  \times\vB\right)\,\, .  
\label{eq:ind_ambipolar}
\ee
MHD of an ion-neutral mixture in the tightly coupled regime behaves
thus as ordinary MHD with an additional dissipative term. The effect
of this term may be estimated by defining an ambipolar Reynolds number
as the comparison of the two terms on the RHS of
Eq.~(\ref{eq:ind_ambipolar}), i.e.
\be
R_{\sm{amb}} \equiv \frac{ v\, L\, \alpha_{\sm{in}}}{(\vAl^i)^2}\approx 
\frac{ v\, L\, \alpha_{\sm{in}}\, X_e}{\vAl^2}
\label{eq:amb_Reynolds_number}
\ee
where
\be 
\vAl^i\approx \vAl/\sqrt{X_e} 
\ee
is the Alfv\'en propagation velocity in the ion-neutral weakly coupled
limit. It may be seen that the condition $R_{\sm{amb}} \gg 1$
(assuming self-consistently $v \sim \vAl$) automatically implies the
tight-coupling condition Eq.~(\ref{eq:drift_vel}). It is thus evident
that MHD with dissipation due to ambipolar diffusion in the
tight-coupling regime may never become viscous due to this ambipolar
``drag''~\footnote{This is with the exception that for particularly
chosen initial conditions, i.e. $\vv_i \neq \vv_n$ exponential damping
of fluid flows on the integral scale may occur.}.  In the language
(JKO98) appropriate to linear MHD this implies that overdamped modes
proportional to the magnetic stresses do not exist.  This is in stark
contrast to MHD with fluid shear viscosity, or with momentum drag due
to a homogeneous background component, where viscous MHD
(i.e. overdamped modes) exist.

These arguments assume the absence of other sources of dissipation.
Consider, for example, shear viscosity due to neutral-neutral scattering
as described by a term $\rho_n\eta \laplace\,\vv_n$ on the RHS of
Eq.~(\ref{eq:Euler_neutral}). 
Assuming viscous MHD due to this term (i.e. the kinetic Reynolds
number Eq.~(\ref{eq:Reynolds_number}) $\Rey \ll 1$) the condition for
tight coupling is modified, and now reads $\eta/L^2\ll
\alpha_{\sm{in}}\,X_e$. Nevertheless, even in this case one finds that
the condition of tight coupling is equivalent to the requirement
$R_{\sm{amb}} \gg 1$. The flow may thus be viscous in the tight
coupling regime, but only due to sources of dissipation other than 
ambipolar diffusion.

Once the flows reach the limit $R_{\sm{amb}} \simle 1$ the neutral
species decouples from the flow. In this limit MHD evolution is
described by Eq.~(\ref{eq:Euler_ion}) with $v_n\to 0$ and
Eq.~(\ref{eq:induction_ion}) completely analogous to MHD with 
free-streaming photons or neutrinos (cf. Sec.~\ref{sec:viscous}) and with a
Reynolds number given by Eq.~(\ref{eq:visc_Reynolds_number}). One may
then show that the flow is viscous due to ambipolar drag.
Only when $\alpha_{\sm{in}}$ is reduced by a further factor of
$\sim \sqrt{X_e}$ (or equivalently, the Alfv\'en crossing time is reduced
by the same factor) does turbulent MHD obtain again. When this happens
typical fluid velocities are $v \approx \vAl^i$, thus increased
with respect to the tight-coupling regime.

It is important to stress the following. The system of equations
Eqs.~(\ref{eq:Euler_ion}), (\ref{eq:Euler_neutral}) and
(\ref{eq:induction_ion}) provides only a proper description of 
MHD in the fluid limit when particle species have mean free paths much
smaller than the scale under consideration. Whereas for the
scales we consider (assuming magnetic fields not too weak) this may be the
case for protons and electrons due to Coulomb scattering, this condition gets
violated at late times for neutral particles. In this limit, i.e.
$l^n_{\sm{mfp}}\gg L$ mixing of neutrals from different regions becomes
significant. Higher moments of the particle distribution $f_n(x,v)$
(with the zeroth moment, density and the first, velocity) become significant
such that a reduction of the Boltzmann equation to the Euler equation is
not anymore adequate. When $l^n_{\sm{mfp}}\gg L$ and $l^i_{\sm{mfp}}\ll L$ 
one may nevertheless describe MHD by the fluid equations
Eqs.~(\ref{eq:Euler_ion}) and (\ref{eq:induction_ion})
for ions, and to a good approximation, assume $\vv_n\approx 0$ due to
mixing of neutrals from different regions.

\section{Evolution of Cosmic Magnetic Fields}
\label{sec:evolution}

In this section we present detailed analytical results for the
evolution of subhorizon 
magnetic fields between an epoch of magnetogenesis
(e.g. the electroweak transition at $T\approx 100\,\GeV$) and the much
later onset of cosmic structure formation (at approximately redshift
$z\approx 10$). Our analysis draws on the general results found in the
previous sections, but includes viscosities as applicable in the early
Universe. In particular, we give coherence length and energy density
as function of cosmic temperature, with generation epoch, magnetic spectral
index, initial magnetic energy density, and helicity left as free
parameters. 
Though the results are fairly straightforward, when applied to the
various regimes in the early Universe (i.e. turbulent and viscous due
to photon- and neutrino- viscosity, respectively) a large
number of expressions emerges. We therefore advise the more
superficially interested reader to skip the third subsection of this section
and proceed to the discussion of results in the next section. 

\subsection{Initial Conditions}
\label{ssec:init}

We define the Fourier transform of the magnetic fields such that the
spatial average of magnetic field strength may be written as follows
\be
<B(x)^2> \equiv \int \di\ln k\, \tilde{B}_k^2
\ee
The spectrum of the magnetic field at the magnetogenesis epoch is
parametrized by
\be
\tilde{B}_{gc}(l_c) = \tilde{B}_{gc} (L_{gc})\, 
\biggl(\frac{l_c}{L_{gc}}\biggr)^{-n/2}
\label{eq:Bfield_spectrum}
\ee

Here, and in what follows, a subscript $g$ denotes quantities at the
magnetogenesis epoch and subscript $c$ refers to {\em comoving}
values. Here comoving lengths are defined as the lengths they would
have at the present epoch (i.e. $l_c(T) = l(T)\, (a(T_0)/a(T))$ where
$a$ is scale factor and $T_0$ present cosmic temperature) and we
define comoving field strength analogously as the field strength it
would have at the present the epoch, if the field would only evolve
according to the requirement of flux conservation (i.e. $B_c(T) =
B(T)\, (a(T)/a(T_0))^2$). Note that $k = 2\pi /l$ and that whereas $l$
denotes an arbitrary scale $L$ always denotes the integral scale
($k^I$ the integral wave vector), i.e. the energy containing, scale.
Our analysis will focus on blue spectra $n > 0$ as appropriate for
magnetic fields generated after an inflationary epoch by a causal
process.  Given these definitions one may compute the magnetic field
energy density at an arbitrary epoch as a function of the temperature
dependent integral scale and the scale factor
\bea
\rho_B (T) & = & \biggl(\frac{a(T)}{a_0}\biggr)^4\, \frac{1}{8\pi}\, 
\int_0^{k^I_c} \di\ln k\,
\tilde{B}^2_{gc} (L_{gc})\,
\biggl(\frac{k_c(T)}{k_{gc}^I}\biggr)^{n} 
\nonumber \\
 & = &
\biggl(\frac{a(T)}{a_0}\biggr)^4\, \frac{1}{8\pi\,n}\, 
\tilde{B}^2_{gc} (L_{gc})\,
\biggl(\frac{L_c(T)}{L_{gc}}\biggr)^{-n}
\eea
It is convenient to define a ratio $r$ between magnetic
energy density and a power of the total radiation entropy density
\be
r \equiv \frac{\rho_B}{s_r^{4/3}}
\label{eq:r_entropy}
\ee
since for constant comoving integral scale (i.e. no dynamic magnetic
field evolution) this ratio stays constant with the expansion of the
Universe. The dynamic (as opposed to geometric) evolution of the field
is therefore more easily deduced from the evolution of $r$. The
quantity $r$ may be related to the ratio of magnetic field energy
density and radiation energy density  $r_{\rho} = \rho_B/\rho_r$ by
\be
r_{\rho} = r \frac{4}{3}\, \biggl(\frac{2\,\pi^2}{45}\biggr)^{1/3}
\frac{g_S^{4/3}}{g_{r}}
\ee
with $g_S$, $g_r$ denoting statistical weight in entropy and radiation,
respectively.
Note here, that $r$ may be converted to average magnetic field
strength
\bea
B(T) & = & 5.72\times 10^{-6}\,\G\,\,\, r^{1/2}(T)\,
\nonumber\\ 
     & & \times\biggl(\frac{g_S}{3.909}\biggr)^{4/3}
     \biggl(\frac{T}{2.351\times 10^{-4}\, {\rm eV}}\biggr)^2
\label{eq:mag_eq}
\eea
such that the comoving (present day) magnetic field strength 
is $B_c = 5.72\times 10^{-6}\,\G$ for $r=1$. (For $r_{\gamma}
\equiv \rho_{B}/\rho_{\gamma} = 1$ the comoving field strength of
$B_c = 3.24\times 10^{-6}\,\G$ results.) The magnetic field strength
given in Eq.~(\ref{eq:mag_eq}) yields an Alfv\'en velocity after the
decoupling of photons of
\be
\vAl = \frac{B}{\sqrt{4\pi\rho}} 
  = 8.86\times 10^5 \, \frac{\cm}{\s}\,  
\biggl(\frac{r}{10^{-10}}\biggr)^{1/2}\,
\biggl(\frac{\Omega{\sm{b}}h^2}{0.02}\biggr)^{-1/2}
\biggl(\frac{T}{0.259\,\eV}\biggr)^{1/2}
\label{eq:va}
\ee
where $g_S = 3.909$ has been assumed and with $\rho = \rho_{\sm{b}}$
in the fully ionized case before recombination as well as in the
partially ionized case in the tightly coupled regime after
recombination (cf. Sec.~\ref{sec:ambipolar}).  Here $\rho_b$ and
$\Omega_b$ denote baryonic density and fractional contribution to the
critical density at the present epoch whereas $h$ is the Hubble
constant in units of 100 ${\rm km\,s^{-1}Mpc^{-1}}$.  We alert the
reader to the distinction between decoupling of photons (i.e.
$l_{\gamma}\simge L$) (typically occurring before recombination) and
recombination itself.  Eq.~(\ref{eq:va}) may be compared to the plasma
speed of sound
\be
v_{\sm{b}} = \sqrt{\gamma \frac{T_{\sm{b}}}{m_{\sm{b}}}} = 
5.99\times 10^5 \, \frac{\cm}{\s}\,\gamma^{1/2}
\biggl(\frac{T_{\sm{b}}}{0.259\,\eV}\biggr)^{1/2}
\label{eq:vb}
\ee
where $\gamma = 10/3$ and $2$ for adiabatic and isothermal compression,
respectively, and where we have neglected corrections due to the presence
of helium. Note that below redshift $z\simle 100$ the baryon temperature falls
more rapidly than the photon temperature, i.e. as $T_{\sm{b}}\sim a^{-2}$.

To determine the integral scale at the generation epoch, $L_{g}$, we
assume that turbulence pertains, such that $L_g$ is obtained by
setting the Hubble rate at the generation epoch equal to the Alfv\'en
eddy turnover rate. This yields 
\bea
L_{gc} &\equiv & L_c(T_g)  =  
\biggl(\frac{2025}{4\pi^7}\biggr)^{1/6}\, \frac{\Mpl}{T_0\, T_g}\,
g_{S0}^{-1/3} \sqrt{n\,r_g} \\
& \simeq & 1.55\times 10^{-4}\,\pc\,\,
   \sqrt{n}\,\left(\frac{r_g}{0.01}\right)^{1/2} \,
   \left(\frac{T_g}{100\,\mbox{GeV}}\right)^{-1} 
\label{eq:length_gc}
\eea
where $\Mpl = 1/\sqrt{G} \approx 1.22\times 10^{19}\,\GeV$ the Planck
mass. In the above, the subscript $0$ denotes quantities evaluated at
the present epoch. 
Finally, we parametrize initial helicity of the field by a dimensionless
number $h_g$
\be
\He_{gc} = h_g\, \He_{\sm{max}}^{gc} \quad \mbox{with} \quad
\He_{\sm{max}}^{gc}\approx \frac{2\,\pi}{n-1}\,  
\tilde{B}^2_{gc}(L_{gc})\, L_{gc}
\label{eq:max_helicity}
\ee
such that $h_g\le 1$.

\subsection{Evolution: The General Picture} 
\label{ssec:general}

The evolution of a stochastic magnetic field in the early Universe is
described by alternating epochs of turbulent MHD and viscous MHD. Here
the latter epochs occur when viscosities due to neutrinos, or photons,
become significant. Such a picture has already been established by
JKO98. With the assumed blue magnetic spectra, the gross features of
magnetic field evolution are described by the growth of the integral
scale. Following the arguments in the Sec.~\ref{sec:turbulence}
the instantaneous integral scale is given by the equality between
cosmic time and eddy turnover time at the scale $L$
\be
\frac{1}{t_{\sm{eddy}}}\approx\frac{v(L)}{L_p(T)} \approx
H(T)\approx\frac{1}{t_{H}} 
\label{eq:evolution}
\ee
holding equally for turbulent and viscous eras. In the above expression 
the subscript $p$ denotes the proper (as opposed to comoving) value of the
integral scale, $v(L)$ is the the fluid velocity on scale $L$, and $H$
is the Hubble constant. The velocities $v$ may be determined from the
Euler equation (Eq.~(\ref{eq:Euler_inc})) by an approximate balance of
either the second and third term on the LHS in the turbulent 
case ($\Rey >1$) or the third term on the LHS and the dissipation term
on the RHS in the viscous case. This yields
\be
v(L)\approx \vAl(L)\, ; \quad\quad \Rey > 1
\ee
in the turbulent case and
\be
v(L)\approx \frac{\vAl^2(L)\, L}{\eta}\, , \quad v(L)\approx 
\frac{\vAl^2(L)}{\alpha\, L}\, , \quad\quad
\Rey < 1
\label{eq:viscous_v}
\ee
in the photon (neutrino) diffusive and free-streaming viscous cases,
respectively. 
Note that the velocities in the viscous case may also be written in a
unifying way as $v(L) \approx \Rey\, \vAl$ with $\Rey < 1$ the Reynolds number
Eq.~(\ref{eq:Reynolds_number}) evaluated with the Alfv\'en velocity,
i.e. $v = \vAl$. 

Eq.~(\ref{eq:evolution}) is to be evaluated with proper quantities.
Since a given scale expands continuously with the Universe,
i.e. $l_p=a\,l_c$, the eddy turnover (relaxation) time on this scale
increases with the expansion of the Universe. This relaxation time
increase is enhanced after the decoupling of photons by an additional
decrease of the Alfv\'en velocity, i.e.  $\vAl\propto
B(T)/\sqrt{4\pi\rho_{\sm{b}}}\propto a^{-1/2}$ (whereas, $\vAl\propto
a^0$ when photons are still coupled to the MHD evolution).  On the
other hand, the Hubble time increases as $t_{H}\propto a^2$ during
radiation domination (RD) and as $\propto a^{3/2}$ during matter
domination (MD).  During turbulent evolution, the combined effect is
such
\be
\frac{t_{\sm{eddy}}}{t_{H}} \approx \frac{L/\vAl}{t_{H}} \propto
\frac{a}{a^2} \propto 1/a 
\quad \mbox{(RD)} \quad 
  \propto\frac{a/1/a^{1/2}}{a^{3/2}} \propto a^0
 \quad \mbox{(MD)}
\label{eq:teddy_evol}
\ee
that during RD larger and larger scales may be processed, i.e.  that
the comoving integral scale $L_c$ may grow as the Universe expands.
In contrast, during MD the ratio between eddy- and Hubble- time stays
constant, permitting only logarithmic growth of $L_c$. This, however,
is only the case while the fluid is turbulent. For sufficiently strong
fields (see below) turbulence recommences right after recombination,
with the fluid before recombination strongly dragged by free-streaming
photons. In the viscous regime, with viscosity provided by photons,
one finds $\eta_{\gamma}\propto a^3$ and $\alpha_{\gamma}\propto
a^{-4}$ (cf. Appendix~\ref{apx:dissipation}). This yields for the
comparison of time scales
\be
\frac{t_{\sm{eddy}}}{t_{H}} \approx \frac{L/v}{t_{H}}
  \propto a 
  \quad \mbox{(photon diffusion)}
   \quad  \propto a^{-5/2} 
  \quad \mbox{(photon free-streaming)}
\label{eq:visc_times}
\ee
where we have assumed radiation domination during the diffusive regime
and matter domination during the free-streaming regime. It may be seen
from Eq.~(\ref{eq:visc_times}), that the integral scale may further
increase during the viscous MHD regime when photons are
free-streaming. On the other hand, one may show quite generally that
an increase during the photon diffusion regime is always
prohibited. Essentially identical conclusions result in the case of
neutrinos.

The following general picture for the evolution of the integral scale
thus emerges. At early times, close to the epoch of magnetogenesis in
the early Universe, the fluid is turbulent and as the Universe expands
the comoving scale where one eddy turnover is possible in cosmic time
is continuously increasing. By the process of a direct cascade the
energy of this (integral) scale $L_c$ may thus be dissipated, leaving
only the tail of the initial magnetic field at scales larger than the
integral scale. The spectrum of the magnetic field is thus described
by
\be
\tilde{B}_{c}(l_c) = \tilde{B}_{gc} (L_{gc})\, 
\biggl(\frac{l_c}{L_{gc}}\biggr)^{-n/2}\,\, ,\quad {l_c \ge L_c}\,\, .
\label{eq:Bspectrumnohel}
\ee

As the Universe cools down shear viscosity due to neutrinos becomes
large, thereby reducing the Reynolds number of the flow.  At the epoch
when the Reynolds number becomes of order unity on the integral scale,
a regime of viscous MHD commences. At this point, a further increase
of $L_c$ is prohibited, since in the diffusive regime the relaxation
time grows more rapid than the Hubble time. Any existing fluid flows
are dissipated, leaving, nevertheless, the magnetic field at scales
beyond $L_{\sm{EOT}}^{\nu}$ intact. Here $L_{\sm{EOT}}^{\nu}$ refers
to the integral scale when $\Rey_{\nu}(L(T_{\sm{EOT}}))$ has decreased
to unity.  This is analogous to the survival of magnetic fields in the
overdamped regime of linearized modes, as discussed in JKO98. Only
some time after neutrinos have decoupled from the fluctuations,
i.e. when $l_{\nu}\gg L$, the integral scale may grow beyond that
given at the epoch of end of turbulence, $L_{\sm{EOT}}^{\nu}$. During
this dissipation of magnetic fields in the viscous free-streaming
regime, the integral scale grows more rapidly then during the
turbulent regime. This more rapid increase (as opposed to a slower
increase in the non-expanding case, cf. Sec.~\ref{sec:viscous}) is
mainly due to the strong temperature dependence of the drag
term. Since the neutrino drag is continuously decreasing, some time
before neutrino decoupling at $T\approx 2.6\,\MeV$ the fluid enters
again a turbulent stage. At this point, the integral scale has grown
to a value, as if the plasma would have not at all gone through a
viscous period. The viscous period thus just delays the dissipation of
magnetic fields. These evolutionary trends are shown in
Figs.~\ref{fig:length_ew}, \ref{fig:energy_ew},
\ref{fig:length_qcd} and \ref{fig:energy_qcd}, which show the growth of
magnetic coherence length, and the decay of magnetic energy density,
for a number of initial conditions.  The evolution of the kinetic
Reynolds number $\Rey$ is also shown for a particular scenario. The
frozen-in state of magnetic fields during the diffusive neutrino
regime with $\Rey \simle 1$ and the first part of the free-streaming
neutrino regime becomes apparent by the plateaus in $L_c$ one finds at
$T\sim 10^7-10^8\,\eV$.

A similar picture results for magnetic field evolution after neutrino
decoupling, but now with neutrinos replaced by photons.  There are,
however, subtle differences. After the electrons and positrons become
non-relativistic, i.e. $T < m_e$, the photon mean free path increases
rapidly. This is particularly true during the period of the $e^{\pm}$
annihilation, i.e. $500\,\keV \simge\, T\,\simge
20\,\keV$~\footnote{This annihilation period reduces the number of
$e^{\pm}$ by a large factor $\sim 10^{10}$ and is completed at
$T\approx 20\,\keV$, where the number of positrons has fallen below
the number of electrons.}.  Therefore, viscous MHD evolution with
photons diffusing commences for a wide parameter space during this
period.  The epoch of viscous MHD with drag provided by free-streaming
photons, which starts some time later, is always ended right at
recombination ($T\approx 0.26\,\eV$).  This, of course, is due to the
virtually instantaneous decrease of $\alpha_{\gamma}$ by a factor
$\sim 10^4$ which is the result of the loss of free electrons during
recombination. In contrast to dissipation due to neutrinos, the
viscous period due to photons thus does not only simply delay the
growth of integral scale (i.e the dissipation). Rather, shortly before
the end of the viscous MHD regime before recombination magnetic field
strengths are genuinely larger and integral scales are smaller as
compared to a scenario where the flow would have stayed turbulent all
along. What happens after recombination depends then on the strength
of the magnetic field at recombination.

After recombination two different potential sources of dissipation
come into play; ambipolar diffusion as well as shear viscosity due to
hydrogen atoms. We assume, for the moment, a turbulent flow,
i.e. $v\approx \vAl$, with resulting conclusions turning out
independent of this assumption.  With the aid of
Eq.~(\ref{eq:evolution}), evaluated shortly after recombination, one
may determine the ambipolar Reynolds number
Eq.~(\ref{eq:amb_Reynolds_number}) to be approximately $R_{\sm{amb}}
\approx 10^4$.  It follows that the hydrogen atoms are tightly coupled
to the flow (cf. Sec.~\ref{sec:ambipolar}).  As this is the case,
viscosity due to neutrals may play a role. One may evaluate the
kinetic Reynolds number $\Rey$ Eq.~(\ref{eq:Reynolds_number}) on the
integral scale at recombination due to hydrogen viscosity by noting
that the Alfv\'en crossing rate on the integral scale shortly before
recombination $(\vAl/L)_{\sm{rec}}$ is given by
$(\sqrt{\alpha_{\gamma}\, H})_{\sm{rec}}$
(cf. Eq.~(\ref{eq:evolution}) and ~(\ref{eq:viscous_v}) applied in the
viscous photon-freestreaming regime shortly before
recombination). When this is done one finds that weak magnetic fields
with small coherence lengths~\footnote{It may be instructive to point
out, that, irrespective of initial magnetic spectra, magnetic field
strength and coherence lengths are correlated, through
Eq.~(\ref{eq:evolution}). One may thus formulate statements in either
of these quantities, with the other given by Eq.~(\ref{eq:evolution})}
are entering a viscous regime due to hydrogen viscosity immediately
after recombination, whereas strong fields do not. Here, the dividing
magnetic field strength is given approximately by $B^{\eta}_c \approx
10^{-13}\,\G$, corresponding to integral scales $L_c\approx 10\,\pc$
(cf. Figs.~\ref{fig:length_ew} and~\ref{fig:energy_ew}). Thus,
fields with $B_c \simle B^{\eta}_c$ are not significantly processed
immediately after recombination.  Only some time later, when neutrals
have decoupled from the flow, they are subject to further processing
(i.e. increase of coherence length). This further increase in $L$
mostly takes place at epochs with redshift $z\simle 100$. The increase
in integral scale then occurs in a viscous regime with drag due to
free-streaming hydrogen atoms (cf. Sec.~\ref{sec:ambipolar}), quite
analogous to the regime shortly before recombination. The flows become
turbulent again only when the Universe is reionized as ambipolar drag
then disappears~\footnote{This assumes that the gas is sufficiently
hot after reionization such that ion viscosity is negligible}.  The
epoch of reionization occurs presumably at $z\simeq 10$, at which
point the integral scale grows virtually instantaneous to a larger
value and stays approximately constant thereafter~\footnote{Note that
this growth of integral scale is associated with dissipation of
magnetic fields into heat and therefore amounts into a small increase
of the baryon temperature}.

In contrast, fields with strength $B\simge B^{\eta}_c$ recommence
turbulence after recombination. The growth of coherence scale during
the recombination epoch is characterized by an almost instantaneous
increase of a factor of order $\sim 5$, associated with dissipation
into heat. Subsequent evolution does only increase the integral scale
at best logarithmically, due to the peculiar red-shifting of Alfv\'en
crossing time and Hubble constant (cf. Eq.~(\ref{eq:teddy_evol})). One
may show that even such fields enter a viscous period later on, with
viscosity first due to diffusing hydrogens in the tight coupling
regime and later due to ion-hydrogen collisions (ambipolar drag) in
the weak ion-hydrogen coupling limit. In any case, the magnetic
coherence scale is not modified much anymore, even after the epoch of
reionization~\footnote{We do not consider evolutionary effects due to
cosmic structure formation in this paper}.  Again, these trends may be
followed in Figs.~\ref{fig:length_ew} -- \ref{fig:energy_qcd}.

\bef
\showone{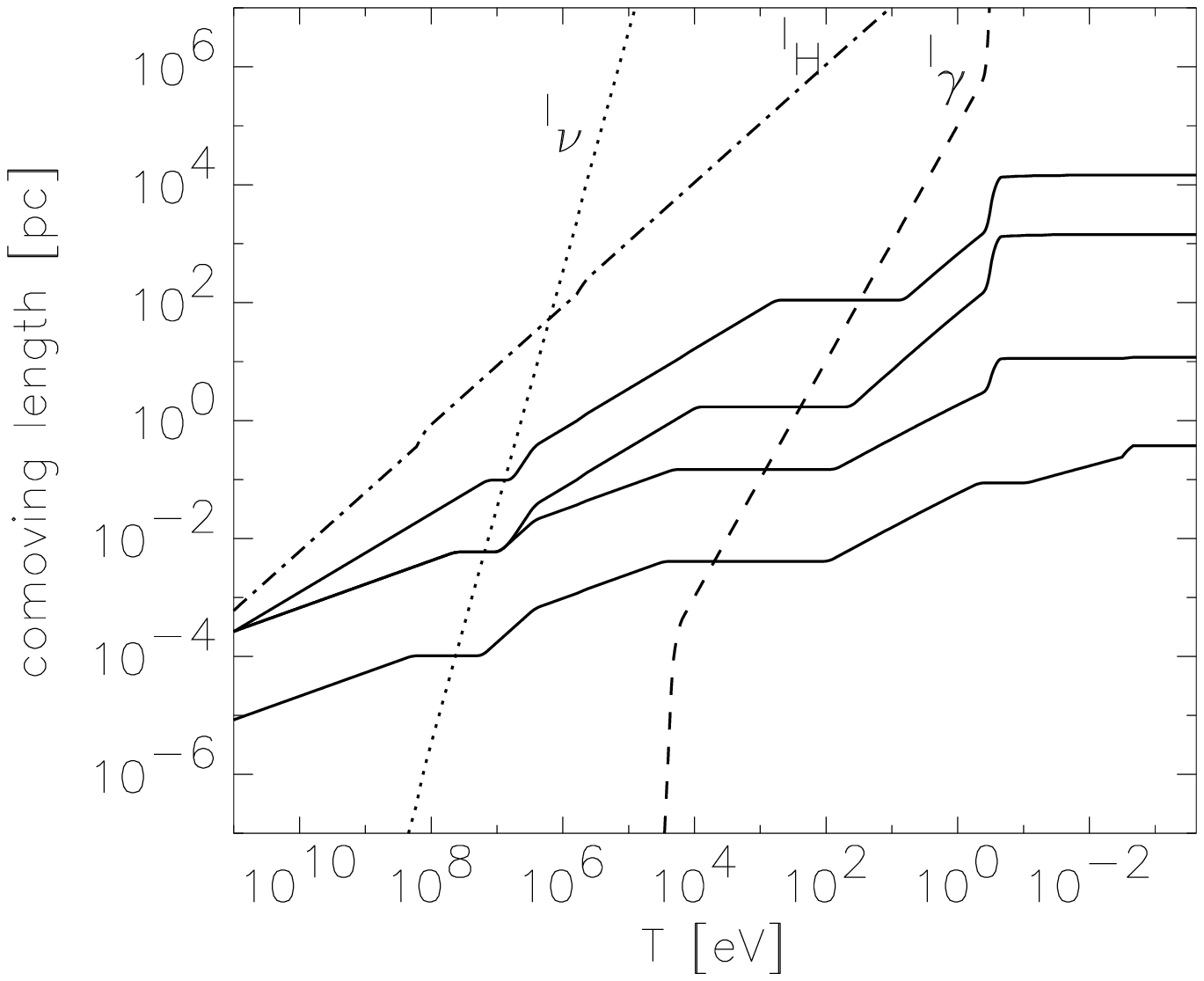}
\caption{The evolution of comoving coherence length for initial
  magnetic field configurations with different spectral indices $n$
  and inital magnetic helicities. Solid lines from top to bottom: (a)
  $h_{\sm{g}} = 1$, $r_{\sm{g}} = 0.01$, (b) $h_{\sm{g}} = 10^{-3}$,
  $n=3$, $r_{\sm{g}} = 0.01$, (c) $h_{\sm{g}} = 0$, $n=3$, $r_{\sm{g}}
  = 0.01$, (d) $h_{\sm{g}} = 0$, $n=3$, $r_{\sm{g}} = 10^{-5}$. The
  labels $l_{\nu}$, $l_{\gamma}$, $l_{\sm{H}}$ refer to the comoving
  mean free paths of neutrinos and photons and the comoving Hubble
  length, respectively. The epoch of magnetogenesis was assumed to
  occur during the electroweak phase transition ($T_g = 100\,\GeV$).}  
\label{fig:length_ew}
\eef

\bef
\showone{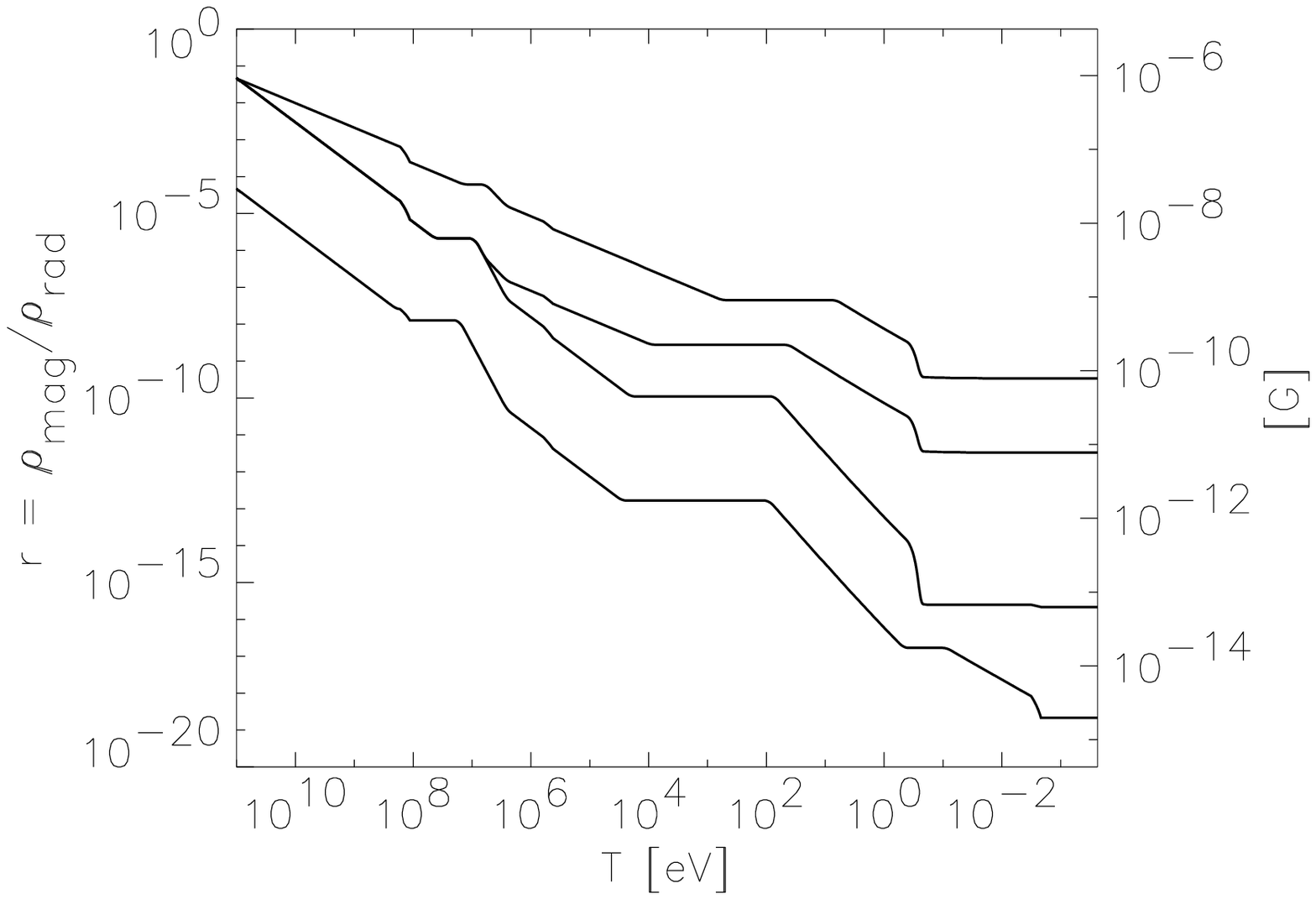}
\caption{The evolution of the relative magnetic energy density $r$
  corresponding to the models shown in Fig.~\ref{fig:length_ew}. Solid
  lines from top to bottom: (a) $h_{\sm{g}} = 1$, $r_{\sm{g}} = 0.01$,
  (b) $h_{\sm{g}} = 10^{-3}$, $n=3$, $r_{\sm{g}} = 0.01$, (c)
  $h_{\sm{g}} = 0$, $n=3$, $r_{\sm{g}} = 0.01$, (d) $h_{\sm{g}} = 0$,
  $n=3$, $r_{\sm{g}} = 10^{-5}$. The epoch of magnetogenesis was
  assumed to occur during the electroweak phase transition ($T_g =
  100\,\GeV$).}
\label{fig:energy_ew}
\eef

\bef
\showone{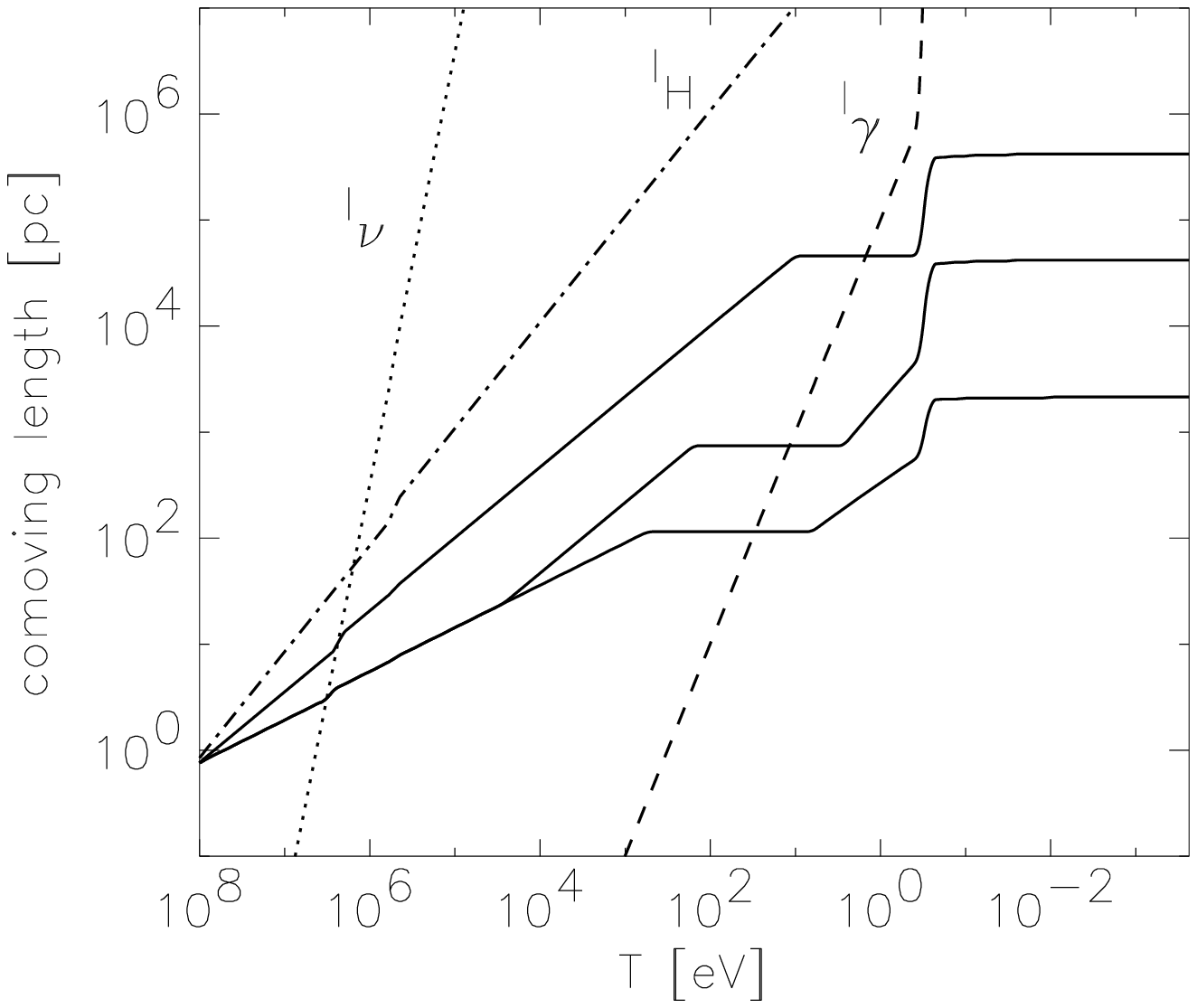}
\caption{The evolution of comoving coherence length for different
  initial magnetic field configurations. Solid lines from top to
  bottom: (a) $h_{\sm{g}} = 1$, $r_{\sm{g}} = 0.083$, $n=3$, (b)
  $h_g=10^{-3}$, $r_{\sm{g}} = 0.083$, $n=3$, (c) $h_g=0$, $r_{\sm{g}}
  = 0.083$, $n=3$. The labels $l_{\nu}$, $l_{\gamma}$, $l_{\sm{H}}$
  refer to the comoving mean free paths of neutrinos and photons and
  the comoving Hubble length, respectively. The epoch of
  magnetogenesis was assumed to occur during the QCD phase transition
  ($T_g = 100\,\MeV$).}
\label{fig:length_qcd}
\eef

\bef
\showone{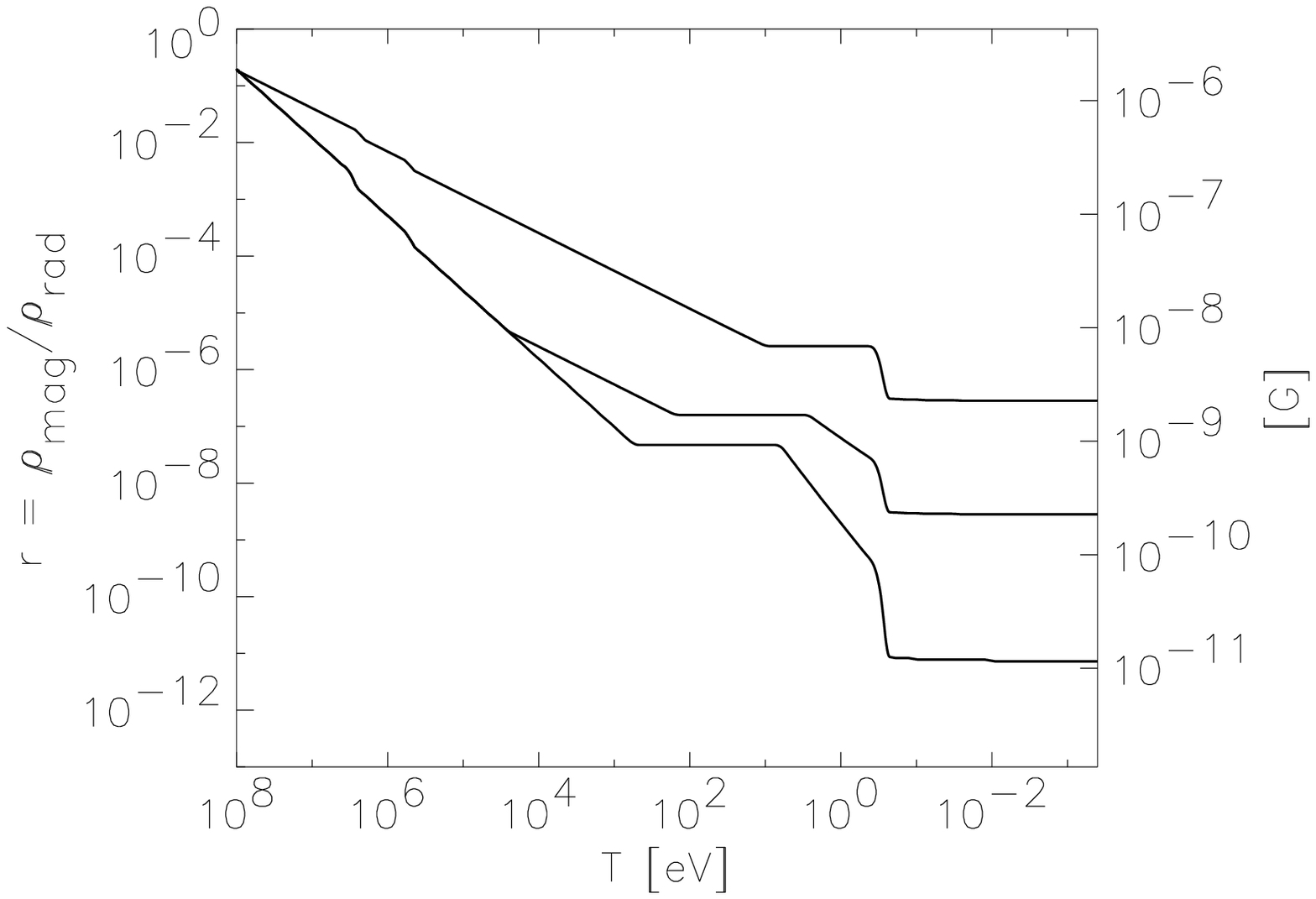}
\caption{The evolution of the relative magnetic energy density $r$
  corresponding to the models shown in Fig.~\ref{fig:length_qcd}. Solid
  lines from top to bottom: (a) $h_{\sm{g}} = 1$, $r_{\sm{g}} =
  0.083$, $n=3$, (b) $h_g=10^{-3}$, $r_{\sm{g}} = 0.083$, $n=3$, (c)
  $h_g=0$, $r_{\sm{g}} = 0.083$, $n=3$. The epoch of magnetogenesis
  was assumed to occur during the QCD phase transition ($T_g =
  100\,\MeV$).}
\label{fig:energy_qcd}
\eef

We have so far assumed that magnetic helicity is negligible. Due to
the high Prandtl numbers in the early Universe helicity is
conserved. Since for typical blue spectra with $n > 1$ non-helical
fields decay more rapidly than helical fields
(cf. Sec.~\ref{sec:turbulence}), initial fields with sub-maximal
helicity $h_g < 1$ will ultimately reach a maximally helical
configuration during the course of field dissipation. Somewhat
oversimplifying it may be understood as the non-helical component
dissipating leaving the fully helical component as a remnant. There
exists a simple criterion of when the fully maximal case is
reached. Using Eq.~(\ref{eq:Bfield_spectrum}) and
Eq.~(\ref{eq:max_helicity}) and the assumption of initial non-helical
evolution, one may show that maximal helicity is reached when the
integral scale has grown to
\be
L^{\rm max}_c = L_{gc}\, h_g^{-{1\over {n-1}}} \quad.
\label{eq:max_length}
\ee

The subsequent evolution of the field is different from the
non-helical case. Whereas processing (i.e. cascading of energy to
smaller scales) on the integral scale and the growth thereof still
proceeds according to the requirement Eq.~(\ref{eq:evolution}), the
required inverse cascade due to the conservation of helicity implies a
transfer of energy from small scales to large scales. The
instantaneous magnetic field spectrum is thus modified compared to
Eq.~(\ref{eq:Bfield_spectrum}) and given by
\be
\tilde{B}_{c}(l_c) = \tilde{B}_{gc} (L_{gc})\,h_g^{1/2}\,
\biggl(\frac{L_c(T)}{L_{gc}}\biggr)^{\frac{n-1}{2}} 
\biggl(\frac{l_c}{L_{gc}}\biggr)^{-n/2}\quad \quad {l_c \ge L_c}
\label{eq:Bfield_spectrum_helic}
\ee
Note that, in contrast to Eq.~(\ref{eq:Bspectrumnohel}), the
prefactor of this spectrum is time dependent (through the temperature
dependence of $L_c$). In accordance with the findings of
Sec.~\ref{sec:turbulence}, the spectrum retains its initial slope $n$
on scales $l_c \ge L_c$. It may be noted that due to the large dynamic
increase of $L_c(T)$ between magnetogenesis and recombination, even
fields with initially very small helicity typically have reached
maximal helicity by recombination.

\subsection{Evolution: Analytic Results}

As discussed above, and observed in Figs.~\ref{fig:length_ew} and
\ref{fig:length_qcd}, the growth of the integral scale before
recombination takes place in different regimes, i.e. turbulent MHD
with neutrino viscosity, viscous MHD with free-streaming neutrinos,
turbulent MHD with photon viscosity, and viscous MHD with
free-streaming photons. Moreover, it is dependent on if maximal
helicity has been reached or not. After recombination strong magnetic
fields ($B_c > B_c^{\eta}$) undergo only slight further evolution in
the turbulent regime, whereas weaker magnetic fields may pass through
an extended viscous hydrogen free-streaming regime.

In the following we give analytic results for the integral scale and
energy density in the different regimes, expressed as functions of the
initial conditions. Here most (but not all) of the notation should be
clear from the definitions in prior sections (e.g. subscripts of $r$,
$S$, $f$, $l$, $q$, $\nu$, $\gamma$, $b$, and $p$ indicate, total
radiation, entropy, particles coupled to the fluid, leptons, 
quarks, neutrinos, photons, baryons, and protons, respectively, 
whereas a subscript $0$ denotes quantities at the present epoch).
One may in principle also derive the transition temperatures,
$T_{\sm{EOT}}$ and $T_{\sm{EOV}}$, at which the fluid passes from a
turbulent state into a viscous one, and vice versa, defined 
by $\Rey\approx 1$. We have nevertheless refrained from doing so, as
the the number of expressions quickly exponentiates, and in some
circumstances (i.e. $0.5\, {\rm MeV} \simge\, T\simge\, 20\, {\rm
  keV}$ due to the $e^{\pm}$ annihilation) closed forms may not be
derived. 

\subsubsection{Evolution before Neutrino Decoupling}

The expressions for the integral scale and energy density during
turbulent MHD evolution before neutrino decoupling are
identical to those before recombination. The
reader is thus referred to Eqs.~(\ref{eq:turbu_ana1})
and~(\ref{eq:turbu_ana2}) , for the
case of submaximally helical fields,
and to~(\ref{eq:turbu_ana3}) and~(\ref{eq:turbu_ana4}), 
for maximally helical fields. 
The expressions for $L_c$ and $r$ in the
viscous neutrino free streaming regime, for submaximally helical
(i.e. $L_c < L^{\sm{max}}_c$) fields are 
\bea
L_c(T) & = & L_{gc}\left[G_1\,
  \left(\frac{1}{G_F^2\,M_{\rm pl}\,T_g^3}\right)
  \right]^{\frac{1}{2+n}} 
  \left(\frac{T}{T_g}\right)^{-\frac{5}{2+n}} 
\label{eq:length_visc_neutrino}
\\
r(T) & = & r_g \left[G_1\,
  \left(\frac{1}{G_F^2\,M_{\rm pl}\,T_g^3}\right)
  \right]^{-\frac{n}{2+n}} 
  \left(\frac{T}{T_g}\right)^{\frac{5 n}{2+n}} 
\label{eq:Bfield_visc_neutrino}
\eea
with 
\be
G_1 \equiv \frac{1}{\zeta(3)} \,
  \left(\frac{49\,\pi^{7}}{405}\right)^{1/2} \,
  \frac{g_r^{5/2}}{g_f\, g_{\nu}\,\left(g_l + g_q\right)}
\ee
Here $L_{gc}$ is given in Eq.~(\ref{eq:length_gc}). As a numerical example,
with index $n=3$ and $g_r = 10.75$, $g_f = 5.5$, $g_{\nu} = 5.25$,
$g_l = 3.5$, and $g_q = 0$ as applicable between QCD-transition and
neutrino decoupling one finds
\bea
L_c (T) & \simeq  & 2.1\times 10^{-2}\, \pc\, \,
   \left(\frac{r_g}{0.01}\right)^{1/2} \,
   \left(\frac{T_g}{100\,\GeV}\right)^{-3/5}\,
   \left(\frac{T}{2.6\,\MeV}\right)^{-1} 
\\
r(T) & \simeq  & 2.1\times 10^{-8}\,  
    \left(\frac{r_g}{0.01}\right) \,
   \left(\frac{T_g}{100\,\GeV}\right)^{-6/5}\,
   \left(\frac{T}{2.6\,\MeV}\right)^{3} 
\eea
For maximally helical fields (i.e. $L > L^{\sm{max}}_c$) the
appropriate expressions are found from Eq.~(\ref{eq:length_visc_neutrino})
and Eq.~(\ref{eq:Bfield_visc_neutrino}) by replacing $n\to 1$, except in
$L^I_{gc}$ which leaves a $\sqrt{n}$ dependence in $L_c$, as 
well as multiplying the RHS of Eq.~(\ref{eq:length_visc_neutrino}) by
$h_g^{1/3}$ and the RHS of Eq.~(\ref{eq:Bfield_visc_neutrino}) by
$h_g^{2/3}$. This yields the following numerical examples 
\bea
L_c (T) & \simeq & 4.8\times 10^{-2}\,\pc\, \,
   \sqrt{n}\,
   \left(\frac{r_g}{0.01}\right)^{1/2} \,
   \left(\frac{h_g}{0.01}\right)^{1/3} \,
   \left(\frac{T_g}{100\,\GeV}\right)^{-1/3}\,
   \left(\frac{T}{2.6\,\MeV}\right)^{-5/3} 
\\
r(T) & \simeq & 3.2\times 10^{-7}\, 
   \left(\frac{r_g}{0.01}\right) \,
   \left(\frac{h_g}{0.01}\right)^{2/3} \,
   \left(\frac{T_g}{100\,\GeV}\right)^{-2/3}\,
   \left(\frac{T}{2.6\,\MeV}\right)^{5/3} 
\eea
where the same parameters as above have been assumed.

\subsubsection{Evolution before Recombination}

The integral scale and magnetic energy during MHD turbulence $\Rey >1$
(equally applicable before neutrino coupling and recombination)
before maximal helicity has been reached (i.e. for $L_c <
L^{\sm{max}}_c$) read  
\bea
L_c (T)  & = &  L_{gc}\,
  \left(\frac{g_S}{g_{r}^{1/2}\, g_f^{1/2}}\right)^{2\over {2+n}}\, 
  \left(\frac{T}{T_g}\right)^{-{2\over {2+n}}}
\nonumber \\
 & = & \left(\frac{2025}{4\pi^7}\right)^{1/6} \,
  \sqrt{n\,r_g} \,
  \frac{\Mpl}{T_g\,T_0} \,
  g_{S0}^{-1/3} \,
  \left(\frac{g_S}{g_{r}^{1/2}\, g_f^{1/2}}\right)^{2\over {2+n}}\, 
  \left(\frac{T}{T_g}\right)^{-{2\over {2+n}}}
\label{eq:turbu_ana1}
\\
r(T) & =  & r_g\, 
  \left(\frac{g_S}{g_{r}^{1/2}\, g_f^{1/2}}\right)^{-{2n\over {2+n}}} \,
  \left(\frac{T}{T_g}\right)^{2n\over{2+n}}\, \label{eq:turbu_ana2}
\eea
A numerical example for $n=3$, $g_r=3.36$, $g_f = 2$, and $g_S =
3.909$ as applicable 
after the $e^{\pm}$-annihilation is given by
\bea
L_c (T) & \simeq & 8.0\times 10^{-2}\,\pc \, \,
   \left(\frac{r_g}{0.01}\right)^{1/2} \,
   \left(\frac{T_g}{100\,\GeV}\right)^{-3/5}\,
   \left(\frac{T}{100\,\keV}\right)^{-2/5} 
\\
r(T) & \simeq  & 3.9\times 10^{-10}\, 
    \left(\frac{r_g}{0.01}\right) \,
    \left(\frac{T_g}{100\,\GeV}\right)^{-6/5}\,
    \left(\frac{T}{100\,\keV}\right)^{6/5} 
\eea
Similar, when maximal helicity has been reached $L_c > L^{\sm{max}}_c$ 
(essentially using the above equations with $n\to 1$ replaced and
inclusion of $h_g$ factors as in the preceding section) one finds
\bea
L_c (T) & =  & L_{gc}\,
  \left(\frac{g_S}{g_{r}^{1/2}\, g_f^{1/2}}\right)^{2/3}\, 
  \left(\frac{T}{T_g}\right)^{-{2/3}}\, 
  h_g^{1/3} \,
\label{eq:turbu_ana3}
\\
r & =  & r_g\,   
  \left(\frac{g_S(T)}{g_{r}^{1/2}\, g_f^{1/2}}\right)^{-2/3}\, 
  \left(\frac{T}{T_g}\right)^{2/3}\,
  h_g^{2/3} \label{eq:turbu_ana4}
\eea
A numerical example for $g_r=3.36$, $g_f = 2$, and $g_S = 3.909$
is given by
\bea
L_c (T) & \simeq  & 4.4\times 10^{-1}\, \pc\, \,
   \sqrt{n}\,
   \left(\frac{r_g}{0.01}\right)^{1/2} \,
   \left(\frac{h_g}{0.01}\right)^{1/3} \,
   \left(\frac{T_g}{100\,\GeV}\right)^{-1/3}\,
   \left(\frac{T}{100\,\keV}\right)^{-2/3} 
\\
r(T) & \simeq & 3.5\times 10^{-8}\, 
    \left(\frac{r_g}{0.01}\right) \,
   \left(\frac{h_g}{0.01}\right)^{2/3} \,
   \left(\frac{T_g}{100\,\GeV}\right)^{-2/3}\,
   \left(\frac{T}{100\,\keV}\right)^{2/3} 
\eea
which is virtually independent of the spectral index $n$.
The expressions during viscous MHD with free streaming photons
(where we assumed $T < 20\,\keV$ as is usually the case) 
for $L_c < L^{\rm max}_c$ are
\bea
L_{c}(T) & =  & L_{gc}\, 
  \left[ G_2\,
  \frac{T_0}{M_{\rm pl}} \,
  \frac{T_0}{\sigma_{\sm{T}}\,n_{b0}\,X_e} \,
  \frac{T_0}{T_g} \right]^{\frac{1}{2+n}}
  \left(\frac{T_g}{T}\right)^{\frac{3}{2+n}}
\label{eq:L_brec_visc} 
\\
r(T) & = & r_g\, \left[G_2\,
  \frac{T_0}{M_{\rm pl}} \,
  \frac{T_0}{\sigma_{\sm{T}}\,n_{b_0}\,X_e} \,
  \frac{T_0}{T_g} \right]^{-\frac{n}{2+n}}
  \left(\frac{T}{T_g}\right)^{\frac{3n}{2+n}} 
\label{eq:B_brec_visc}
\eea
with
\be
G_2 \equiv
  \left(\frac{\pi^3}{45}\right)^{1/2} \,
  g_S^2\, g_r^{-1/2}\, R_r^{1/2}\,
\ee
and where $\sigma_{\sm{T}} = 8\pi\,\alpha^2/3\,m_e^2 \approx 6.65\times
10^{-25}\,\cm^2$ is the Thomson cross section. Here, $R_r \equiv
\rho_r/(\rho_r + \rho_{\sm{DM}})$ accounts for a significant contribution
of dark matter to the Hubble expansion shortly before recombination
and $X_e$ is the ionization fraction ($X_e\approx 1$ before
recombination and $X_e\approx 10^{-4}$ after). As a numerical example for
$n = 3$, $\Omega_{\sm{b}}h^2 = 0.02$, and $X_e=1$ one finds
\bea
L_c (T) & \simeq & 4.0\, \pc \, \,
   \left(\frac{r_g}{0.01}\right)^{1/2} \,
   \left(\frac{R_r}{0.235}\right)^{1/10}\,
   \left(\frac{T_g}{100\,\GeV}\right)^{-3/5}\,
   \left(\frac{T}{0.259\,\eV}\right)^{-3/5} 
\\
r(T)  & \simeq & 3.1\times 10^{-15}\, 
    \left(\frac{r_g}{0.01}\right) \,
   \left(\frac{R_r}{0.235}\right)^{-3/10}\,
   \left(\frac{T_g}{100\,\GeV}\right)^{-6/5}\,
   \left(\frac{T}{0.259\,\eV}\right)^{9/5} 
\eea
where $T\simeq 0.259\,\eV$ corresponds to the temperature at
recombination and $R_r \simeq 0.235$ for $\Omega_{\sm{tot}}h^2 = 0.15$.
Similarly, for maximally helical fields $L > L_{\sm{max}}^h$ one finds
the analytic expressions from those for the submaximal case, as above,
by simply replacing in Eqs.~(\ref{eq:L_brec_visc}) and (\ref{eq:B_brec_visc})
$n\to 1$, except in $L_{gc}$, as well as adding a factor
$h_g^{1/3}$ in Eq.~(\ref{eq:L_brec_visc}) and a factor
$h_g^{2/3}$ in Eq.~(\ref{eq:B_brec_visc}).
Numerical examples are
given by 
\bea
L_c (T) & \simeq & 0.3\, \kpc\, \, \sqrt{n}\,
   \left(\frac{r_g}{0.01}\right)^{1/2} \,
   \left(\frac{h_g}{0.01}\right)^{1/3} \,
   \left(\frac{R_r}{0.235}\right)^{1/6}\,
   \left(\frac{T_g}{100\,\GeV}\right)^{-1/3}\,
   \left(\frac{T}{0.259\,\eV}\right)^{-1} \\
r(T)      & \simeq & 5.2\times 10^{-11}\, 
    \left(\frac{r_g}{0.01}\right) \,
   \left(\frac{h_g}{0.01}\right)^{2/3} \,
   \left(\frac{R_r}{0.235}\right)^{-1/6}\,
   \left(\frac{T_g}{100\,\GeV}\right)^{-2/3}\,
   \left(\frac{T}{0.259\,\eV}\right) 
\eea

\bef
\showone{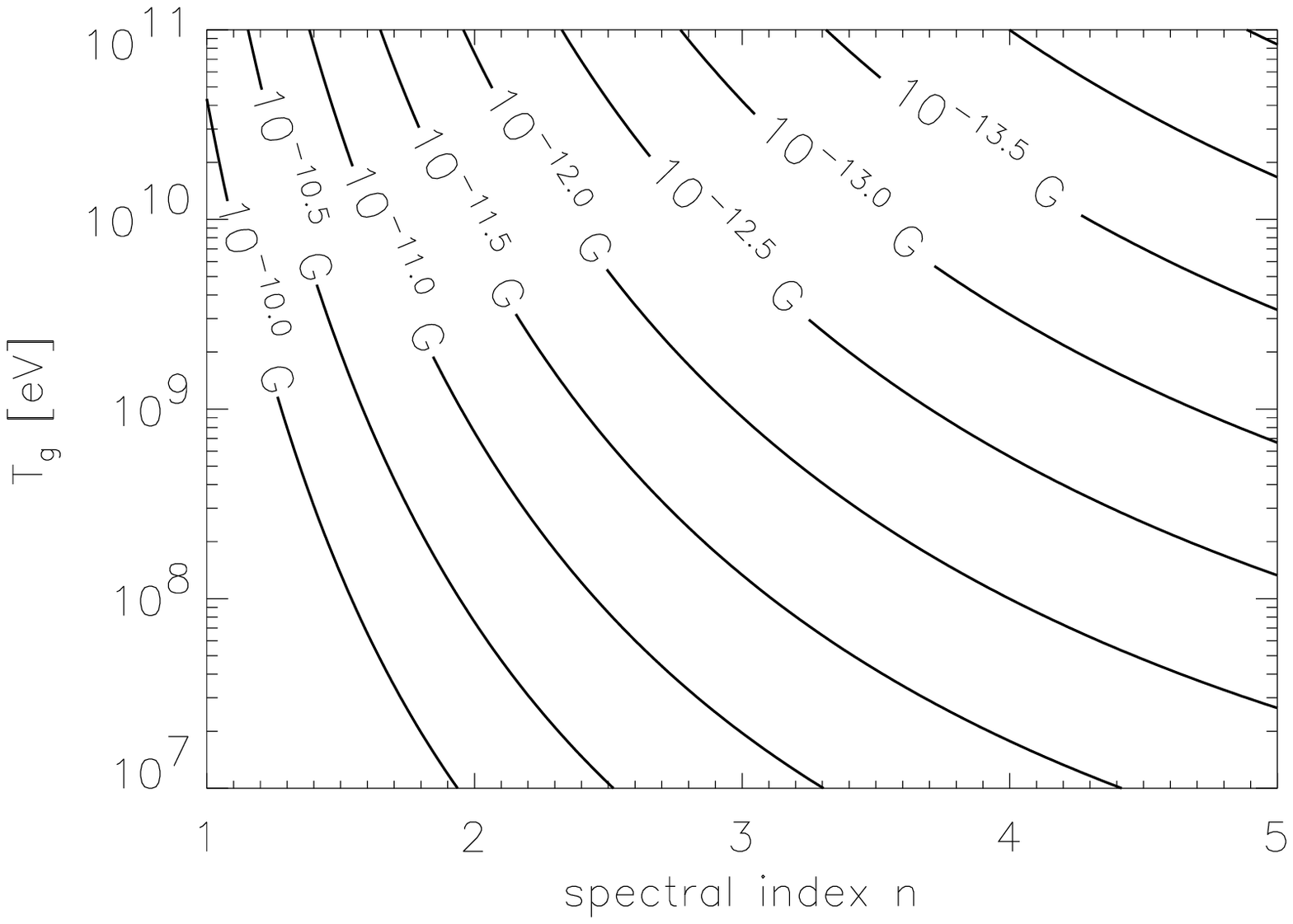}
\caption{Shows the final magnetic field strength, $B(T_0)$, for
  submaximal (i.e. $L < L^{\sm{max}}$) magnetic fields in the
  ($n$, $T_{\sm{g}}$) parameter space. Here, the initial magnetic field
  strength is $r_{\sm{g}} = 0.01$. Results for different $r_{\sm{g}}$ may
  be obtained by scaling the field strength by $(r_{\sm{g}}/0.01)^{1/2}$.}
\label{fig:cont_mag_n_Tgen}
\eef

\bef
\showone{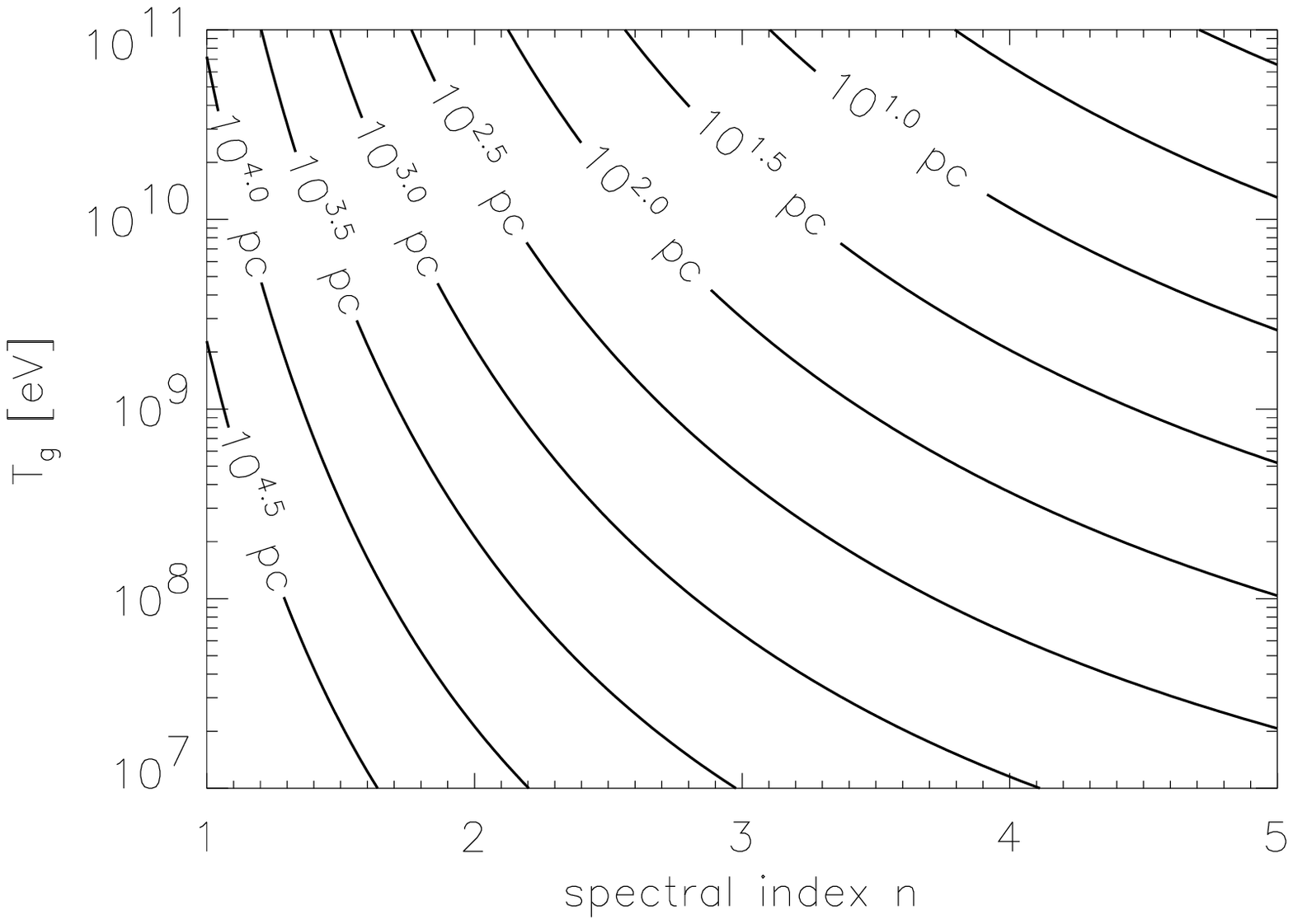}
\caption{Shows the final coherence length, $L(T_0)$, of submaximal
  magnetic fields (i.e. $L < L^{\sm{max}}$) in the ($n$, $T_{\sm{g}}$)
  parameter space. Here, the initial magnetic field strength is
  $r_{\sm{g}} = 0.01$. Results for different $r_{\sm{g}}$ may
  be obtained by scaling the coherence length by $(r_{\sm{g}}/0.01)^{1/2}$}
\label{fig:cont_Lc_n_Tgen}
\eef

\subsubsection{Evolution after Recombination}

In the turbulent regime after recombination the quantities of interest
for $L_c < L^{\sm{max}}_c$ are given by  
\bea
L_c & = & L_{gc} \,
 \left(\frac{\Omega_{\gamma}}{\sqrt{3\,\Omega_{\sm{tot}}\,\Omega_{\sm{b}}}} \, 
       g_{S0} \, 
       \ln(a/a_{\sm{rec}}) 
 \right)^{2\over {2+n}} \,
   \left(\frac{T_g}{T_0}\right)^{{2\over {2+n}}} 
\label{eq:Lc0nohel}
\\
r & = & r_g\, 
  \left(\frac{\Omega_{\gamma}}{\sqrt{3\,\Omega_{\sm{tot}}\,\Omega_{\sm{b}}}}\,
    g_{S0} \,
    \ln(a/a_{\sm{rec}})
  \right)^{-{2n\over {2+n}}} \,
  \left(\frac{T_0}{T_g}\right)^{2n\over{2+n}}
\eea
including a mild logarithmic growth factor $\ln(a/a_{\sm{rec}})$. 
Here $\Omega_{\gamma}$, $\Omega_{\sm{b}}$, and
$\Omega_{\sm{tot}}$ are the present day fractional contributions to the
critical density of CMBR photons, baryons, and total matter,
respectively. (In the derivation of this expression the contribution
of radiation to the total density after recombination has been
neglected. This induces about $\sim 10\%$ error in $r$ and $5\%$ in
$L_c$ immediately at recombination for the values below, but is
asymptotically correct).
Numerical examples for $n=3$,
$g_{S0}=3.909$, $\Omega_{\sm{tot}}h^2 = 0.15$,  $\Omega_{\sm{b}}h^2 =
0.02$, $\Omega_{\gamma}h^2 = 2.48\times 10^{-5}$ are
\bea
L_c (T) & \simeq & 12\, \pc \,\,
   \left(\frac{r_g}{0.01}\right)^{1/2} \,
   \left(\frac{T_g}{100\,\GeV}\right)^{-3/5} 
\label{eq:Lc0exnohel}
\\
r(T)  & \simeq & 1.1\times 10^{-16}\, 
    \left(\frac{r_g}{0.01}\right) \,
   \left(\frac{T_g}{100\,\GeV}\right)^{-6/5} 
\\ 
B_c(T) & \simeq & 6.0\times 10^{-14} \, \G \, \,
    \left(\frac{r_g}{0.01}\right)^{1/2} \,
   \left(\frac{T_g}{100\,\GeV}\right)^{-3/5}
\label{eq:Bc0exnohel}
\eea
where we have also evaluated the comoving field strength via
Eq.~(\ref{eq:mag_eq}). For the above quantities we neglected the
factor ${\rm ln}(a/a_{\sm{rec}})$ (as for most fields the period of
turbulent MHD after recombination is rather short) 
Similarly, for $L > L_{\rm max}^h$ (when having attained
maximal helicity) one finds
\bea
L_c & = & L_{gc} \,
  \left(\frac{\Omega_{\gamma}}
         {\sqrt{3\,\Omega_{\sm{tot}}\,\Omega_{\sm{b}}}} \, 
        g_{S0}\,
        \ln(a/a_{\sm{rec}}) 
  \right)^{2/3} \,
  \left(\frac{T_g}{T_0}\right)^{2/3} \, 
  h_g^{1/3} 
\label{eq:Lc0hel}
\\
r   & = & r_g \, 
  \left(\frac{\Omega_{\gamma}}
           {\sqrt{3\,\Omega_{\sm{tot}}\,\Omega_{\sm{b}}}}\,
        g_{S0}\,
        \ln(a/a_{\sm{rec}})
  \right)^{-2/3} \, 
  \left(\frac{T_0}{T_g}\right)^{2/3} \,
  h_g^{2/3}
\eea
with numerical examples (with the input numerical values as above) given by
\bea
L_c (T) & \simeq & 1.9\, \kpc\,\, 
   \sqrt{n} \,
   \left(\frac{r_g}{0.01}\right)^{1/2} \,
   \left(\frac{h_g}{0.01}\right)^{1/3} \,
   \left(\frac{T_g}{100\,\GeV}\right)^{-1/3}
\label{eq:Lc0exhel}
\\
r(T) & \simeq & 8.1\times 10^{-12}\, 
   \left(\frac{r_g}{0.01}\right) \,
   \left(\frac{h_g}{0.01}\right)^{2/3} \,
   \left(\frac{T_g}{100\,\GeV}\right)^{-2/3}
\\
B_c(T) & \simeq & 1.6\times 10^{-11}\,\G\, \,
    \left(\frac{r_g}{0.01}\right)^{1/2} \,
   \left(\frac{h_g}{0.01}\right)^{1/3} \,
   \left(\frac{T_g}{100\,\GeV}\right)^{-1/3}
\label{eq:Bc0exhel}
\eea
Note that there is only a residual dependence on spectral index $n$.

\bef
\showone{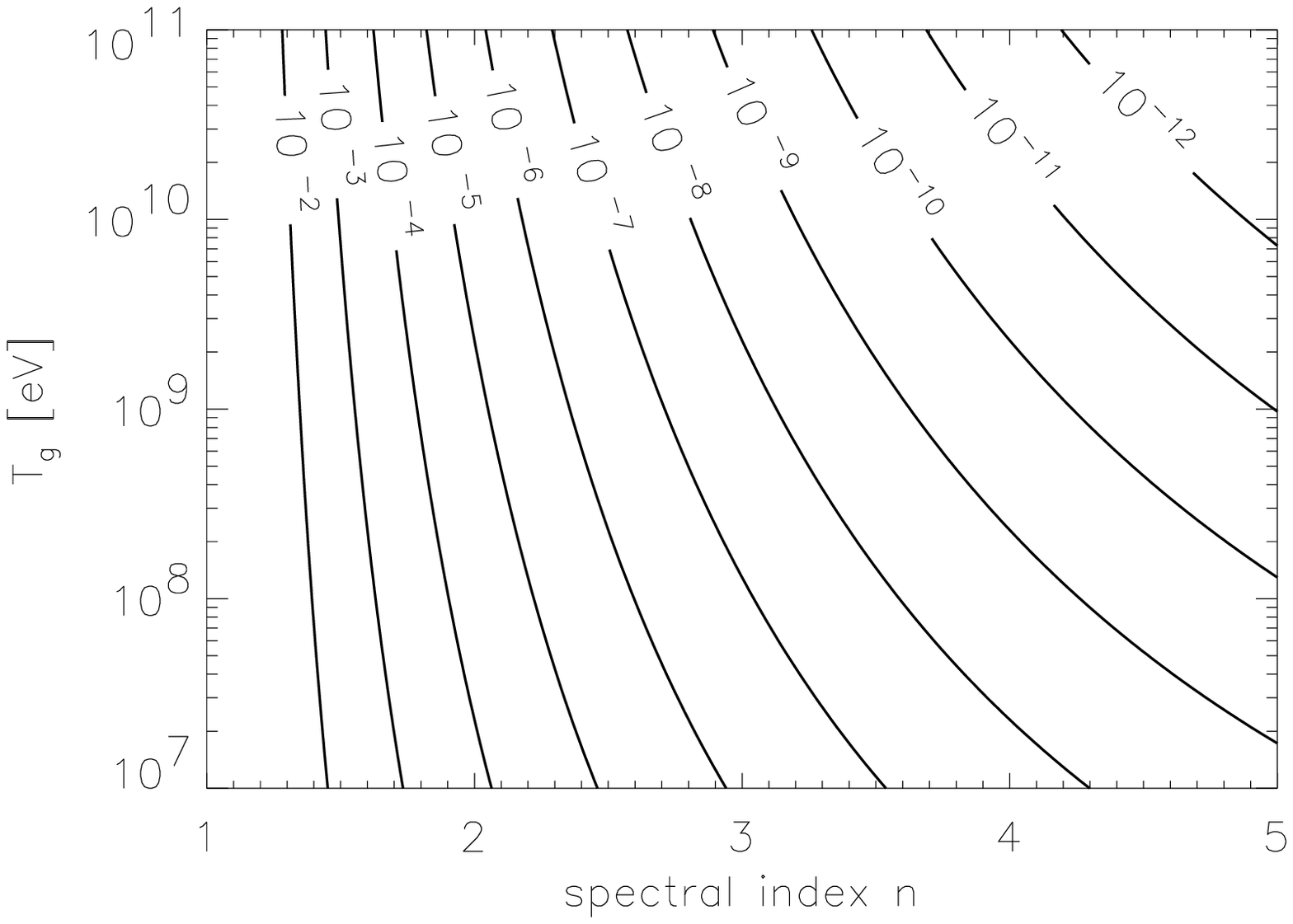}
\caption{Shows the minimal relative initial magnetic helicity,
  $h_{\sm{g}}$, necessary for a magnetic field to have become maximal
  helical (i.e. $L \simge L^{\sm{max}}$) at the present epoch. 
  For instance, initial magnetic
  fields generated at $T_{\sm{g}} = 100\,\MeV$ with $n = 3$ become maximal
  helical if $h_{\sm{g}} \simge 1.2\times 10^{-7}$. Note, this
  condition is independent of the initial magnetic field strength,
  $r_{\sm{g}}$.}
\label{fig:cont_h_n_Tgen}
\eef

\bef
\showone{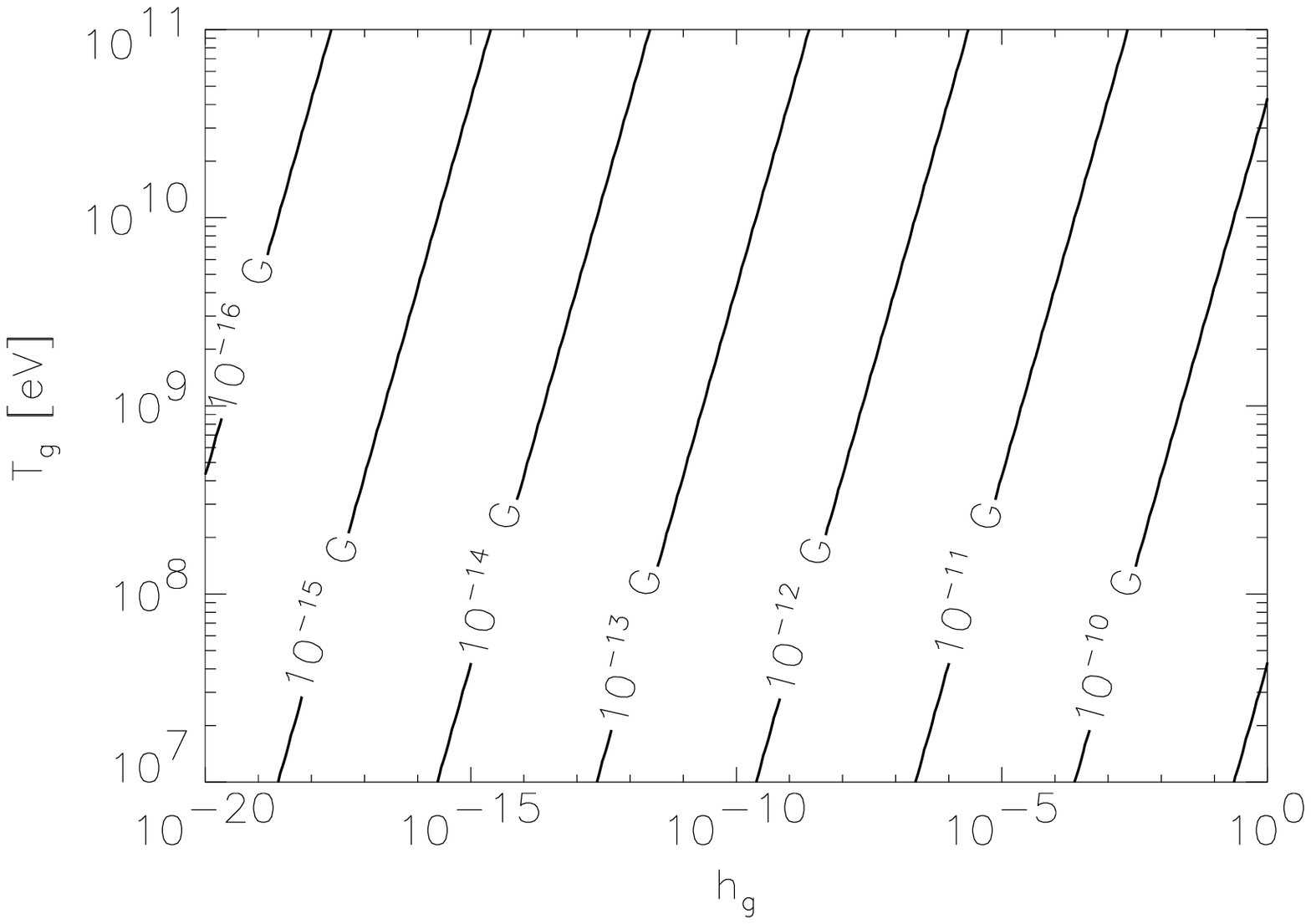}
\caption{Shows the final magnetic field strength, $B(T_0)$, for
  magnetic fields which have become maximal helical (i.e. $L \simge
  L^{\sm{max}}$) in the ($h_{\sm{g}}$, $T_{\sm{g}}$) parameter
  space. Here, the initial magnetic field strength is $r_{\sm{g}} =
  0.01$, and for different $r_{\sm{g}}$ results scale with
  $(r_{\sm{g}}/0.01)^{1/2}$. Note the absence of a dependence on spectral
  index $n$.}
\label{fig:cont_Bmh_h_Tgen}
\eef

\bef
\showone{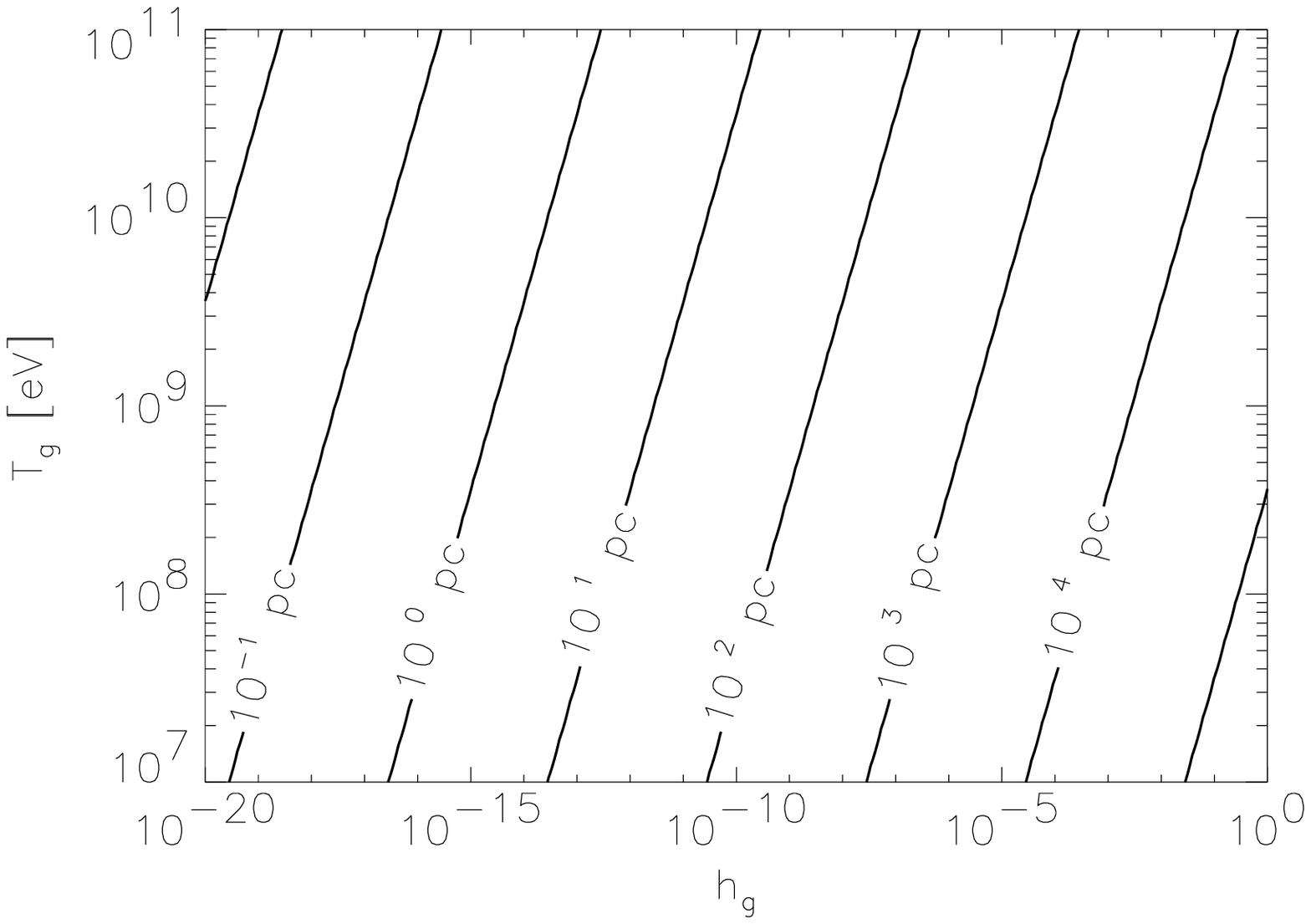}
\caption{Shows the final coherence length, $L(T_0)$, for
  magnetic fields which became maximal helical (i.e. $L \simge
  L^{\sm{max}}$) in the ($h_{\sm{g}}$, $T_{\sm{g}}$) parameter
  space. Here, the initial magnetic field strength is $r_{\sm{g}} =
  0.01$, and for different $r_{\sm{g}}$ results may be rescaled
  by $(r_{\sm{g}}/0.01)^{1/2}$.}
\label{fig:cont_Lmh_h_Tgen}
\eef

\section{Summary and Discussion}
\label{sec:summary}

The detailed numerical and analytical examination presented in the
previous chapters has lead to a surprisingly simple picture concerning
the gross features of cosmic magnetic field evolution, in particular,
the evolution of magnetic coherence scale and energy density.  The
growth of the coherence scale is described by the simple {\em
causality} relation
\be
v(L)/L\approx H(T)
\label{eq:causality}
\ee
independent if ocurring in high kinetic Reynolds number $Re\gg 1$
turbulent flow with $v\approx \vAl$ or during the multiple epochs of
viscous ($Re\ll 1$) MHD evolution with $v\ll \vAl$ (with $v$ given by
Eq.~(\ref{eq:viscous_v})) and independent of the helical properties of
the fields. In particular, non-linear (direkt) cascading of magnetic
energy to the dissipation scale always occurs on that scale.
Remaining magnetic energy densities after evolution from very high
temperature to an epoch with temperature $T$ are then simply given, in
the sub-maximal case, by all the initial magnetic energy present on
scales $l\simge L^>$, with $L^>$ the as yet largest length scale
having been processed during prior evolution, and in the maximally
helical case, by conservation of helicity density
(see Eq.~(\ref{eq:Hel_by_E})) with the field sitting on scales $l\simge
L^>$. Quantities of particular interest to cosmology are the
anticipated present-day coherence length and magnetic field strength
given particular initial conditions immediately after the epoch of
magnetogenesis. We have shown that whereas strong magnetic fields
$B\gg B_c^{\eta}\approx 10^{-13}\,\G$ are essentially undergoing no
further evolution (i.e. growth of $L^{>}$) after recombination weak
fields $B\simle B_C^{\eta}$ do. In either case, after the later epoch
of reionization the distinction between strong and weak disappears
such that any reasonable strength field is turbulent at present. The
present day field strength are then simply obtained by applying
Eq.~(\ref{eq:causality}) with $v\approx \vAl$ today.  This yields the
correlation
\be
B_0 \approx 5\times 10^{-12}\,\G\,
\biggl(\frac{L_c}{\rm kpc}\biggr)
\label{eq:B0}
\ee 
between magnetic field strength and magnetic correlation
length~\footnote{Note that Eq.~(\ref{eq:B0}) displays a weak spectral
index dependence, i.e. $1/\sqrt{n}$ on the RHS, and we have assumed
$n=3$ in Eqs.~(\ref{eq:B0}) and (\ref{eq:Brec}) below}.  The magnetic
correlation length itself is given by the initial conditions after the
epoch of magnetogenesis (cf. Eq.~(\ref{eq:Lc0nohel}) and
Eq.~(\ref{eq:Lc0hel}) for the sub-maximal helical- and maximally
helical- case, respectively, with example values for an $n=3$ spectrum
given in Eq.~(\ref{eq:Lc0exnohel}) -- Eq.~(\ref{eq:Bc0exnohel}) and
for fields which during the course of evolution have become maximally
helical in Eq.~(\ref{eq:Lc0exhel}) -- Eq.~(\ref{eq:Bc0exhel})).  Both
quantities are shown for non-helical fields as a function of spectral
index and magnetogenesis temperature in
Figs.~\ref{fig:cont_mag_n_Tgen} and \ref{fig:cont_Lc_n_Tgen}.

\bef
\showone{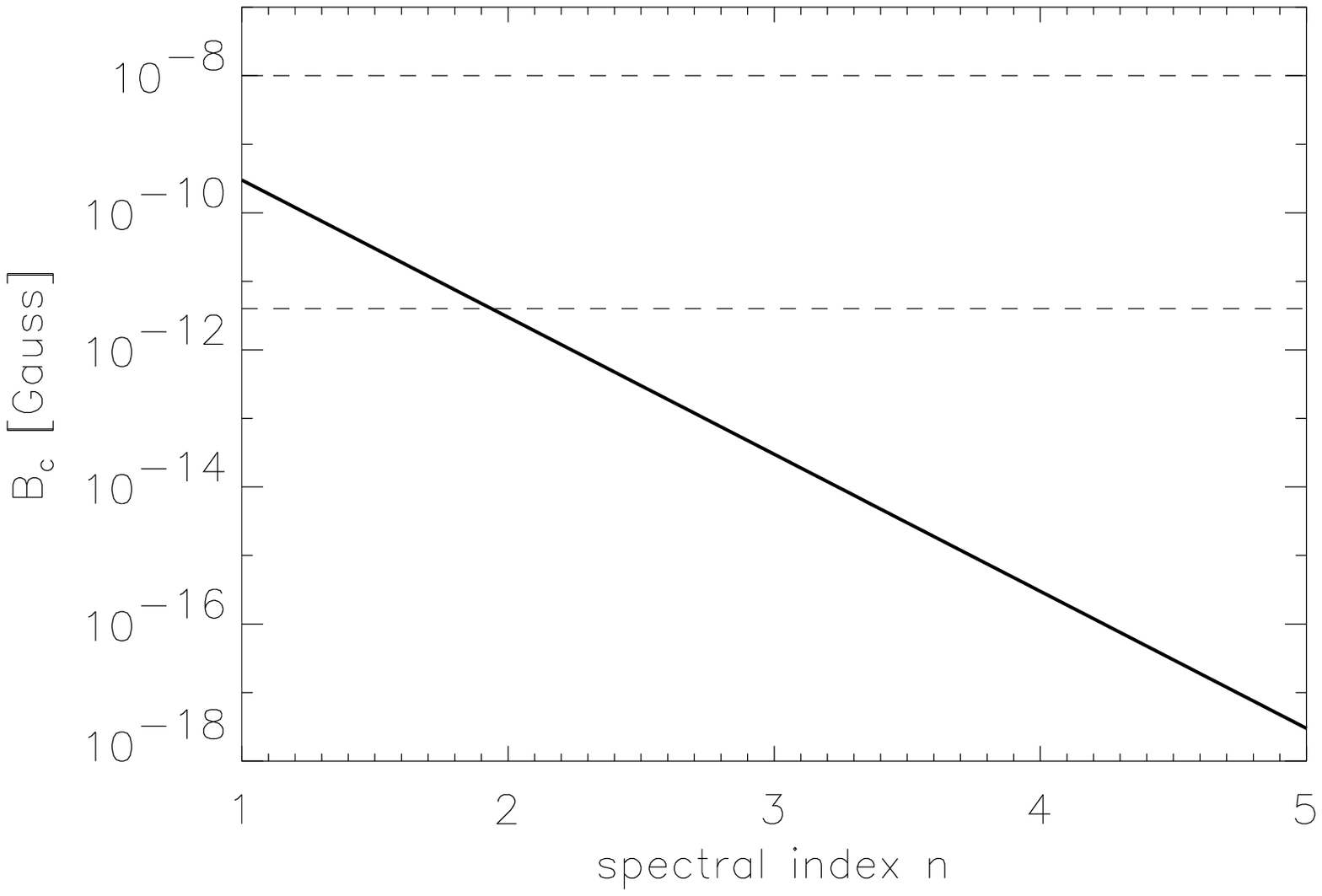}
\caption{Exclusion plot on the comoving magnetic field strength on $L = 10 \,
  \Mpc$ (equally applicable to the present epoch and recombination) as 
  a function of magnetic field spectral index. The solid lines shows
  the upper limit on $B_c(10{\rm Mpc})$ from an excessive distortion
  of the CMBR blackbody spectrum by magnetic field dissipation at
  redsfift $z = 3\times 10^6$~\cite{jeda00}, whereas the horizontal lines
  (from top to bottom) show the upper limits from present-day
  intergalactic Faraday rotation measurements~\cite{kim91} and the possible
  overproduction of cluster magnetic fields (see text), respectively.}
\label{fig:B_by_n}
\eef

Concerning the evolution of maximally helical fields we have found and
reverified prior work~\cite{chris01} on the intriguing property of
self-similar evolution of such fields. In particular, though maximal
helical fields keep their initial spectral index on large scales
(i.e. for $L\simge L^{>}$) the amplitude of the large-scale tail of
magnetic fluctuations is subject to a seemingly ``acausual''
amplification on scales which are far larger than the distance an
Alfv\'en wave may travel (cf. Figs.~\ref{fig:e_k_turb_h}
and~\ref{fig:E_k_visc_fs_n5_h}). The large-scale magnetic field
spectrum for maximally helical fields is given by
Eq.~(\ref{eq:Bfield_spectrum_helic}), whereas
Eq.~(\ref{eq:Bspectrumnohel}) describes that of submaximally helical
fields.  We have numerically disputed the claimed effect
(e.g.~\cite{field00,vach01}) that maximally helical fields do not
excite fluid motions and are therefore not subject to viscous damping
(cf. Sec.~\ref{sec:helical_fields}). Rather, only due to the
excitation of fluid motions the magnetic correlation length of
maximally helical fields may continuously grow during the evolution of
the early Universe. It is important to note that due to the large
dynamic increase of $L^{>}$ between the epoch of magnetogenesis and
the present (and the associated large dissipation of magnetic energy)
a field with a minute amount of initial helicity typically evolves
into a maximally helical field at present.  This happens when the
magnetic correlation length has grown beyond that given in
Eq.~(\ref{eq:max_length}).  Fig.~\ref{fig:cont_h_n_Tgen} shows the
amount of initial helicity, $h_g$, (cf. Eq.~(\ref{eq:max_helicity}))
as a function of magnetic spectral index required to reach a maximally
helical state at present.  Completely helical fields may thus not
necessarily be considered an unlikely remnant of the early Universe.

The evolution of magnetic fields during epochs with intermediate
redshifts $z\approx 10^3 - 10^7$ is described by turbulent evolution
at higher redshifts, followed by a viscous MHD period without further
growth of $L^{>}$, and a viscous MHD period with comparatively rapid
growth of $L^{>}$ with viscosity due to free-streaming photons
(cf. Figs.~\ref{fig:length_ew} and \ref{fig:length_qcd}).  For
essentially all interesting magnetic field strengths and spectra the
plasma is in this last state shortly before recombination, allowing
for the prediction of a correlation 
\be 
  B_{\rm rec}\approx 8\times 10^{-11}\,\G\, \biggl(\frac{L_c}{\rm
    kpc}\biggr) 
\label{eq:Brec}
\ee
between magnetic field strength and correlation length shortly before
recombination. Eq.~(\ref{eq:Brec})
is obtained via Eq.~(\ref{eq:causality}) and noting that 
$v\approx \vAl^2/(\alpha_{\gamma}L)$ during viscous photon free-streaming.
The correlation length to be employed in Eq.~(\ref{eq:Brec}) may only
be derived when the initial conditions shortly after the magnetogenesis
scenario are known, i.e. via Eq.~(\ref{eq:L_brec_visc}) and the comments
further below concerning maximally helical fields, whereas the instantaneous
spectra are again given by Eqs.~(\ref{eq:Bspectrumnohel}) and
~(\ref{eq:Bfield_spectrum_helic}).

The correlation in Eq.~(\ref{eq:Brec}) is almost identical to that one
expects from a linear analysis (JKO98). In contrast, for the
formulation of limits on primordial magnetic fields due to magnetic
field dissipation at redshift $z\approx 2.5\times 10^6$ and the
concomitant production of spectral $\mu$ distortions in the CMBR,
Ref.~\cite{jeda00} have employed the results of a linear analysis
leading to the claim that fields of $3\times 10^{-8}\,\G$ on scales of
$400\,\pc$ are disallowed. Though the limiting field strength does not
change when non-linear evolution is considered, as it is an energetic
constraint, the comoving length scale does.  This is due to the bulk
of energy being not contained on the dissipation scale (at $\sim
400\,\pc$) but rather on the integral scale given by applying
Eq.~(\ref{eq:B0}) at $z\approx 2.5\times 10^6$. However,
coincidentally the change is only mild, moving the limiting scale from
$400\,\pc$ up to $1\,\kpc$, since for such field strength the flow at
$z\approx 2.5\times 10^6$ is only mildly turbulent (i.e. $Re\approx
100$) and since in the viscous regime both treatments almost
coincide. When $B\simge B_c^{\eta}$ the cosmic recombination process
is associated with an almost instantaneous jump in the magnetic
correlation length. How large this jump is than depends on the
magnetic field spectral index determined during the magnetogenesis
epoch.  It would be interesting to examine at what field strength the
associated energy dissipation could impact on the recombination
process itself, especially in light of fields with strength $B\simge
6\times 10^{-11}\,\G$ being able to produce small-scale density
perturbations as then $\vAl\simge v_b$ (cf. Eqs.~(\ref{eq:mag_eq}),
(\ref{eq:va}), and (\ref{eq:vb})).

With the advances in high-precision CMBR anisotropy observations the
interest in putative signals due to primordial magnetic fields has
immensely risen. Essentially all current magnetic field induced CMBR
anisotropy examinations assume fields of strengths $\sim 10^{-10} -
10^{-9}\,\G$ on scales roughly the Silk scale, $L\approx
10\,\Mpc$. This is done, of course, since for much weaker fields the
signal is hardly observable and when moving to much smaller scales not
only are satellite missions like WMAP and Planck not able to resolve
these but also are signals naively expected to be reduced due to the
thickness of the last scattering surface $\sigma\approx 10\,\Mpc$. By
inspection of Eq.~(\ref{eq:Brec}) it is clear that the scale of $\sim
10\,\Mpc$ may not be the integral scale but rather a scale much
beyond. We argue here that unless substantial primordial magnetic
fields have their origin during an inflationary phase a search for
primordial magnetic fields of $\sim 10^{-9}\, \G$ on $10\,\Mpc$ seems
futile (at best controversial). This is due to a number of
reasons. First, with fields which are causally generated during, for
example, early cosmic phase transitions such strength on these scales
are impossible to reach. This is of course due to the smallness of the
Hubble scale in the very early Universe and since due to causality the
spectrum must be sufficiently blue.  Second, due to the blueness of
the spectrum of causally generated magnetic fields, limits on smaller
scales are easily violated.  Fig.~\ref{fig:B_by_n} shows the maximum
possible magnetic field strength on the scale $10\,\Mpc$ as a function
of magnetic field spectral index. Fields which are above this value
produce CMBR spectral $\mu$ distortions in excess of those
observed~\cite{jeda00}.  It is seen that only for unrealistically
small spectral indices $B\sim 10^{-9}\,\G$ on $10\,\Mpc$ may be
reached. Finally, a direct constraint on the scale $\sim 10\,\Mpc$ may
be applied when the low-redshift collapse of a magnetized plasma to a
cluster is considered~\cite{DBL99} It is found that
pre-cluster-collapse fields in of $4\times 10^{-12}\,\G$
(corresponding to the authors $10^{-9}\,\G$ at redshift $z=15$) are
sufficient to reproduce observed Faraday rotation measures in present
day clusters. Larger fields seem to overproduce the Faraday rotation
measure and should therefore be ruled out.

When trying to detect primordial magnetic fields a more promising and
realistic alley should be the search for CMBR anisotropies on very
small scales, in particular, on scales $\sim 10\,\kpc$ (corresponding
to multipoles $l\sim 10^6$), possibly close or only slightly above the
integral scale, rather than the canonical $10\,\Mpc$ (multipoles
$l\sim 10^3$). This is due to the existence of viable scenarios
producing fields of interesting amplitude $\sim 10^{-9}\,\G$ on such
scales and further such fields evading constraints from CMBR spectral
distortions and observed cluster magnetic fields.  Moreover, though
this scale is much below the width of the last scattering surface, the
expected signal is not necessarily small.  In particular, for small
scales $l\simle 1\,\Mpc$ the magnetic field induced peculiar
velocities are $v\approx \vAl^2/(\alpha_{\gamma}l)$. CMBR temperature
fluctuations follow $\delta T/T\propto v$ and for scales $l < \sigma$
are additionally suppressed by $\sqrt{l/\sigma}$ due to the thickness
of the last scattering surface. Combing these factors one finds
$\delta T/T\propto l^{-n-1/2}$ for $l\simge L$, thus an increasing
signal with decreasing scale. Here $n$ is the spectral index of the
primordial magnetic field. Primordial magnetic fields should therefore
leave their strongest CMBR signal on small scales $l\sim 10^6$.  It
remains to be seen if contamination of the primordial CMBR
anisotropies by foregrounds may pose serious problems to such
observations.

Last but not least, we have already challenged in a prior
publication~\cite{bj03} the long-standing and wide-spread belief that
cluster magnetic fields may not be entirely of primordial origin. It
is typically argued that causal magnetogenesis scenarios (as, for
example, due to local processes during the QCD- or electroweak
transitions) yield only weak magnetic fields $\simle 10^{-20}\,\G$ on
the (pre-collapse) scale of a cluster of galaxies.  Since during
cluster collapse further magnetic field amplification by only modest
factors $10^3-10^5$ due to magnetic flux conservation result, starting
from $10^{-20}\,\G$ implies that one is still far from the observed
$\mu\G$ fields in clusters, requiring further field amplification
processes such as a dynamo.  The problem with this argument is that, a
priori, it is not clear if the initial magnetic fields have to reside
on the cluster scale itself, or if magnetic field energy density
contained on much smaller scales may, during the collapse be
transferred to the cluster scale. In fact, the only numerical
simulation of the collapse of a magnetized plasma to a cluster to
date~\cite{DBL99} seems to indicate an approximate independence of the
final result on initial magnetic coherence length, with final cluster
Faraday rotation measures only dependent on the pre-collapse magnetic
energy density (required are $B\approx 4\times 10^{-12}\,\G$).  It is
currently not clear by what mechanism magnetic energy may inverse
cascade from small scales to large scales during the cluster
collapse. But if indeed it does, cluster magnetic fields could be
entirely primordial, since magnetic fields of $\sim
10^{-12}-10^{-11}\,\G$ on approximately $\sim \kpc$ scales are
possible by either having magnetogenesis occur late, during the QCD
phase transition, and/or magnetogenesis scenarios which generate a
very small amount of initial helicity (cf. Eqs.~(\ref{eq:Lc0nohel}),
(\ref{eq:Lc0hel}), and (\ref{eq:B0})).  It is interesting to note that
such a scenario also lead to a ``prediction'' of magnetic field
strength and amplitudes in voids, far from galaxies. Fields in such
environments are presumably not affected by magnetic fields in
galactic outflows and could be, in the optimistic case observable by
future technology~\cite{lemo97,bert02,plaga95}.

\acknowledgments We acknowledge technical assistance and emotional
support by T.~Abel, A.~Kercek, and M.~Mac Low to undertake this study.

\appendix
\section{The MHD equations in Minkowski space}
\label{apx:general_MHD}

We are using Gaussian natural units, i.e. $\mu_0 = 1 = \epsilon_0$,
and $c = 1$, for solving the MHD equations. On a static background in
the Newtonian and non-relativistic limit 
the MHD equations are given by (see e.g. \cite{jack75})
\bea
\dfdt{\rho} + \diver\left(\vv\,\rho\right) & = & 0 
\label{eq:apx_matter}
\\
\dfdt{\vv} + \left(\vv\cdot\nabla\right)\vv 
 & = & 
 - \frac{1}{\rho}\,\nabla\,p 
 - \frac{\vB\times\left(\curl\vB\right)}{4\pi\,\rho}
 - \nabla\,\Phi 
\label{eq:apx_Euler}
\\
\dfdt{\vB} & = & \curl\left(\vv\times\vB\right) 
+ \frac{1}{4\pi\sigma}\nabla^2\vB
\label{eq:apx_induction}
\\
\triangle\,\Phi & = & 4\pi\,\rho  \quad,
\label{eq:apx_Poisson}
\eea
where $\rho$, $\vv$, $p$, $\vB$, $\Phi$, and $\sigma$ are matter density, fluid
velocity, thermal pressure, magnetic field, gravitational
potential, and electrical conductivity, respectively. 
The above equations have to be closed by an
equation of state.

\section{Conformal properties of the MHD equations in an expanding
  Universe} 
\label{apx:FRW_MHD}

In the following we assemble the MHD equations in the FRW universe
with the scale factor $a$ and the Hubble parameter $H$
for (a) relativistic MHD, i.e. when photons are still coupled to the plasma
on the scale of the magnetic fluctuations, $l_{\gamma}\ll L$, and (b)
for non-relativistic MHD in the opposite limit $l_{\gamma}\gg L$. 
In both limits, the equation of state in the early Universe is well
approximated by being isothermal due to incompressibility in the limit (a)
and due to the efficiency of electron-photon Thomson scattering and the
associated cooling in the limit (b). We then show how in both limits
the MHD equations in the FRW background may be essentially reduced
to those in Minkowski space, when appropriate scalings with scale factor
of the physical quantities are introduced. For further details on the
derivation of the equations we refer the reader to JKO98.
To lowest non-trivial order in $1/\sigma$, the 
relativistic MHD
equations  are
\be
\dfdt{\rho}
 + \frac{1}{a}\,\nabla\cdot\Big(\left(\rho + p\right)\,\vv \Big)
     + 3\,H\,\left(\rho + p\right)
= 0
\label{eq:exp_e_cons}
\ee
\bea
 \left(\dfdt{}
   + \frac{1}{a}\,\left(\vv\cdot\nabla\right) + H \right) \vv
 + \frac{\vv}{\rho + p}\,\dfdt{p}
 + \frac{1}{a}\,\frac{\nabla\,p}{\rho + p}
 + \frac{1}{a}\left(\frac{\vB\times\left(\curl\vB\right)}
         {4\pi\,\left(\rho + p\right)}\right)
    & = & \nonumber \\
  \frac{\eta}{a^2}\,\left(\laplace\,\vv
      + \frac{1}{3}\nabla\left(\diver\vv\right) \right)\;,
 & &
\label{eq:exp_Euler_eq}
\eea
\be
\left(\dfdt{} + 2\,H\right)\vB =
\frac{1}{a}\,\curl\left(\vv\times\vB\right) 
 + \frac{1}{4\pi\sigma a^2}\nabla^2\vB\;,
\label{eq:exp_induction}
\ee
where we assumed that $\rho_{\sm{em}} \ll \rho = \rho_{\sm{fluid}}$
(here $\rho$ refers to internal energy density for radiation)
and we kept only terms of the lowest order in $v/c$. The shear
viscosity $\eta$ 
is given by \cite{wein72}
\be
\eta = \frac{4}{15}\frac{\pi^2}{30}\,g_t\,T^4\,l_{\sm{mfp}}/(\rho + p)
\quad,
\label{eq:shear_viscosity}
\ee
where $g_t$ is the number of relativistic degrees of freedom of
the particles with the longest mean free path $l_{\sm{mfp}}$.
Using the following the rescaled variables
(e.g. \cite{brand96b,subr98,enqv98}): 
\be
\begin{array}{rclcrclcrcl}
\trho & \equiv & \rho\,a^4     & \quad
\tp   & \equiv & p\,a^4        & \quad
\tvB  & \equiv & \vB\,a^2
\\
\tvv  & \equiv & \vv           & \quad
\tT   & \equiv & T\,a          & \quad
\tilde{\eta}  & \equiv & \eta\,a^{-1} & \quad
\\
d\tit & \equiv & dt\,a^{-1} & \quad
\tilde{\sigma} & \equiv & \sigma a
\end{array}
\label{eq:rad_rescaling}
\ee
the MHD equations~(\ref{eq:exp_e_cons}) -- (\ref{eq:exp_induction}) in
the radiation dominated universe (i.e. $a \propto t^{-1/2}$ and
$p = \rho/3$) can be written as:
\be
\dfdtt{\trho}
  + \nabla\cdot\Big(\left(\,\trho+\tp\,\right)\tvv\Big) = 0
\label{eq:rad_energy_cons}
\ee
\bea
\left(\dfdtt{} + \left(\tvv\cdot\nabla\right)\right)\,\tvv
   + \frac{\tvv}{\trho + \tp}\,\dfdtt{\tp}
   + \frac{\nabla\,\tp}{\trho + \tp}
   + \frac{\tvB\times\left(\curl\tvB\right)}
       {4\pi\left(\,\trho + \tp\,\right)}
 & = &
\nonumber \\
\tilde{\eta}
 \left(\laplace\,\tvv + \frac{1}{3}\nabla\left(\diver\tvv\right)
\right)
\label{eq:rad_Euler}
\eea
\be
\dfdtt{\tvB} =
  \curl\left(\tvv\times\tvB\right)  
+ \frac{1}{4\pi\sigma}\nabla^2\tvB\;,\quad,
\label{eq:rad_induction}
\ee
A similar rescaling transformation can be done in the matter dominated (MD)
regime (i.e. $a \propto t^{3/2}$ and $p \ll \rho$) using {\em
  super comoving variables} \cite{kerc01}: 
\be
\begin{array}{rclcrclcrcl}
\trho & \equiv & \rho\,a^3     & \quad
\tp   & \equiv & p\,a^4        & \quad
\tvB  & \equiv & \vB\,a^2
\\
\tvv  & \equiv & \vv\,a^{1/2}  & \quad
\tilde{\eta}  & \equiv & \eta\,a^{-1/2}  & \quad
d\tit & \equiv & dt\,a^{-3/2}
\\
\tH   & \equiv & a^{3/2}\,H & \quad
\tilde{\sigma} & \equiv & \sigma a^{1/2} 
\end{array}
\label{eq:matter_rescaling}
\ee
The transformations yield almost the form of the ordinary non-relativistic
MHD equations (cf. Eqs.~(\ref{eq:apx_matter}) --
\ref{eq:apx_induction}):
\bea
\dfdtt{\trho} + \nabla\cdot (\trho\,\tvv) & = & 0  \quad,
\label{eq:matter_conservation}
\\
\dfdtt{\tvv} + (\tvv\cdot\nabla)\,\tvv
+ \frac{1}{\trho}\,\nabla\,\tp
+ \frac{1}{4\pi\trho}\,\tvB\times(\curl\tvB)
& = & -\tvs \quad,
\label{eq:matter_Euler}
\\
\dfdtt{\tvB} - \nabla\times(\tvv\times\tvB) & = & 
\frac{1}{4\pi\sigma}\nabla^2\tvB\;,\quad.
\label{eq:matter_induction}
\eea
Here, the dissipation term is
\bea
\tvs & = & \half\,\tH\,\tvv
 - \tilde{\eta}
\,\left(\laplace\,\tvv
      + \frac{1}{3}\nabla\left(\diver\tvv\right) \right)
\label{eq:tildediss}
\eea
where $\rho$ is again matter density. Here the term
$\half\,\tH\,\tvv$ in Eq.~(\ref{eq:tildediss}) represents the
only difference to the MHD equations in Minkowski space and may
be interpreted as a drag term. In particular, fluid momentum dissipation due to
a homogeneous photon background with $l_{\gamma}>>L$, i.e. photon drag, 
is described by the addition of a term $-\alpha v$ on the RHS of 
Eq.~(\ref{eq:exp_Euler_eq}) due to free-streaming photons
(and the dropping of the terms proportional
to shear viscosity $\eta$ due to diffusing photons). In the scaled
variables this leads to the following dissipation term
\be
\tvs = \left(\half\,\tH\ + \tilde{\alpha}\right)\tvv
\ee
with $\tilde{\alpha} = a^{3/2}\alpha$.

\section{Dissipation of Energy and Helicity}
\label{apx:dissipation}

With a homogeneous density and pressure distribution the total energy
density is given by 
\be
E = \frac{\rho +p}{2}\frac{1}{V}
\int_{V}\,\di^3x\, \left(\vv^2 + \vvA^2\right)\, ,
\label{eq:total_energy}
\ee
where $\vvA = \vB/\sqrt{4\pi(\rho+p)}$ is the Alfv\'en velocity. The
magnetic helicity density is 
\be
\He \equiv \frac{1}{V}\int_V\,\di^3x\,\vA\cdot\vB \, ,
\label{eq:mag_helicity}
\ee
where $\vA$ is the vector potential, i.e. $\vB = \curl\vA$.
Using the MHD equations of Appx.~\ref{apx:general_MHD} the time
evolution of the above quantities is given by (up to surface terms)
\bea
V\,\frac{\di E}{\di t} & = & -\eta\left(\rho + p\right)\,
                     \int_V \di^3 x\,\left(\nabla\times\vv\right)^2
                    -\frac{1}{(4\pi)^2\sigma}\int_V \di^3 x\,
                     \left(\nabla\times\vB\right)^2\, ,
\label{eq:energy_decay}
\\
V\,\frac{\di \He}{\di t}  & = & -\frac{1}{2\pi\sigma}\,
    \int_V \di^3 x\,\vB\cdot(\curl\vB)\, ,
\label{eq:helicity_decay}
\eea
In the case when fluid momentum dissipation occurs by free-streaming
particles rather than diffusing particles, i.e. $\eta\,\laplace\,\vv
\to -\alpha\,\vv$, the first integrand (and prefactor) of the RHS of
Eq.~(\ref{eq:energy_decay}) must be replaced by
$\alpha\,\rho\,\vv^2$. Note that in the case of {\em ideal} MDH
(i.e. $\sigma \to \infty$) the helicity~(\ref{eq:mag_helicity}) becomes
a conserved quantity.

In the radiation dominated regime the shear viscosity is given by
(cf. also JKO98)
\be
\eta  = \frac{1}{5}\frac{g_t}{g_f}\,l_{\sm{mfp}}\, ,
\ee
with $l_{\sm{mfp}}$ being the mean free path of neutrinos or
photons with statistical weight $g_t$, and where $g_f$ denotes
the statistical weight of the total fluid energy radiation density. 
Here, the photon mean free path, $l_{\sm{mfp}}^{\gamma} =
1/(\sigma_{\sm{T}}\, n_e)$, as measured in comoving units, is
\be
l^{\gamma}_{\sm{mfp},c} \approx 1.77\,\Mpc\, X_e^{-1}\, 
   \left(\frac{\Omega_{\sm{b}}h^2}{0.02}\right)^{-1} \, 
   \left(\frac{T}{0.26\,\eV}\right)^{-2}\, , 
\label{eq:photon_mfp}
\ee
for $T \simle 20\keV$. For the neutrino mean free path we assume
$l_{\sm{mfp}}^{\nu} \simeq 1/(G_F^2\, T^2\,(n_l +n_q))$ (with the Fermi
constant $G_F \approx 1.1663\times 10^{-5}\,\GeV^{-2}$ and the number
densities $n_l$ and $n_q$ of relativistic leptons and quarks,
respectively). We find at $T = 2.6\,\MeV$ the comoving value
\be
l^{\nu}_{\sm{mfp},c} \simeq 6.7\,\pc\, 
    \left(\frac{g_l+g_q}{8.75}\right)^{-1}
    \left(\frac{T}{2.6\,\MeV}\right)^{-4}\, ,
\label{eq:nu_mfp}
\ee
where $g_l = (7/8)\,10$ and $g_q = 0$ has been assumed. Note that
when using Eq.~(\ref{eq:nu_mfp}) for the computation of $\alpha_{\nu}$
care has to be taken to not only include
scattering but also neutrino annihilation (cf.~\cite{jeda94}).
On the other hand, neutrino self-scattering does not contribute such
that, below the QCD-transition, $g_l$ should be effectively reduced
to $3.5$ (only $e^{\pm}$.
We use the following drag coefficients for neutrinos and
photons, respectively~\cite{jeda94,peeb65}
\bea
\alpha_{\nu} & \simeq &
   \frac{g_{\nu}}{g_f}\,
   \frac{1}{l_{\sm{mfp}}^{\nu}}\, ,
\label{eq:drag_nu}
\\
\alpha_{\gamma} & \simeq & \frac{4}{3}\, 
    \frac{\rho_{\gamma}}{\rho_{\sm{\sm{b}}}} \, 
    \frac{1}{l^{\gamma}_{\sm{mfp}}} 
  \approx 
    \frac{4}{3}\,X_e\,
    \frac{\sigma_{\sm{T}}\rho_{\gamma}}{m_p}\, ,
\label{eq:drag_gamma}
\eea
with $\sigma_{\sm{T}}$ the Thomson cross section, $m_p$ the proton mass,
and $\rho_{\gamma}$, $\rho_{b}$ denoting photon- and baryon- density,
respectively.
In the high temperature regime ($1\,\MeV \simle T \simle m_W$) the
electrical conductivity $\sigma$ is given by
\cite{ahon96}  
\be
0.76\,T \; \simle \; \sigma \; \simle \; 6.7\,T \quad,
\label{eq:conductivity_high}
\ee
where the larger value refers to the upper temperature bound. 
At temperatures below the electron mass the conductivity becomes
\cite{jack75} 
\be
\sigma = \frac{\alpha\,n_e\,\tau_c}{m_e} \simeq
\frac{m_e}{\alpha\,\ln\Lambda} 
\left(\frac{2}{\pi}\frac{T}{m_e}\right)^{3/2} \quad,
\label{eq:conductivity_low}
\ee
where $\tau_c$ is the mean time between two collisions, $\Lambda =
\frac{1}{6\pi^{1/2}} \frac{1}{\alpha^{1/2}}
\left( \frac{m_e^3}{n_e} \right)^{1/2} \left( \frac{T}{m_e} \right)$,
and $\alpha$, $m_e$, and $n_e$ are fine structure constant, electron mass
and electron density, respectively.
The magnetic Prandtl number $P_m$
which gives the relative importance of the kinetic and magnetic
diffusion is very large in the early universe,
\be
P_m = 4\pi\,\eta\,\sigma \simeq 
 2.9\times 10^8\,\left(\frac{\keV}{T}\right)^{3/2}  \,
\ee
for $T < m_e$. This allows to neglect the dissipation of magnetic
field energy due to finite conductivity.
We use the ion-neutral momentum transfer at low temperatures
given by~\cite{Draine:1983}
\be
\alpha_{\sm{in}} \approx 3.2\times 10^{-9}\, \sec^{-1} \, \,
\biggl(\frac{n_H}{\rm cm^{-3}}\biggr)
  = \alpha_{\sm{in}} \, \Omega_{\sm{b}}h^2\,
\biggl(\frac{a}{a_0}\biggr)^{-3}
\label{eq:ion_neutral_drag}
\ee
where $\alpha_{\sm{in}} \equiv 4.4\times 10^{-14}$s$^{-1}$
(note $\alpha_{in} \neq\alpha_{ni}$),
$n_H\approx n_{\sm{b}}$ denotes hydrogen density, and $a_0$ is the present day
scale factor. 
The hydrogen mean free path $l^H_{\sm{mfp}}$
after recombination is determined
by hydrogen-hydrogen elastic scattering. Scattering on electrons (or protons)
may be neglected in computing $l^H_{\sm{mfp}}$ due to the small degree of
ionization (i.e. $X_e\approx 4\times 10^{-4}$).
Assuming a temperature independent~\footnote{Elastic atomic scattering 
  cross sections usually increase slightly with decreasing temperature}, 
cross section which is approximately 
$\sigma_{HH}\approx\pi a_{\sm{B}}^2$, where $a_{\sm{B}} = 5.29\times
10^{-9}\,\cm$ is the Bohr radius one finds
\be
l^{H}_{\sm{mfp},c} \approx 9.9\times 10^{-3}\,\pc\,
\left(\frac{\sigma_{HH}}{10^{-16}\,\cm^2}\right)^{-1}
\left(\frac{T}{0.259\,\eV}\right)^{-2}
\ee
for the comoving mean free path. Shear viscosity due to hydrogen-hydrogen
elastic scattering may be estimated by
\be
\eta_{HH}\sim \frac{1}{3}\, v_H^{\sm{th}}\, l^H_{\sm{mfp}} \approx
8.0\times 10^{18}\,\frac{\cm^2}{\sec}\, 
\left(\frac{\sigma_{HH}}{10^{-16}\,\cm^2}\right)^{-1}
\left(\frac{T}{0.259\,\eV}\right)^{-5/2}
\ee
where $v_H^{\sm{th}} = \sqrt{3\,T/m_p}$ denotes hydrogen thermal velocity.
Finally we estimate the mean free path of electrons in the plasma.
After recombination, and with an effective cross section
$\sigma\sim (\alpha/T_e)^2$~\cite{Shu:1982}, where $\alpha$ is the
fine structure constant and $T_e$ the electron temperature we find
\be
l^{e}_{\sm{mfp},c} \sim 10^{-2}\,\pc
\ee
between the epochs with redshift $z\approx 1100$ and $z\approx 100$. Note
that the comoving mean free path is independent of temperature. Below
redshift $z\approx 100$ the mean free path decreases even further due
to a more rapid decrease in electron temperature than photon temperature.
We take an constant ionization fraction after recombination of
\be
X_e\approx 4\times 10^{-4}
\ee
neglecting residual dependencies on $\Omega_{\sm{b}}$ and $\Omega$.

\section{Generation of Helical Fields}
\label{apx:helicity}

To excite a stochastic magnetic field with or without initial
helicity we choose a coordinate system in $k$-space useful for helical
fields with the orthogonal unit vectors
$\{\ve_+, \ve_-,\hat{\vk}\}$ (see e.g. \cite{jack75}).
By expanding the Fourier transformed vector potential $\Ah$ in this
basis, i.e.
\be
\Ak = A^+_{\vk}\,\ve_+ + A^-_{\vk}\,\ve_- + A^k_{\vk}\,\hat{\vk}
\ee
one obtains the magnetic field in the new basis 
\be
\Bk = -i\,\vk\times\Ak = - k\left(\,A^+_{\vk}\,\ve_+ -
          A^-_{\vk}\,\ve_-\right) \quad.
\label{eq:bk_new}
\ee
With this set of basis vectors the magnetic field spectra are the
given by (cf. also \cite{chris01})
\be
|\Bk|^2 = k^2 \left(\,|A^+_{\vk}|^2 + |A^-_{\vk}|^2 \,\right) \quad,
\label{eq:bk2_from_ak}
\ee
whereas the magnetic helicity becomes
\be
\He = \frac{1}{(2\pi)^3}\,\int\,\di^3k\, \Ak^{\ast}\cdot\Bk
  = \frac{1}{(2\pi)^3}\,\int\,\di^3k\,H_{\vk}
\label{eq:helicity_by_abk}
\ee
with
\be
H_{\vk} \equiv \Ak^{\ast}\cdot\Bk = 
 - k\,\left(\,|A^+_{\vk}|^2 - |A^-_{\vk}|^2 \,\right) \quad.
\label{eq:hk}
\ee
Note, that this choice of coordinate system 
reflects also that the helicity (\ref{eq:helicity_by_abk}) is a well
defined physical quantity as it is gauge independent, i.e. independent
of $A^k_{\vk}$.
To ensure a real vector potential $\vA(\vx)$ and from that a real
magnetic field $\vB(\vx)$ the $A^{\pm}_{\vk}$ have to
fulfill the relation 
\be
(A^{\pm}_{\vk})^{\ast} = -A^{\pm}_{-\vk} \quad.
\ee
The magnetic helicity can be of either sign but the magnitude
$|H_{\vk}|$ is limited by the relation 
\be
|H_{\vk}| \leq k^{-1}\,|\Bk|^2 \quad.
\ee
A magnetic field is said to be maximally helical if the equals sign in
the above equation holds. 
From the relations (\ref{eq:bk2_from_ak}) and (\ref{eq:hk}) it can be
seen that the strength of the magnetic field can be chosen
independently of the magnetic helicity (in this approach one can
consider either $(A^+_{\vk}\,,A^-_{\vk})$ or $(B_{\vk}\,,H_{\vk})$ as
independent variables). This allows to excite stochastic magnetic
fields with arbitrary helicity. 
To excite stochastic magnetic fields with a fractional helicity we
choose
\be
A^{-}_{\vk} \equiv \sqrt{f}\,A^{+}_{\vk} \quad,
\ee
where $f \in [0,1]$. This convention leads to
\be
|H_{\vk}| = \frac{1}{k} \, |\Bk|^2 \,
  \frac{1-f}{1+f} \quad.
\label{eq:frac_helicity}
\ee
for the magnitude of the helicity in terms of the magnetic field.
From equation~(\ref{eq:frac_helicity}) it can be seen that
$(1-f)/(1+f)$ is the fraction of the maximal helicity
magnitude $H_{\sm{max}} = k^{-1}\,|\Bk|^2$. This can be used to adjust
the magnetic helicity to an arbitrary magnitude.

Note, that the choice of exciting magnetic fields with a fractional
helicity~(\ref{eq:frac_helicity}) is not unique. This particular
choice just reduces the amplitude of the helicity spectra (compared
to that of the maximal helicity spectra) by a factor of
$(1-f)/(1+f)$. For non-maximal helicity it is also possible that the
helicity spectra $H_{\vk}$ do not follow the spectra of the magnetic
field $|\Bk|^2$, but are rather independently distributed in $k$-space. The
particular choice of the implementation of a fractional helicity
may influence the evolution of magnetic fields. 

\bef
\showone{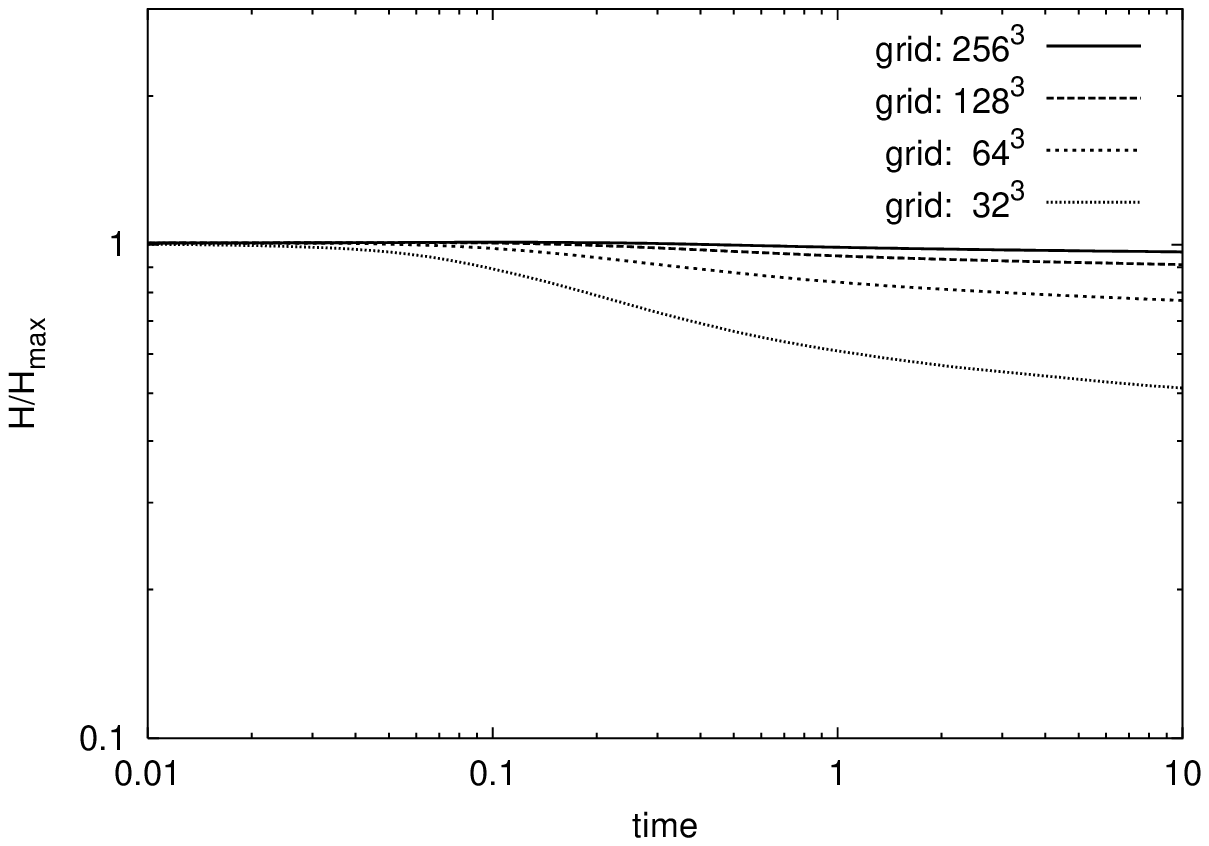}
\caption{Time evolution of the magnetic helicity $\He$ in the case of
  a maximal helical magnetic field. The loss of helicity is due to
  magnetic diffusion, which is solely due to numerical diffusion which
  can be seen by the resolution study.}
\label{fig:h_by_time}
\eef

\section{Numerical Methods}
\label{apx:numerics}

We performed the numerical simulation using
ZEUS3D~\cite{stone92a,stone92b,clar94}~\footnote{Informations and the
code to download are available from the Laboratory for Computational
Astrophysics (LCA) at
http://zeus.ncsa.uiuc.edu/lca\_intro\_zeus3d.html.}.  All simulations
were performed with periodic boundary conditions. This mimics an
infinitely large volume, where the surface integrals of the MHD
variables around the entire box vanish exactly.  Furthermore, we
extended the code for the purposes of our studies.  We used Gaussian
random field for the initial fluctuations of the magnetic components
with {\em zero mean}. To ensure a divergence-free magnetic field this
is done by exciting modes of the vector potential $\Ak$ in $k$-space
in the following way: The complex vector potential $\Ak =
(\hat{A}^1_{\vk},\hat{A}^2_{\vk},\hat{A}^3_{\vk})$ is generated by \be
\hat{A}^i_{\vk} = |\hat{A}^i_{\vk}|\,e^{i\,\varphi_{\vk}} \qquad i \in
[1,2,3] \quad,
\label{eq:apx_vpotential}
\ee
where the amplitudes $|\hat{A}^i|$ are randomly selected using a
Gaussian distribution, i.e.
\be
P(|\hat{A}^i_{\vk}|) =
  \frac{1}{\sqrt{2\pi}\,\sigma_{\vk}} \,
  \exp{\left\{-\frac{|\hat{A}^i_{\vk}|^2}{2\,\sigma_{\vk}^2}\right\}}
  \quad,
\label{eq:apx_Gauss}
\ee
and the phases $\varphi_{\vk}$ are randomly selected with an uniform
distribution from the interval $[0,2\pi]$. The amplitudes are related
to the variance $\sigma_{\vk}$ by 
\be
|\hat{A}^i_k|^2 \propto \sigma_k^2 \propto k^n \quad,
\label{apx:spectra}
\ee
where we assumed an isotropic universe, i.e. $\hat{A}^i_{\vk} =
\hat{A}^i_k$. These modes were excited up to a cut-off $k_c$.  The
initial stochastic velocity field is generated in the same way as the
initial magnetic field described above. In addition one can either
generate the stochastic velocity field $\vv$ by using
equation~(\ref{eq:apx_Gauss}) directly, or one can generate a {\em
divergence-free} velocity field by first exciting a vector potential
$\vA$ and then computing $\vv = \curl\vA$. The latter avoids strong
density perturbations.

\end{document}